\documentclass[manuscript,acmsmall,screen,dvipsnames,table]{acmart} %Previous Template
\acmJournal{TOSEM}
\acmVolume{X}
\acmNumber{X}
\acmArticle{X}
\acmMonth{8}

\acmISBN{978-1-4503-XXXX-X/18/06}

\usepackage{paralist}
\usepackage{graphicx}
\usepackage[inkscapelatex=false]{svg}
\usepackage{xspace}
\usepackage{booktabs}
\usepackage{braket}
\usepackage{amsmath}
\usepackage[most,breakable]{tcolorbox}
\usepackage{hyperref}
\usepackage{tikz}
\usetikzlibrary{arrows.meta,patterns}
\usepackage{subcaption}
\usepackage{multirow}
\usepackage[inline]{enumitem}

\usepackage[ruled,vlined,linesnumbered]{algorithm2e}
\usepackage{algorithmic}

\usepackage{amsthm}
\theoremstyle{definition}
\newtheorem{definition}{Definition}

\usepackage{xcolor} % Required for color support

\definecolor{myblue1}{HTML}{CFE1F2}
\definecolor{myblue2}{HTML}{B6D4E9}
\definecolor{myblue3}{HTML}{93C4DE}
\definecolor{myblue4}{HTML}{6AADD5}
\definecolor{myblue5}{HTML}{4A97C9}
\definecolor{myblue6}{HTML}{2D7DBB}
\definecolor{myblue7}{HTML}{1764AB}

\usepackage{lipsum}
\usetikzlibrary{quantikz2, positioning, arrows.meta, shapes.misc}
\hypersetup{
colorlinks = true, % false: boxed links; true: colored links
hidelinks,
linkcolor=black, % color of internal links
citecolor=black, % color of links to bibliography
urlcolor=black % color of external links
}

\newcommand{\easytodetect}{\textit{easy-to-detect}\xspace}
\newcommand{\highorder}{higher order\xspace}
\newcommand{\firstorder}{first order\xspace}
\newcommand{\secondorder}{second order\xspace}
\newcommand{\numMutations}{\ensuremath{n}\xspace}

\newcommand{\R}{\ensuremath{R}\xspace}
\newcommand{\rx}{\ensuremath{\R_x}\xspace}
\newcommand{\ry}{\ensuremath{\R_y}\xspace}
\newcommand{\rz}{\ensuremath{\R_z}\xspace}
\newcommand{\U}{\ensuremath{U}\xspace}

\newcommand{\addrxparam}{\ensuremath{Add\_R_{x}}\xspace}
\newcommand{\addryparam}{\ensuremath{Add\_R_{y}}\xspace}
\newcommand{\addrzparam}{\ensuremath{Add\_R_{z}}\xspace}
\newcommand{\adduparam}{\ensuremath{Add\_U}\xspace}
\newcommand{\replaceuparam}{\ensuremath{Replace\_U}\xspace}

\newcommand{\effectmeasure}{\ensuremath{\mathit{EffectMeasure}}\xspace}
\newcommand{\detectionRatio}{\ensuremath{\mathit{DetectionRatio}}\xspace}

\newcommand{\quantumcircuit}{\ensuremath{\mathit{C}}\xspace}
\newcommand{\mutantcircuit}{\ensuremath{\mathit{C'}}\xspace}

\newcommand{\initcirc}{\ensuremath{\mathit{i}}\xspace}
\newcommand{\measurecirc}{\ensuremath{\mathit{m}}\xspace}

\newcommand{\individual}{\ensuremath{\mathit{I}}\xspace}

\newcommand{\mutOp}{\ensuremath{\mathit{M}}\xspace}
\newcommand{\mutationposition}{\ensuremath{\mathit{P}}\xspace}

\newcommand{\outputstate}{\ensuremath{\mathit{o}}\xspace}
\newcommand{\OutputStates}{\ensuremath{\mathit{O}}\xspace}

\newcommand{\probabilityorigin}{\ensuremath{\mathit{D_{\quantumcircuit}}}\xspace}
\newcommand{\probabilitymutant}{\ensuremath{\mathit{D_{\mutantcircuit}}}\xspace}

\newcommand{\completeTS}{\ensuremath{\mathit{CTS}}\xspace}
\newcommand{\testsuite}{\ensuremath{\mathit{TS}}\xspace}
\newcommand{\testcase}{\ensuremath{\mathit{tc}}\xspace}
\newcommand{\testcasedetecting}{\ensuremath{\mathit{tcd}}\xspace}
\newcommand{\numtestcases}{\ensuremath{\mathit{\#tc}}\xspace}

\newcommand{\tcdetectionsetfom}{\ensuremath{\testcasedetecting_{\mathit{fom}}}\xspace}
\newcommand{\tcdetectionsethom}{\ensuremath{\testcasedetecting_{\mathit{hom}}}\xspace}
\newcommand{\Detect}{\ensuremath{\mathit{Detect}}\xspace}
\newcommand{\Hellinger}{\ensuremath{\mathit{Hellinger}}\xspace}
\newcommand{\StatisticalTest}{\ensuremath{\mathit{StatisticalTest}}\xspace}
\newcommand{\Distance}{\ensuremath{\mathit{Distance}}\xspace}

\newcommand{\muskit}{Muskit\xspace}
\newcommand{\qmutpy}{QMutPy\xspace}
\newcommand{\randomsearch}{RS\xspace}
\newcommand{\geneticalgorithm}{GA\xspace}
\newcommand{\hillclimbing}{HC\xspace}
\newcommand{\oneplusone}{(1+1)-EA\xspace}

\newcommand{\approach}{QUMUG\xspace}

\title{Search-Based Generation of Undetected Quantum Circuit Mutants}
\author{Eñaut Mendiluze Usandizaga}
\orcid{0009-0007-3315-1664}
\email{enaut@simula.no}
\affiliation{
\institution{Simula Research Laboratory \& Oslo Metropolitan University}
\city{Oslo}
\country{Norway}
}
\author{Thomas Laurent}
\orcid{0000-0002-0953-774X}
\email{tlaurent@tcd.ie}
\affiliation{
\institution{RI Lero \& Trinity College Dublin, School of Computer Science and Statistics}
\city{Dublin}
\country{Ireland}
}
\author{Paolo Arcaini}
\orcid{0000-0002-6253-4062}
\email{arcaini@nii.ac.jp}
\affiliation{
\institution{National Institute of Informatics}
\city{Tokyo}
\country{Japan}
}
\author{Shaukat Ali}
\orcid{0000-0002-9979-3519}
\email{shaukat@simula.no}
\affiliation{
\institution{Simula Research Laboratory \& Oslo Metropolitan University}
\city{Oslo}
\country{Norway}
}

\begin{document}

\begin{abstract}
% Quantum software testing has gained increasing attention in recent years, with q
Quantum mutation analysis is emerging as an essential technique for evaluating test suites due to the limited availability of real faulty quantum programs. However, existing quantum mutation analysis tools use fixed gate-based mutations, resulting in mutants that are easy to detect, which reduces their effectiveness in assessing the quality of test suites. We propose \approach, a search-based approach for generating challenging mutants by utilising parameterisable quantum gates. \approach employs search algorithms to optimise mutation parameters and find non-equivalent mutants passing a given test suite. In our evaluation over 30 quantum programs, \approach produced mutants that are three times more challenging than the mutants generated by existing tools. Among the four evaluated search algorithms, the genetic algorithm was the most effective, generating an average of 494 undetected mutants per program with a 99.67\% success rate and 94.3\% non-equivalent ratio. The generated mutants demonstrated their effectiveness by requiring the addition of five times more test cases to the test suite than the mutants generated by existing tools. We also analysed the behaviour of \highorder mutants in quantum circuits, and showed that while \firstorder mutations are more effective for enhancing the test suite, \highorder mutants highlight the need for new unique test cases.% that would not be considered by using only \firstorder mutants.
\end{abstract}



\maketitle

\section{Introduction}\label{sec:intro}

Quantum computing is a developing field that holds the potential to address complex problems faster than traditional classical computing~\cite{quantumSpeedup}. 
However, quantum computing also introduces new challenges, particularly in ensuring the reliability of quantum software~\cite{qseRoadmapTOSEM2025,zhao2021LandscapesAndHorizons,quantumTestingRoadmapTOSEM2025}. For instance, quantum programs often exhibit non-deterministic behaviour, and their internal states cannot be observed without collapsing the quantum state.
As a result, there is a pressing need for innovative software testing techniques that are tailored to the unique characteristics of quantum computing, such as superposition and entanglement~\cite{nielsen2010quantum,yanofsky2008quantum}.

In recent years, various software testing techniques have been developed for quantum software~\cite{quantumTestingRoadmapTOSEM2025,delaBarrera2022,QST_SOTA,stInQuantumWorldIEEEComputer2026}. 
It is essential to evaluate the effectiveness of these techniques, which requires the establishment of testing benchmarks using faulty programs~\cite{qbugs,zhao2021bugs4q}. Unfortunately, such benchmarks are scarce in the realm of quantum computing.
% However, such benchmarks remain scarce in the domain of quantum computing. To address this lack of benchmarks, mutation analysis has been employed with the objective of generating testing benchmarks~\cite{Mendiluze2021,QmutPy}.
To mitigate this scarcity, researchers have turned to mutation analysis to systematically generate benchmarks of faulty programs~\cite{Mendiluze2021,QmutPy}. For quantum software, mutation analysis involves generating ``mutants'' of quantum programs by applying gate-based operations, resulting in modified versions that introduce specific faults. This technique enables the creation of a larger set of comprehensive benchmarks for assessing quantum software testing techniques.

In a previous study~\cite{MendiluzeUsandizaga2025}, we performed an analysis of the mutants generated by \muskit~\cite{Mendiluze2021}. We found that \muskit's mutation operators introduce changes that have too big of an effect on the quantum programs outputs. This leads to mutants that behave completely differently from the original program, making them \easytodetect, with most being caught by even a single test case. Therefore, mutation scores are usually very high, not because test suites are particularly effective, but because the mutants themselves are trivial to detect.
% Consequently, test suites tend to achieve unrealistically high mutation scores, which no longer reflect their true quality. 
Moreover, in the context of mutation testing, since these \easytodetect mutants are already covered by existing tests, they do not encourage the development of new or stronger test cases~\cite{MutationOrigin}.

%This inflates mutation scores, making them appear high not because the test suites are particularly effective, but because the mutants themselves are trivial to detect.
%This means that most test suites achieve a high mutation score, and 
%that the mutation score does not reflect the strength of a test suite. 
%Furthermore, in the context of mutation testing, since all mutants have already been detected, these mutants do not contribute to the development of new test cases for the test suite~\cite{MutationOrigin}.

In this paper, we introduce {\it Quantum Undetected MUtant Generation} (\approach), an approach that directly addresses these limitations. Unlike quantum mutation tools available in the literature (i.e., \muskit and \qmutpy), \approach focuses on generating mutants that remain undetected by a test suite. By doing so, we are able to generate mutants that represent faults not covered by the current test suite, revealing a deficiency in it.

We propose novel mutation operators utilising parameterisable quantum gates, which allow us to control the magnitude of the applied effect. 
%Previous quantum mutation tools have employed parameterisable gates as well, but instead of adjusting the parameters, they used fixed default values. 
These gates offer a large range of parameters, providing us with a wide variety of mutants; however, they also create a very large search space to explore. To address this issue, we utilise search-based algorithms to fine-tune these parameters and to determine where in the circuit the mutations should be applied. 
%This approach enables us to generate a set of mutants that contribute to the improvement of the test suite. 
Furthermore, we introduce the possibility to incorporate multiple mutations within the same circuit to generate \highorder mutants, a concept widely used in classical mutation analysis and that has demonstrated potential in test suite assessment~\cite{empiricalfirstandsecondorder}.
%with its superior cost-benefit ratio and reduced the possibility of generating equivalent mutants~\cite{empiricalfirstandsecondorder}.

%a concept widely used in classical mutation analysis and has shown the ability to generate new types of faults or show relations between them, but has never before been applied in quantum mutation analysis.

We evaluate \approach using a total of 30 quantum programs that vary in type (six different algorithms) and number of qubits (3 to 7). 
To perform a comprehensive evaluation of \approach and validate the generated mutants, we simulate a mutation testing scenario: we generate mutants based on a given test suite and evaluate whether they require the inclusion of new tests to be detected. We compare the performance of four different search algorithms: genetic algorithm (\geneticalgorithm), hill climbing (\hillclimbing), 1+1 evolutionary algorithm (\oneplusone), and random search (\randomsearch), in finding mutants requiring new test cases.

The evaluation demonstrates that \geneticalgorithm achieved the best results for generating mutants that are not detected by the test suite given as input and that are not equivalent: it generated an average of 494 mutants per program, with a 94\% non-equivalent ratio. It achieved a 99.67\% success rate across all programs, failing to generate non-equivalent mutants that pass the test suite only once out of 300 executions. Moreover, the results indicate that \approach, using any of the four search algorithms, generated more challenging mutants than the baselines, with mutants detected by three times fewer test cases than the ones generated by the baselines. \geneticalgorithm showed the best results in terms of generating mutants that contribute to the improvement of the test suite, requiring a median of five different test cases to detect all of them. In contrast, the mutants generated by the baseline tools could be detected by simply adding one new test case to the test suite.

We also analysed the behaviour of \highorder mutants for quantum programs, showing that while they contained, in proportion, fewer equivalent mutants than \firstorder mutants, they were less challenging (i.e., they were detected by more test cases). The generated \highorder mutants required fewer test cases to be detected, contributing less to the improvement of the test suite. Still, the \highorder mutants generated by \approach were better than the mutants generated by the tools in the literature. We also studied interactions between the mutations introduced into the same circuit, which showed that most \highorder mutants were generated by combining mutations in different circuit positions: affecting different qubits or gates. The analysis of test cases detecting \highorder mutants showed that detection is not always directly related to the test cases detecting each individual mutation. The results show that interactions between mutations can either increase or decrease a mutation's effect and even make it detectable in cases where they would not be on their own. This suggests that, as in classical mutation testing, there could be subtle \highorder quantum mutants which represent unique faults that can not be represented with \firstorder mutants, and thus need the inclusion of specific test cases to detect them.

To summarise, the contributions of this paper are:
\begin{inparaenum}[(1)]
\item \approach, a novel approach that introduces new parametric mutation operators capable of generating more challenging mutants, which represent faults not detected by the given test suite, thereby contributing to its improvement.
\item The first implementation of \highorder mutants for quantum programs, leading to interesting relationships among mutations and proving that they require different test cases to be detected compared to \firstorder mutants.
\end{inparaenum}
Through these contributions, we provide researchers and developers with a novel approach for enhancing their test suites by offering a wider, more challenging variety of mutants that represent a broader range of faults.

The paper is structured as follows. Sections~\ref{sec:background} and \ref{sec:relatedwork} present the background and related work, respectively. We present the methodology of the approach in Section~\ref{sec:methodology}, followed by the experiment design in Section~\ref{sec:experiment}. Finally, we introduce the experimental results in Section~\ref{sec:results} and provide a more detailed discussion in Section~\ref{sec:discussion}. We conclude the paper in Section~\ref{sec:Conclusionsandfuture}.

\section{Background}\label{sec:background}
This section introduces some essential concepts to understand the paper. First, Section~\ref{subsec:backgroundQuantum} presents fundamental concepts of quantum computing, including quantum circuits and quantum software testing. Next, Section~\ref{subsec:backgroundMutation} covers the principles of mutation analysis, while Section~\ref{subsec:backgroundQuMutation} explains how mutation analysis adapts to the quantum realm.

\subsection{Quantum Computing}\label{subsec:backgroundQuantum}
Quantum computing leverages the principles of quantum mechanics to perform calculations significantly faster than classical computers, known as {\it quantum speedup}~\cite{quantumSpeedup}. Quantum computers use {\it quantum bits} (or {\it qubits}) as their basic units for computations. Qubits exhibit unique properties, such as {\it superposition} and {\it entanglement}, which enable calculations to be performed in ways that classical bits cannot. Quantum entanglement refers to the ability of qubits to be interconnected regardless of the distance between them~\cite{yanofsky2008quantum}. This implies that, by measuring the state of one qubit, the state of another one can be directly determined. 
Superposition refers to the ability of a qubit to exist in multiple states simultaneously~\cite{nielsen2010quantum}. Unlike a classical bit, which can only be either 0 or 1 at any given time, a qubit can remain in a superposition state of both 0 and 1. 
By leveraging these quantum principles, quantum programs can process information more efficiently.
A commonly used notation represents the state of a qubit as $\ket{\psi} = \alpha \ket{0} + \beta \ket{1}$, where $\alpha$ and $\beta$ are complex probability amplitudes~\cite{nielsen2010quantum}.
In order to be able to interpret the state of a qubit in a classical way, the qubit must be measured, which collapses its state into one of the computational basis states $\ket{0}$ or $\ket{1}$. This collapse destroys the original superposition, meaning the qubit is no longer in a quantum state.

\begin{figure}[!tb]
\centering
\includegraphics[width=0.25\linewidth]{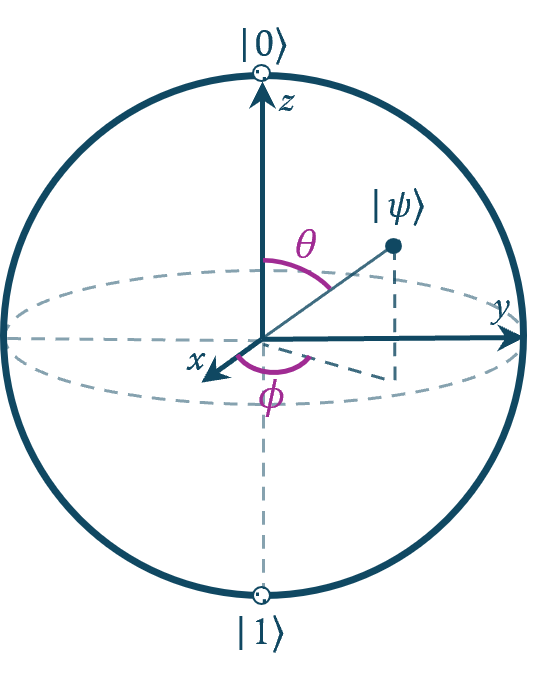}
\caption{Bloch sphere qubit representation}
\Description{Bloch sphere qubit representation}
\label{fig:blochSphere}
\end{figure}

Figure~\ref{fig:blochSphere} shows a {\it Bloch sphere}, a common visual representation of the state of a qubit~\cite{nielsen2010quantum}.
The Bloch sphere is defined with three axes, labelled $X$, $Y$, and $Z$, each corresponding to a different possible measurement base. The measurement basis specifies the axis along which the state is projected during this process~\cite{nielsen2010quantum}. The default measurement base is the $Z$-basis, where the measurement distinguishes between $\ket{0}$ and $\ket{1}$. By applying appropriate single-qubit rotations before measurement, we can effectively change the measurement axis to align with the $X$- or $Y$-basis. In this representation, the position of the qubit’s state, $\ket{\psi}$, is parametrised by three angular variables: $\theta$, $\phi$, and $\lambda$. The angle $\theta$ defines the position along the sphere’s vertical axis, determining the relative weighting between the computational basis states $\ket{0}$ and $\ket{1}$. The angle $\phi$ specifies the rotation around the sphere’s equator, thereby setting the relative phase between these basis states. The third parameter $\lambda$ corresponds to an overall phase rotation. Each of these parameters controls a distinct type of rotation or ``flip'' on the Bloch sphere.

\subsubsection{Quantum Circuits}
\begin{figure}[!tb]
\centering
\includegraphics[width=0.6\linewidth]{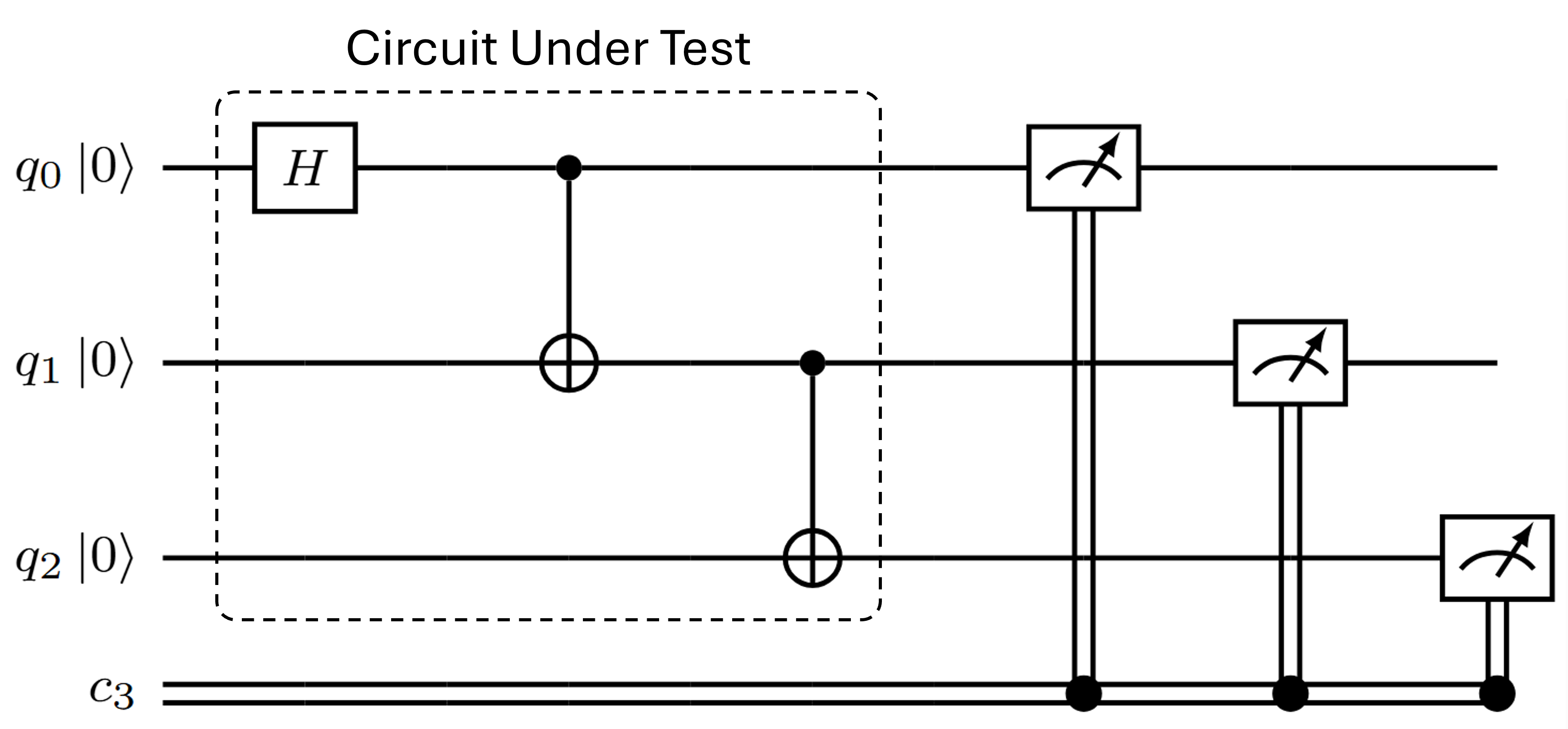}
% \includesvg[width=0.6\linewidth]{basicCircuit.svg}
% \begin{tikzpicture}
% \node (circuit) at (0,0) {
% \begin{quantikz}
% \lstick{$q_0\ket{0}$} & \gate{H} & \qw & \ctrl{1} & \qw & \qw & \qw & \meter{} & \qw & \qw \\
% \lstick{$q_1\ket{0}$} & \qw & \qw & \targ{} & \qw & \ctrl{1} & \qw & \qw & \meter{} & \qw \\
% \lstick{$q_2\ket{0}$} & \qw & \qw & \qw & \qw & \targ{} & \qw & \qw & \qw & \meter{} \\
% % Spacer for visual separation (optional)
% \lstick{$c_3$} & \cw & \cw & \cw & \cw & \cw & \cw & \cwbend{-3} & \cwbend{-2} & \cwbend{-1}
% % \lstick{$c_{[0\text{-}2]}$} & \cw & \cw & \cw & \cw & \cw & \cw & \cwbend{-3} & \cwbend{-2} & \cwbend{-1}
% \end{quantikz}
% };
% \end{tikzpicture}
\caption{Quantum circuit representation of a GHZ Entangled state of three qubits}
\Description{Quantum circuit representation of a GHZ Entangled state of three qubits}
\label{fig:QC}
\end{figure}

{\it Quantum programs} are typically represented as {\it quantum circuits}, which describe the sequence of quantum operations applied to qubits. These operations are implemented through {\it quantum gates}.
Quantum gates can be understood as rotations on the qubit’s Bloch sphere, allowing qubits to change between different states~\cite{nielsen2010quantum}. 
Figure~\ref{fig:QC} shows an example of a quantum circuit consisting of three qubits, $q_0$, $q_1$, and $q_2$. 
As shown in the figure, quantum gates may operate on a single qubit ({\it single-qubit gates}) or on multiple qubits ({\it multi-qubit gates}). For instance, a Hadamard gate, represented as $H$ in the figure, is a single-qubit gate commonly used at the beginning of quantum circuits to create superpositioned states. The controlled-NOT ($CX$) gate, indicated by a dot and a cross in the figure, is a two-qubit operation that acts as a conditional gate, where the state of one qubit defines whether the operation in the second qubit is performed. By combining these gates, the circuit in Figure~\ref{fig:QC} shows the representation of a Greenberger–Horne–Zeilinger (GHZ) entanglement state, a maximally entangled state, where the state of each individual qubit is entirely dependent on the states of the other qubits~\cite{ghz}. The circuit also includes a group of classical bits, referred to as $c3$ at the bottom part of the figure, known as {\it classical registers}.

At the end of the circuit, the {\it measurement operations} are shown, represented by a square with a semicircle and an arrow that connects the qubits to the classical registers. These operations collapse the quantum state and translate it into the classical registers present in the circuit. 

This work uses specific parametric gates that allow us to control the desired effect in the circuit.
To manipulate the orientation of a qubit’s state, quantum circuits employ single-qubit {\it rotational gates} (\R-gates): $\rx(\theta)$, $\ry(\theta)$, and $\rz(\phi)$, which rotate the qubit around the $X$, $Y$, and $Z$ axes of the Bloch sphere, respectively. By combining such rotations, any arbitrary single-qubit transformation can be achieved. In addition to the rotational gates, there is also the {\it universal gate} \U, which can represent any single-qubit gate. The \U gate consists of three parameters $\theta$, $\phi$, and $\lambda$; by setting specific values to the three parameters, it is capable of applying a specific rotation to the Bloch sphere. When constructing a circuit, we typically use predefined gates that represent specific versions of these parametric gates mentioned. Table~\ref{tab:gates} lists some of the most common single-qubit gates and their translation to a combination of \R-gates, and to a \U gate~\cite{RGatesConversion}. 
\begin{table}[!tb]
\centering
\caption{Standard quantum gates and their representation as \R-gates and a \U gate}
\label{tab:gates}
\setlength{\tabcolsep}{10pt}
\begin{tabular}{@{}ccc@{}}
\toprule
\multirow{2}{*}{Single-qubit gate} & \multicolumn{2}{c}{Equivalent representation as}\\
\cmidrule{2-3}
& \R-gates & \U gate\\
\midrule
$H$ & $\rx(\pi)\cdot \ry(\frac{\pi}{2})$ & $\U(\frac{\pi}{2},0,\pi)$ \\
$X$ & $\rx(\pi)$ & $\U(\pi,-\frac{\pi}{2},\frac{\pi}{2})$\\
$Y$ & $\ry(\pi)$ & $\U(\pi,0,0)$\\
$Z$ & $\rz(\pi)$ & $\U(0,0,\pi)$\\
$T$ & $\rz(\frac{\pi}{4})$ & $\U(0,0,\frac{\pi}{4})$\\
$S$ & $\rz(\frac{\pi}{2})$ & $\U(0,0,\frac{\pi}{2})$\\ 
$Sdg$ & $\rz(-\frac{\pi}{2})$ & $\U(0,0,-\frac{\pi}{2})$\\ 
\bottomrule
\end{tabular}
\end{table}

\subsubsection{Quantum Software Testing}\label{sec:backgroundQST}
Software testing for classical systems is a well-established field. However, in the quantum realm, the need for effective testing approaches remains, as quantum software introduces challenges that classical testing techniques cannot directly address. 
Indeed, many methods in classical software testing rely on checking the internal state of a program or copying its state, which, in the case of quantum software, cannot be done due to the collapse of the quantum state~\cite{yanofsky2008quantum}. To address these issues, new software testing techniques have been developed in recent years with the goal of testing quantum software~\cite{quantumTestingRoadmapTOSEM2025,delaBarrera2022,QST_SOTA,genTestsQPSSBSE2021,metamorphic,mutation-based,pauliStringsASE2024,testingQuantumICST2021,CTQuantumQRS2021,qucatASE23tool,QuraTestASE2023,WangICST2021,LongTOSEM2024,projectionBased,back1996evolutionary,honarvar2020property}.

When testing quantum software, we refer to the section of the circuit where the main computations are performed as the {\it circuit under test}. Typically, this section is located right after the qubit initialisation and before measurements, representing the core of the quantum algorithm. Figure~\ref{fig:QC} highlights the circuit under test in our example circuit, which consists of the $H$ gate on $q_0$ and two $CX$ gates connecting all three qubits.

The input of a quantum circuit is considered to be the initialisation of the qubits, as presented in the literature~\cite{quitoASE21tool}. 
A qubit can be initialised in either a {\it classical} or a {\it quantum} state. 
To initialise a qubit in a classical state, it is sufficient to set its initial state to either 0 or 1 when creating the quantum circuit. The compiler will then translate this classical state using the appropriate quantum gates. In classical initialisation, we consider the possible initialisations based on the binary combinations of all qubits, totalling \(2^{\#\text{Qubits}}\). For quantum initialisation, a separate quantum circuit must be constructed to generate the desired quantum state. This circuit is then introduced before the circuit under test. By doing this, the desired quantum state is achieved prior to executing the circuit under test~\cite{QuraTestASE2023}.

After all the gates in the circuit have been executed, each qubit is measured to determine the output of the circuit. The output of a quantum circuit is represented as a distribution of all possible outcomes obtained at the end of the circuit. This distribution shows the probabilities of each possible measurement outcome when the circuit is executed. It is generated by running the circuit multiple times (i.e., running multiple \textit{shots}) with the same input and recording each of the outcomes, which indicates how often each possible output state occurs. The resulting output distribution is then compared with the oracle output distribution to assess the circuit's correctness.

\subsection{Mutation Analysis}\label{subsec:backgroundMutation}
{\it Mutation analysis} is a technique widely used in classical software testing to evaluate testing methods. Essentially, mutation analysis involves injecting artificial faults in the programs under test and determining whether tests can detect those faults~\cite{MutationOrigin}. 

Mutation analysis employs several operators to systematically modify the program under test, thus creating faulty versions known as {\it mutants}~\cite{MutationOrigin}. Each of the mutants can have a different number of mutations, i.e., modifications, introduced in the program. Mutants with only one modification in the program are referred to as \firstorder, while \highorder mutants have more than one~\cite{MutationSurvey}. These mutants serve as a benchmark of faulty programs that can be used to assess the effectiveness of the tests.

The ability of a test suite to identify faults is typically measured by the {\it mutation score}, which indicates the proportion of detected mutants out of all mutants~\cite{MutationOrigin}. A higher mutation score reflects a better test suite, as it is capable of detecting more faults. It is important to note that the mutation score depends not only on the test suite itself but also on the mutants used to compute it and their quality. Therefore, different mutation tools implementing different mutation operators could return different mutation scores for the same test suite~\cite{AmmannICST2014,KintisSCAM2016,LaurentICST2017}.

Mutants are typically classified based on whether the tests can detect them. Inside the category of undetected mutants, we can find {\it equivalent mutants}, which represent a significant challenge in mutation analysis. Even though these mutants have been syntactically altered, they exhibit identical behaviour to the original program for all inputs, making them undetectable. Equivalent mutants pose a problem as they misleadingly decrease the mutation score, making test suites appear worse than what they actually are, and require considerable developer time to analyse~\cite{MutationSurvey,schuler2013covering}.

The detected mutants can be classified into different groups. On one hand, we have clusters of {\it duplicated mutants}, which, although they may contain different changes, behave the same. As a result, detecting one of these mutants will also result in the detection of all others in the cluster. If a large group of duplicated mutants is detected, it can artificially increase the mutation score without reflecting a real increase in fault detection capability.

{\it Subsumed mutants} do not behave exactly the same but are always detected when another mutant, i.e., a \textit{subsuming mutant}, is detected. Subsumed mutants also inflate the mutation score artificially, as the faults they represent are already represented by the subsuming mutant.

\subsection{Quantum Mutation Analysis}\label{subsec:backgroundQuMutation}
Similar to other software testing techniques, mutation analysis also needs adaptation for quantum software due to its unique characteristics. To facilitate this adaptation, two quantum mutation analysis tools have been developed in the literature, \muskit~\cite{Mendiluze2021} and \qmutpy~\cite{QmutPy}.

\begin{figure}[!tb]
\centering
% \includesvg[width=0.6\linewidth]{QuantumOperators.svg}
\begin{tikzpicture}
 \node (circuit) at (0,0) {
 \begin{quantikz}
 \lstick{$q_0\ket{0}$} & \qw & \qw & \ctrl{1} & \qw & \qw & \qw & \meter{} & \qw & \qw \\
 \lstick{$q_1\ket{0}$} & \qw & \qw & \targ{} & \qw & \ctrl{1} & \qw & \qw & \meter{} & \qw \\
 \lstick{$q_2\ket{0}$} & \qw & \qw & \qw & \qw & \targ{} & \qw & \qw & \qw & \meter{} \\
 \lstick{$c_3$} & \cw & \cw & \cw & \cw & \cw & \cw & \cwbend{-3} & \cwbend{-2} & \cwbend{-1}
 \end{quantikz}
 };
 % Remove CX gate
 \draw[myblue7, very thick] (-0.3, 0.4) -- (-0.06, -1); % Top-left to bottom-right
 \draw[myblue7, very thick] (-0.06, 0.4) -- (-0.3, -1); % Top-left to bottom-right
 % Replace H gate
 \node[draw=myblue7, thick, minimum size=0.55cm, text=myblue7] at (-1.2,2.2) (Sgate) {\textit{S}};
 \node[draw=black, fill=white, thick, minimum size=0.55cm, text=black] at (-2.3,1.27) (Hgate) {\textit{H}};
 \draw[{Latex}-{Latex}, draw=myblue7, thick] (Sgate.west) -- (Hgate.north);
 % Add T gate
 \node[draw=myblue7,dotted, fill=white, thick, minimum size=0.55cm, text=blue] at (-2.3,-0.7) (Space) {};
 \node[draw=myblue7, fill=white, thick, minimum size=0.55cm, text=myblue7] at (-2.3,-2) (TGate) {\textit{T}};
 \draw[{Latex}-, draw=myblue7, thick] (Space.south) -- (TGate.north);
 %Positions
 \node[text=myblue7, font=\small] at (-2.65,1) {1};
 \node[text=myblue7, font=\small] at (-1.75,0.1) {2};
 \node[text=myblue7, font=\small] at (-0.4,-0.9) {3};
 \node[text=myblue7, font=\small] at (0.9,1) {4};
 \node[text=myblue7, font=\small] at (2.1,0) {5};
 \node[text=myblue7, font=\small] at (3.3,-1) {6};
\end{tikzpicture}
\caption{Quantum Circuit Operators}
\Description{Quantum circuit operators}
\label{fig:operators}
\end{figure}
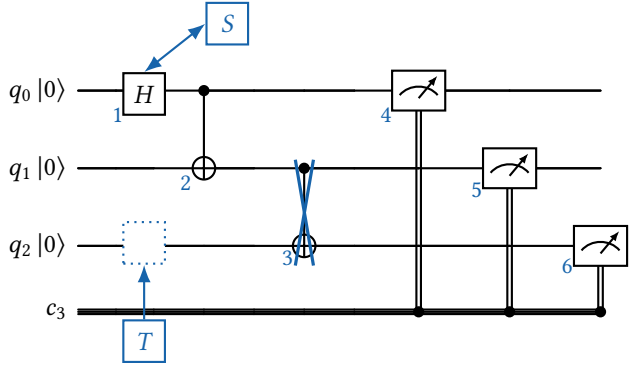

These tools treat quantum circuits as the original programs and modify them by applying gate-based operations. Both tools perform changes in the circuit by either adding a new gate, removing an existing gate, or replacing an existing gate with another one. Figure~\ref{fig:operators} illustrates examples of the application of these mutation operators to the circuit shown in Figure~\ref{fig:QC}.
The numbers in the figure indicate the positions where a mutation can be applied in the circuit; for the addition operator, the position is given by the gate following the gap where the mutation will be introduced.
The \textit{Add} operator introduces a new quantum gate to the circuit at a specific location, in this case introducing a \textit{T} gate in qubit $q_2$, position 3. The \textit{Remove} operator eliminates a gate already present in the circuit, as it is the case in the figure for the \textit{CX} gate connecting qubits $q_1$ and $q_2$ in position 3. The \textit{Replace} operator substitutes an existing gate with another; the figure shows the substitution of the \textit{H} gate in qubit $q_0$, position 1, with an \textit{S} gate.

\section{Related Work}\label{sec:relatedwork}
In this section, we review related work. First, we discuss prior research on both classical and quantum mutation analysis in Section~\ref{subsec:relatedMutation}. Then, we explore other relevant studies that utilise quantum circuit mutation operators in Section~\ref{subsec:relatedQuantumMutation}.

\subsection{Classical Mutation Analysis}\label{subsec:relatedMutation}
Mutation analysis has been widely studied in classical software engineering as an effective technique for assessing test suite quality and improving fault detection~\cite{MutationOrigin}. Surveys by Jia and Harman~\cite{MutationSurvey} and by Papadakis et al.~\cite{MutationSurvey2} have highlighted its relevance, emphasising both its strong fault-revealing capability and its role in evaluating test adequacy. These studies also identify key challenges, such as the high computational cost of executing mutants and the problem of equivalent mutants, which have driven continued research into more efficient mutation analysis techniques. A recent empirical study by Sánchez et al.~\cite{mutationtestingpractice} demonstrates the practical relevance of mutation analysis in real-world development environments. The authors conducted a qualitative study with open-source developers and reported high satisfaction with mutation analysis, highlighting its usefulness in improving test quality.

As shown in the literature review by Silva et al.~\cite{SILVA201719}, several approaches have already been developed for classical software to generate mutants using search-based techniques. Their analysis shows that while the genetic algorithm is the most popular and widely evaluated method, specialised algorithms often demonstrate better stability and convergence rates in specific contexts. We decided to use a genetic algorithm in this work because it serves as a well-established and widely accepted baseline for search-based optimisation, allowing us to focus on evaluating the proposed mutation operators for quantum programs. For instance, the work by Schwarz et al.~\cite{schwarz2011breeding} tries to address the limitations of traditional mutation testing by proposing a genetic algorithm to generate high-impact, undetected mutants. Similarly to \approach, Schwarz et al. prioritise maximising the ``unique impact'' of undetected mutations to identify deficiencies in test suites. Their evaluation demonstrated that an evolutionary approach can increase the percentage of undetected mutants while simultaneously increasing their average impact by a factor of five. In our case, we focus on finding undetected mutants for a given test suite by applying a novel parametric mutation operator for quantum circuit mutants and using search algorithms to tune the mutations.

When referring to \highorder mutants, Papadakis and Malevris~\cite{empiricalfirstandsecondorder} empirically compared first and second order mutation strategies to address the high computational and manual costs of mutation analysis. Their study of industrial C programs revealed that while using all \firstorder mutants remains the most effective choice for fault detection, \highorder mutation strategies achieve a superior cost-benefit ratio. The authors found that \highorder mutation reduces the number of equivalent mutants by 80\% and requires 30\% fewer test cases with a moderate drop in fault detection capabilities compared to \firstorder mutation. Recognising the potential of \highorder mutants demonstrated in classical software for test suite assessment, we aimed to explore how these mutants would behave in the quantum domain.

When evaluating the difficulty of detecting a mutant, Visser~\cite{whathard} considers various factors affecting mutant detectability through an exhaustive analysis of the input space. Unlike previous studies that relied on fixed test suites and manual analysis, Visser's approach eliminates test-suite bias, allowing for a more accurate calculation of the ``kill ratio'' of mutants. In our case, as explained better in Section~\ref{subsec:ts}, we use a comprehensive test suite that includes all possible classical inputs to quantum circuits, as well as an equal number of quantum inputs. We also consider the three output measurement basis, resulting in what we believe to be a thorough test suite for assessing the difficulty of mutant detection.

\subsection{Quantum Mutation Analysis}\label{subsec:relatedQuantumMutation}
Current research in quantum mutation analysis has focused on generating mutants by introducing gate-based modifications to quantum circuits. 
Mendiluze Usandizaga et al.~\cite{Mendiluze2021} introduced \muskit, a mutant generator for quantum circuits that creates mutants by adding, removing, or replacing gates within the circuit, regardless of the gate type. On the other hand, \qmutpy, developed by Fortunato et al.~\cite{QmutPy,QMutPy2,QMutPy3}, is a quantum circuit mutation tool built upon the classic Python mutation tool, MutPy. In addition to the mutation operators introduced by \muskit, \qmutpy also offers two new mutation operators related to measurement gates. \qmutpy generates mutants based on the syntactic equivalence of gates, which, as the authors state, helps reduce the overall number of generated mutants. Both tools aim to generate mutants to evaluate quantum software testing techniques and to apply mutations to Qiskit code. 
%\muskit functions as an independent tool with two parts, one generating mutants directly from the circuit and saving them as separate circuits for future use, and the other being responsible for executing the test cases on them, utilising pre-defined test cases. In contrast, \qmutpy requires the user to provide test cases for the circuits to perform both mutant generation and execution.

Mutants generated by these tools are often redundant and easy to detect, providing little help when assessing test suites. \approach, on the contrary, aims to generate mutations that are undetected by a given test suite, thereby forcing improvements to the test suite and making the resulting mutations more valuable for testing. To that end, we present new parametric mutation operators based on those provided by previous tools. To mutate a circuit, \approach adds or replaces parametric gates and tunes them by using a search algorithm to find the optimal mutants for our goal. Also, compared to current tools, \approach can generate \highorder mutants, which have not yet been considered for quantum circuits.

Mendiluze Usandizaga et al.~\cite{MendiluzeUsandizaga2025} conducted an empirical evaluation of current literature on mutants, analysing the behaviour of over 700,000 mutants across 385 quantum circuits. The authors examined how various characteristics of the circuits, algorithms, and mutations influenced the detection of these mutants, as well as the interactions between different characteristics. This study provides valuable insights into mutant behaviour in quantum circuits. 
%The authors also introduced a new metric to assess a mutant's survival capability, called the ``survival rate''. This metric indicates how many mutants sharing the same characteristics pass a given test case.
%In our case, contrary to Mendiluze Usandizaga et al., when assessing the generated mutants, we base our difficulty in detecting a mutant on the number of test cases that can detect them, as our focus is on the test suite provided by the user. 

In our case, instead of assessing the mutants themselves and analysing the effects of individual mutation characteristics, as Mendiluze Usandizaga et al. do, we propose \approach, an approach to generate mutants that are undetected by a test suite. Also, as their study is based on the mutants generated by \muskit, there is no analysis on \highorder mutants, which we provide as part of our study.

Some works focus on other aspects of mutation analysis. For instance, Wang et al.~\cite{mutation-based} present MutTG, a mutation-based test generation approach for finding the minimum number of test cases that maximise the number of detected mutants. The authors define mutant difficulty based on the number of test cases that detect a mutant. While this concept is related to mutant detectability, our objective focuses on generating surviving mutants that are not detected by the existing test suite, whereas Wang et al. focus on generating test cases that maximise mutant detection.
%The authors define the difficulty of detecting a mutant based on the number of test cases that detect it. While our technique for measuring how challenging it is to detect a mutant is similar to the one used in MutTG, the objective of \approach is to identify and generate mutants not detected by the test suite, opposite to Wang et al., where their objective is to find the tests that detect most of the mutants. 
We believe that \approach could be complementary by offering more challenging mutants to assess the effectiveness of generated test suites.
%We consider that \approach could be complementary, as our mutants could help strengthen their test generation by generating more difficult mutants.

Kölle et al.~\cite{MutationGASynthesis} propose a quantum circuit synthesis approach that utilises a genetic algorithm and mutates quantum circuits using gate-based operations. In this case, the authors utilise the same addition, removal, and replacement operations as \muskit and \qmutpy, and introduce a new type: the swap operation. The mutation operations are not used directly to generate mutants for test assessment, but rather to help the search algorithm find optimised circuits. Authors conclude that the ``swap, addition, deletion'' strategy optimises quantum circuits most efficiently, producing near-optimal results and reducing circuit depth.

Ishimoto et al.~\cite{MutationBasedFL} applied quantum circuit mutations as a fault localisation strategy for quantum circuits. The authors utilised \qmutpy as the mutation tool to generate mutants and assigned a suspiciousness score to each statement in the code based on how test case detection changes when that statement is mutated. The authors demonstrate that quantum-specific mutation operators are more effective at localising faults in quantum circuits than classic mutation operators. 
%However, mutation-based fault localisation struggles when challenged with real quantum circuit faults, as the approach again focuses only on \firstorder mutations, making it hard to identify patterns in the code.

Mendiluze Usandizaga et al.~\cite{quantumRepICST2026} presented QRep, an automated quantum circuit repair approach by gate prioritisation. The authors implemented gate-based modifications, similar to mutations described in the literature, to faulty circuits using an iterative approach that prioritises quantum gates. The authors define a suspiciousness score based on how each gate's modification affects the circuit, and rank the gates by their likelihood of being faulty. The results show that QRep repairs 70\% of faults in circuits with up to 13 qubits, offering greater scalability than other methods in the literature. This study highlights that gate-based modifications can be applied to both quantum circuit fault localisation and repair. The authors also mention that future work may involve multi-gate modifications to the circuit to enable repair of additional faults.

\approach focuses solely on mutation analysis, with the main objective of generating mutants that contribute to the improvement of the test suite. However, as we present five new operators for generating mutations, we believe these operators could also be useful in other quantum software applications. Also, as mentioned in the work by Ishimoto et al.~\cite{MutationBasedFL}, the inclusion of \highorder mutations could help in several domains apart from mutation analysis and testing.

\section{\approach: Proposed search-based quantum mutants generator}\label{sec:methodology}

\approach is motivated by the fact that the quantum circuit mutations proposed in the literature drastically affect the circuit's behaviour. These alterations lead to substantial changes in the circuit's output, making any differences easy to observe, regardless of the tests employed. The majority of the tests trigger the fault, making the mutants trivial or \easytodetect. 
Based on this observation, we aim to create mutants that would produce minor effects on the circuit, so that they would require a stronger test suite to be detected. 
In this section, we will discuss the methodology and implementation of \approach.

\subsection{Overview}\label{subsec:approach}

We propose a new type of parameterisable mutation operators. 
Similar to existing quantum mutation tools \muskit and \qmutpy, these operators generate mutants by applying gate-based modifications; however, rather than using predefined or fixed gates, \approach inserts parameterisable gates into the circuit.
These gates allow us to control the magnitude of the quantum phase applied to a state, providing control over the strength of the mutation.

Another positive aspect of this type of mutation is the diversity it provides. The tools available in the literature are restricted by fixed angle values, resulting in a limited number of mutants. In our case, since the parameters adjusted in the gates are floating-point values, this presents infinite possibilities for creating new mutants that have not been previously explored, each with a specific phase shift. However, this also leads to an infinite number of mutants, which makes their execution infeasible. Therefore, we introduce a search-based algorithm to find the optimal parameters for each mutation. This strategy reduces the number of mutants and discards those that introduce drastic changes to the circuit's behaviour.

In addition, we explore the possibility of introducing multiple mutations within the same quantum circuit, thereby generating \highorder quantum mutants. To the best of our knowledge, there has not been a method for applying this concept in quantum circuits, even though it is a widely used strategy in classical software testing~\cite{MutationSurvey}. 
By generating \highorder mutants, we also expand our search space, as we could explore the interactions of different mutations in the same or different qubits.

\subsection{Formal Definitions}\label{subsec:perliminaries}
We now introduce the notation and fundamental concepts used in the remainder of the section.

\begin{definition}[Circuit Under Test]
As introduced in Section~\ref{sec:backgroundQST}, the \emph{circuit under test} represents the main computational component consisting of a sequence of quantum gates that capture the internal logic of the quantum computation. In our study, mutations are applied exclusively to the circuit under test. In the following, for the sake of brevity, we will call the circuit under test simply as \emph{circuit}. We define our \emph{original circuit} as \quantumcircuit, and the \emph{mutated circuit} as \mutantcircuit, which is the circuit obtained by applying one or more mutation operations to \quantumcircuit, resulting in a modified circuit structure. The initialisation and measurement components that define the execution context of \quantumcircuit and \mutantcircuit are handled separately through test cases (see Definition~\ref{def:testCase}).
\end{definition}

\begin{definition}[Test Case]\label{def:testCase}
A test case, denoted by $\testcase$, is an element that defines the execution context for \quantumcircuit. Formally, a \testcase is defined as a pair $\testcase = (\initcirc, \measurecirc)$, where \initcirc denotes the circuit initialisation applied prior to executing \quantumcircuit, and \measurecirc denotes the measurement basis used after execution to obtain the resulting output (e.g., $X$, $Y$, or $Z$ basis).
\end{definition}

\begin{definition}[Test Suite]
We define a test suite $\testsuite = \{\testcase_1, \ldots, \testcase_k\}$ as a set of test cases used to evaluate a circuit. In our context, \testsuite is used to determine whether a mutated circuit \mutantcircuit passes or fails the evaluation criteria by executing both \quantumcircuit and \mutantcircuit under the same \testsuite.
\end{definition}

\begin{definition}[Output Distribution]
The output distribution of a circuit \quantumcircuit for a test case \testcase is the probability distribution over measurement outcomes obtained by executing the circuit in the context specified by \testcase. Formally, it represents the probabilities of observing each possible measurement outcome \outputstate among all possible measurement outcomes \OutputStates:
\[
\probabilityorigin(\testcase)
=
\left\{
\Pr(\outputstate \mid \quantumcircuit,\testcase)
\;\middle|\;
\outputstate \in \OutputStates
\right\}
\qquad \text{where} \;\;
\sum_{\outputstate \in \OutputStates}
\Pr(\outputstate \mid \quantumcircuit,\testcase)=1.
\]
The original output distribution, denoted by $\probabilityorigin(\testcase)$, is obtained by executing the original circuit \quantumcircuit multiple times (shots) in the context defined by a test case $\testcase$. The mutant output distribution, denoted by $\probabilitymutant(\testcase)$, is obtained by executing a mutant circuit \mutantcircuit in the same context defined by \testcase.
\end{definition}

\begin{definition}[Mutant Detection]\label{def:mutantDetection}
Mutant detection refers to the process of determining whether \mutantcircuit can be distinguished from \quantumcircuit under a given \testcase.
For each test case $\testcase \in \testsuite$, a statistical test is applied to compare the original output distribution $\probabilityorigin(\testcase)$ with the mutant output distribution $\probabilitymutant(\testcase)$. A mutant is considered to be detected by a test case \testcase if the resulting p-value of the statistical test is below a predefined significance threshold $\alpha$, indicating a statistically significant difference between the two distributions.
\[
%\Detect(\probabilityorigin(\testcase), \probabilitymutant(\testcase))
\Detect(\quantumcircuit, \mutantcircuit, \testcase)
=
\begin{cases}
1, & \text{if }\quad \StatisticalTest(\probabilityorigin(\testcase),\probabilitymutant(\testcase)) < \alpha,\\
0, & \text{otherwise}.
\end{cases}
\]
\end{definition}

While mutant detection determines whether a statistically significant difference exists between the output distributions of the original and mutated circuits, it provides only a binary indication of their difference. To better assess these differences, we use a distance metric to measure the distance between the two distributions. This continuous measure helps detect subtle behavioural changes in mutants.

\begin{definition}[Distance Metric]\label{def:distanceMetric}
In our context, a \emph{distance metric} quantifies the similarity between two probability distributions, considering both the differences in their probabilities and the shapes of their distributions. We denote the distance for circuits \quantumcircuit and \mutantcircuit in the context specified by \testcase as $\Distance(\quantumcircuit, \mutantcircuit, \testcase)$. It is used to measure the distance between the original output distribution $\probabilityorigin(\testcase)$ and the mutant output distribution $\probabilitymutant(\testcase)$. It ranges from $0$ to $1$, where $0$ indicates matching distributions and $1$ indicates completely different distributions.
\end{definition}

% The Hellinger distance quantifies the similarity between two probability distributions, considering both the differences in their probabilities and the shapes of their distributions.
% %The Hellinger distance between two probability distributions \(\probabilityorigin\) and \(\probabilitymutant\) in the context specified by \testcase is defined as:
% In our context, we define the Hellinger distance for circuits \quantumcircuit and \mutantcircuit in the context specified by \testcase as:
% %
% \begin{equation*}
% %\Hellinger(\probabilityorigin(\testcase), \probabilitymutant(\testcase))
% \Hellinger(\quantumcircuit, \mutantcircuit, \testcase)
% = \frac{1}{\sqrt{2}} \sqrt{\sum_{\outputstate \in \OutputStates} \left(\sqrt{\Pr(\outputstate \mid \quantumcircuit,\testcase)} - \sqrt{\Pr(\outputstate \mid \mutantcircuit,\testcase)}\right)^2}
% \end{equation*}

% It is used to measure the distance between the original output distribution $\probabilityorigin(\testcase)$ and the mutant output distribution $\probabilitymutant(\testcase)$. It ranges from $0$ to $1$, where $0$ indicates matching distributions and $1$ indicates completely different distributions. 

\subsection{Mutation Operators}\label{subsec:mutOP}
In \approach, we use the three basic rotational gates (\rx, \ry, and \rz) along with the universal gate (\U). We chose these gates as they allow us to generate mutants with minimal impact on the circuit. 
Rotational gates enable us to specify the desired phase shift of a qubit in a specific direction without altering other directions. This capability allows us to specify the magnitude of the change applied to the qubit. The universal gate, on the other hand, can represent any single-qubit gate by applying specific angle shifts to all three phases simultaneously. We present five parameterisable mutation operators, listed below:
\begin{itemize}
\item \addrxparam: it inserts an \rx gate parameterised by $\theta$, introducing a rotation around the $X$-axis.
\item \addryparam: it inserts an \ry gate parameterised by $\theta$, introducing a rotation around the $Y$-axis.
\item \addrzparam: it inserts an \rz gate parameterised by $\phi$, introducing a rotation around the $Z$-axis.
\item \adduparam: it inserts a \U gate parameterised by $\theta, \phi, \lambda$, allowing for single-qubit state transformations.
\item \replaceuparam: it replaces an existing single-qubit gate with a \U gate parameterised by $\theta, \phi, \lambda$.
\end{itemize}

We chose to focus on adding a new gate rather than removing or replacing an existing one, except for the \U gate. This is because removing or replacing an existing gate can significantly affect the original circuit, leading to major alterations. Replacing a gate with a parameterised \U gate is included as an option since the \U gate can mimic the behaviour of any single-qubit gate. This option allows a close approximation of the replaced gate, which could lead to subtle deviations if tuned correctly.

\subsection{Individual Representation}\label{subsec:individual}
In \approach, an individual \individual represents a mutant \mutantcircuit derived from \quantumcircuit, encoded as an array of mutation parameters that are applied to \quantumcircuit to generate the resulting mutated circuit \mutantcircuit. In the case of a single mutation (i.e., a \firstorder mutant), the individual will consist of a single array that represents the five parameters needed to generate a mutation:
\begin{equation*}
I_{\mathit{firstOrder}} = [\mutationposition, \mutOp, \theta, \varphi, \lambda]
\end{equation*}

These parameters include Position \(\mutationposition \in \{1, \ldots, |\quantumcircuit|\}\), which indicates the location in \quantumcircuit where the mutation is applied; mutation operator $\mutOp \in \{\addrxparam,$ $\addryparam,$ $\addrzparam,$ $\adduparam,$ $\replaceuparam\}$, where \mutOp represents a possible mutation selected from the set of available mutation operators; and three phase values \(\theta, \varphi, \lambda \in [-180, 180]\), which represent the floating-point parameters controlling the phase settings. In the case of generation of \highorder mutants, the individual will consist of several arrays, depending on the desired number of mutations, each containing the same five variables: 
\begin{equation*}
I_{\mathit{highOrder}} = [[\mutationposition_1, \mutOp_1, \theta_1, \varphi_1, \lambda_1]\ldots[\mutationposition_{\numMutations}, \mutOp_{\numMutations}, \theta_{\numMutations}, \varphi_{\numMutations}, \lambda_{\numMutations}]]
\end{equation*}

In the following, with an abuse of notation, we will indicate an individual directly with the obtained mutant \mutantcircuit.

\subsection{Fitness Function}\label{subsec:fitnessfunc}
The fitness function measures the impact of mutations on the original circuit. The goal is to identify mutants that pass the test suite given in input (\testsuite) while still causing a behavioural change in the circuit. We first define \testcasedetecting as the number of test cases that detect the mutant \mutantcircuit:
\begin{equation}\label{eq:tcd}
\testcasedetecting(\mutantcircuit, \testsuite) =
%\sum_{\testcase \in \testsuite} \Detect(\probabilityorigin(\testcase), \probabilitymutant(\testcase))
\sum_{\testcase \in \testsuite} \Detect(\quantumcircuit, \mutantcircuit, \testcase)
\end{equation}

For each mutant, the sum of the distances of all test cases is also calculated, providing insight into how close the mutant is to passing the test cases. This is defined as \effectmeasure and is used in the fitness function to help guide the search towards the desired direction. 
\begin{equation*}
\effectmeasure(\mutantcircuit, \testsuite) =
\sum_{\testcase \in \testsuite} \Distance(\quantumcircuit, \mutantcircuit, \testcase)
\end{equation*}

Therefore, the fitness function to be minimised is defined as follows:
\begin{equation}\label{eq:fitnessFunction}
f(\mutantcircuit) =
\begin{cases}
\testcasedetecting(\mutantcircuit, \testsuite) \times \#\testcase + \effectmeasure(\mutantcircuit, \testsuite) & \text{if } \testcasedetecting(\mutantcircuit, \testsuite) > 0 \\
\#\testcase - \effectmeasure(\mutantcircuit, \testsuite) & \text{if } \testcasedetecting(\mutantcircuit, \testsuite) = 0
\end{cases}
\end{equation}

The fitness function is defined utilising two cases, each consisting of a method with two terms. This first case is applied when at least one test case is able to detect the mutant (i.e., $ \testcasedetecting(\mutantcircuit, \testsuite) > 0$) and starts by multiplying the number of tests that detect the mutant by the total number of tests in the given test suite (i.e., $\testcasedetecting(\mutantcircuit, \testsuite) \times \#\testcase$).
By doing so, the function assigns more relevance to each test case that passes, increasing its impact on the fitness function.
To guide the search towards the passing of the test cases, $\effectmeasure(\mutantcircuit, \testsuite)$ is added, providing some more insight into the effect of the introduced mutations.

The second case in the fitness function is utilised if all the test cases in the provided test suite pass, (i.e., $\testcasedetecting(\mutantcircuit, \testsuite) = 0$). 
In this case, the objective is to maximise the effect of the mutation, ensuring that the mutant is not equivalent, even after successfully passing all the tests in the given suite. 
To achieve this, $\effectmeasure(\mutantcircuit, \testsuite)$ is subtracted from the total number of given test cases $\#\testcase$, that way ensuring that the fitness function remains a minimisation function and will always produce a smaller value than the best outcome of the first case.

\section{Experiment Design}\label{sec:experiment}

In this section, we explain how we plan to evaluate \approach. First, in Section~\ref{subsec:RQ}, we define the research questions we aim to answer and outline the methods we will use to address them. Then, in Sections~\ref{subsec:subjectSystems} and \ref{subsec:ts}, we present the subject systems and the test suite \testsuite that will be used by \approach. In Section~\ref{subsec:validation}, we define how we assess the equivalence of mutants. In Section~\ref{subsec:comparedapproach}, we present the approaches compared in our experiments: Section~\ref{subsubsec:searchalg} describes the search algorithms used by \approach, and Section~\ref{subsubsec:baselines} the mutation approaches from the literature used as baselines. Then, in Section~\ref{subsec:experimentSetup}, we define the experimental setup used in our experiment. In Sections~\ref{subsec:metrics} and \ref{subsec:statistics}, we present the metrics and statistical tests used to evaluate the approach, respectively. Finally, in Section~\ref{subsec:execution}, we describe the execution environment utilised in the experiment. All detailed experiment results, code, and data are available in the online repository~\cite{ZenodoRepository}.

\subsection{Research Questions (RQs)}\label{subsec:RQ}
The objective of \approach is to generate mutants that represent faults not detected by the provided test suite, to help improve it. 
To assess both the effectiveness of the proposed generation strategy and the characteristics that influence the detectability of these mutants, we define the following research questions:
\begin{description}
\item \textbf{RQ1}: {\it How good is \approach in generating non-equivalent, undetected quantum circuit mutants that contribute to the improvement of the test suite?}
\begin{description}
\item \textbf{RQ1.1}: {\it How many non-equivalent mutants that pass the user test suite are generated?}
\item \textbf{RQ1.2}: {\it How challenging to be detected are the generated mutants?}
\item \textbf{RQ1.3}: {\it Do the generated mutants contribute to the inclusion of more test cases in the test suite?}
\end{description}

In this RQ, we assess whether \approach can generate mutants that are not detected by the test suite given as input and are not equivalent, and whether these mutants are more challenging and contribute more to improving the test suite than mutants generated with existing tools. This allows us to evaluate the practical value of the mutation generation strategy.

For that, in RQ1.1, we compare the ability of each search approach to generate undetected, non-equivalent mutants and compare them with those generated by tools in the literature. Then, in RQ1.2, we further analyse the generated mutants in terms of how challenging they are to detect compared to those generated by the tools in the literature. Finally, in RQ1.3, we assess how much the generated mutants contribute to improving the test suite, evaluating the need to add new test cases to the test suite given in input.

\item \textbf{RQ2}: {\it What is the impact of \highorder mutation on the generation and detection of quantum circuit mutants?}

\begin{description}
\item \textbf{RQ2.1}: {\it How do \highorder mutants compare to \firstorder mutants?}
\item \textbf{RQ2.2}: How do mutations in the same circuit interact with each other in \highorder mutants?
\item \textbf{RQ2.3}: {\it How do interactions between mutations affect test case detection of \highorder mutants?}
\end{description}

In RQ2, we examine the role of \highorder mutation in quantum mutation analysis. Here, we aim to determine whether combining multiple mutations results in mutants that are more or less challenging, and whether interactions among mutations lead to unique behavioural changes in quantum circuits.

To answer this question, we utilise the mutants generated by the best search algorithm in RQ1 and, first, in RQ2.1, we compare \firstorder mutants with \highorder mutants using the metrics from RQ1 to assess their effectiveness. Then, we further analyse the interactions of the mutations in \highorder mutants in terms of mutation characteristics and their combination in RQ2.2, as well as the test case detection for each mutation in RQ2.3.
\end{description}

\subsection{Subject Systems}\label{subsec:subjectSystems}

For our evaluation, we selected six quantum algorithms, each having five quantum circuit implementations that range from three to seven qubits, totalling 30 quantum circuits. We decided to use a minimum of three qubits in our circuits because one of the algorithms (VQE) cannot be represented with fewer qubits. We imposed an upper limit due to resource constraints. These circuits were sourced from MQT Bench~\cite{quetschlich2023mqtbench}, a widely recognised benchmark suite for quantum circuits.
The selection of these circuits was guided by the different output types outlined in~\cite{MendiluzeUsandizaga2025}, which categorise quantum algorithms into two groups: {\it dominant output} and {\it diverse output}. Dominant output algorithms aim to identify the quantum state with the highest probability, while diverse output algorithms focus on the distribution and relationships between the probabilities of multiple states. We chose three quantum algorithms from each category, ensuring a diverse set of circuits that represent various types of quantum circuits. The specific characteristics of the selected quantum circuits are summarised in Table~\ref{tab:program_characteristics}.
\begin{table}[!t]
\centering
\caption{Characteristics of the original benchmarks (i.e., quantum algorithms).}
\label{tab:program_characteristics}
\setlength{\tabcolsep}{5pt}
%\scriptsize
\begin{tabular}{@{}llccccc@{}}
\toprule
Algorithm & Output type & \# qubits & \# gates & \# single-qubit gates & \# multi-qubit gates & Depth\\
\midrule
\texttt{ae} & Dominant & 3-7 & 15-53 & 9-25 & 6-28 & 12-36\\
\texttt{ghz} & Diverse & 3-7 & 4-8 & 1 & 3-7 & 4-8 \\
\texttt{qft} & Diverse & 3-7 & 8-32 & 3-7 & 5-25 & 7-15 \\
\texttt{qpe} & Dominant & 3-7 & 9-36 & 5-13 & 4-23 & 7-16 \\
\texttt{vqe} & Dominant & 3-7 & 14-34 & 9-21 & 5-13 & 8-12 \\
\texttt{w-state} & Diverse & 3-7 & 10-26 & 5-13 & 5-13 & 7-15 \\
\bottomrule
\end{tabular}
\\
\smallskip
\begin{flushleft}
{\footnotesize Amplitude Estimation (\texttt{ae}); Greenberger-Horne-Zeilinger state (\texttt{ghz}); Quantum Fourier Transform (\texttt{qft}); Quantum Phase Estimation (\texttt{qpe}); Variational Quantum Eigensolver ( \texttt{vqe}); W-State (\texttt{w-state}).}
\end{flushleft}
\end{table}

\subsection{Test Suite}\label{subsec:ts}
The test suite \testsuite used in this study is defined by combining different quantum circuit initialisations and output measurement bases (see Definition~\ref{def:testCase}). First, we randomly generate 20\% of all possible classical input initialisations for each qubit number. This number was chosen based on preliminary experiments showing that a smaller number of inputs resulted in test suites being too weak. Moreover, this percentage was chosen to avoid excessive time consumption during the experiments, as the test suite needs to be executed several times during the search. Then, as the number of quantum initialisations can be infinite, we choose to generate a number of quantum initialisations equivalent to the number of classical initialisations generated. 

To generate these quantum states, we utilised a method proposed by Ye et al.~\cite{QuraTestASE2023}. Among the various methods suggested in~\cite{QuraTestASE2023} for creating quantum states with template-based quantum circuits, we opted for the UCNOT approach, as, according to their evaluation, it is the most effective for detecting mutants. This method uses a template of \U gates and $CX$ gates to connect all qubits and generate random quantum states. 

By combining both classical and quantum inputs, we achieve a number of possible initialisations proportional to the number of qubits in the circuit: $2^{\#Qubits} \times 0.4$. These inputs are not specific to any particular quantum algorithm but depend on the number of qubits in the circuit. Algorithms that use the same number of qubits will share identical inputs for test cases.

Regarding the circuit's output, we decided to increase the variability of our measurement approach. As each qubit can be measured in three different bases ($X$, $Y$, and $Z$), we decided to include all three bases in our test suite. This adds another layer of variability to our test cases. Consequently, each of the inputs mentioned earlier will produce three possible outputs, one for each base. As a result, the total number of test cases is $2^{\#\mathit{Qubits}} \times 0.4 \times 3$. 

To address the inherent uncertainty in quantum computing, each test was executed several times for a given number of shots. We performed a number of shots proportional to the number of qubits, calculated as $2^{\#\mathit{qubits}} \times 2$. This approach enabled us to obtain a good estimate of the output distribution.

\subsection{Non-equivalent Mutant Validation}\label{subsec:validation}
The objective of \approach is to generate mutants that represent faults not covered by \testsuite. Therefore, in the evaluation of \approach, it is necessary to distinguish between mutants that pass the test suite because they represent uncovered faults and mutants that are equivalent to the original circuit. Equivalent mutants do not provide useful information about the test suite, as their behaviour cannot be distinguished from the original circuit.

To assess mutant equivalence, we extend \testsuite to create a more {\it comprehensive test suite}, denoted as \completeTS. Since \completeTS is exclusively utilised for the external evaluation of the generated mutants and is not included in the search process, its execution time does not limit the approach. Therefore, we include all possible classical input initialisations left and the same number of quantum initialisations, also measured across all three bases. As a result, \completeTS is five times larger than \testsuite.

A generated mutant is considered non-equivalent if it is detected by at least one test case from \completeTS. In other words, if the comprehensive test suite is able to distinguish the mutant from the original circuit, the mutant is considered to exhibit an observable behavioural difference and, therefore, represents a fault not covered by the original test suite \testsuite. This validation process allows us to identify mutants that pass \testsuite while still representing behaviours that could be detected by additional testing.

\subsection{Compared Approaches}\label{subsec:comparedapproach}
In this section, we describe the approaches that will be compared to assess the effectiveness of \approach. First, in Section~\ref{subsubsec:searchalg}, we outline the four search algorithms that will be utilised by \approach. Then, in Section~\ref{subsubsec:baselines}, we discuss the state-of-the-art mutant generators that we will use as baselines for \approach.

\subsubsection{Search Algorithms used in \approach}\label{subsubsec:searchalg}
In this work, we evaluate the performance of \approach using four different search algorithms: Genetic Algorithm (\geneticalgorithm), which is included as it is one of the most widely used evolutionary algorithms in the literature; Hill Climbing (\hillclimbing), which we include as a local search strategy; 1+1 Evolutionary Algorithm (\oneplusone), which serves as a simple evolutionary baseline; and Random Search (\randomsearch) as a baseline without any learning or guided search process, providing a reference point for comparison. Each of these algorithms represents a distinct search strategy, allowing us to assess the behaviour of \approach under diverse optimisation dynamics.

Regarding the search operators, we needed to adjust some commonly used ones to fit our encoding and search. In the crossover step for \geneticalgorithm, the parents are combined using a personalised crossover operator designed specifically for our solution representation, a variation of the commonly used single-point crossover. Since individuals have different settings based on the number of mutations desired, we decided to implement a specific version of single-point crossover for each type. In cases where multiple mutations are desired, i.e., \highorder mutants, the individuals are split in a way that each mutation remains intact. Practically, this means that the points selected during the crossover operation can only occur between arrays, without splitting any specific sub-array and modifying individual mutations. For \firstorder mutations, since they consist solely of one mutation, the crossover point is chosen between the variables of that single mutation and combined with another to create a new mutation.

Regarding the mutation operator (for the three search algorithms that utilise it, i.e., \geneticalgorithm, \hillclimbing and \oneplusone), we utilised a normalised mutation operator that modifies the individual by selecting a random value from a normalised distribution. It is important to note that all external components remained consistent in all search algorithms, and the generated mutants were evaluated in the same manner.

\subsubsection{State-of-the-art Mutant Generators}\label{subsubsec:baselines}
We compare \approach with two existing approaches discussed in the literature \muskit~\cite{Mendiluze2021} and \qmutpy~\cite{QmutPy}. Both approaches systematically generate quantum circuit mutants by applying gate-based operations, with the sole objective of generating all possible mutants. Our primary objective is to assess how \easytodetect their mutants are, and to determine whether \approach can generate more challenging mutants. To achieve this, we generated mutants using both state-of-the-art approaches. Table~\ref{tab:MutantsBaselines} displays the total number of mutants obtained from each mutation generator when applied to the same selected quantum circuits.
\begin{table}[!t]
\centering
\caption{Total number of mutants produced by each baseline tool}
\label{tab:MutantsBaselines}
\setlength{\tabcolsep}{10pt}
\begin{tabular}{@{}lccc@{}}
\toprule
Algorithm & \muskit & \qmutpy & Total\\
\midrule
\texttt{ae} & 4150 & 2120 & 2535\\
\texttt{ghz} & 665 & 580 & 1245\\
\texttt{qft} & 2407 & 1754 & 4161\\
\texttt{qpe} & 2691 & 1998 & 4689\\
\texttt{vqe} & 2955 & 1560 & 4515\\
\texttt{w-state} & 2205 & 1380 & 3585\\
\midrule
Total & 15073 & 9392 & 24465\\
\bottomrule
\end{tabular}
\end{table}

For our evaluation, we select only the mutants that pass the test suite \testsuite given in input, as these are the mutants the user would be interested in. Then, we evaluate whether the remaining mutants are detected by \completeTS to determine if these mutants are equivalent or represent faults that were not detected by \testsuite. Then, both tools are compared against the search-based approaches utilising \approach.

\subsection{Experimental Setup}\label{subsec:experimentSetup}
To conduct our experiments, we need to establish several key parameters. Initially, we conducted a preliminary experiment using a small set of circuits with up to five qubits~\footnote{The results of this preliminary experiment can be found in the external repository containing all the results~\cite{ZenodoRepository}.}. This preliminary experiment allowed us to observe the behaviour of \geneticalgorithm and to fine-tune its parameters accordingly. We employed a grid search algorithm to identify the optimal parameters for \geneticalgorithm operators, which indicated that the optimal parameters for both \firstorder and \highorder mutants include a population size of 100, with a mutation probability (\textit{Mutpb}) of 0.5 and a crossover probability (\textit{Cxpb}) of 0.5. To maintain consistency in our experimental design, we allocated a budget of 100 generations for each execution. For \geneticalgorithm, this means 100 generations with a population of 100. To ensure a fair comparison with the other algorithms, we based the experiments on the number of evaluations, allocating a maximum of 10,000 evaluations (100 generations multiplied by 100 individuals in a population) for each search algorithm.

Also, a preliminary assessment of the mutation operators revealed that the vast majority of the selected angles for the non-detected mutants found in the search fell within the range of -0.5 to 0.5 for all the gate parameters. This observation allowed us to reduce the search space from the full 360 degrees to a range of 1 degree, specifically from -0.5 to 0.5, as the remaining values seemed to have too much impact and were therefore irrelevant to our case.

In this study, we limited the number of mutations included in our \highorder mutants to two, thereby creating \secondorder mutants. We made this decision due to time and resource constraints, and we believe that incorporating two mutations into the circuit adequately captures the potential for interactions among multiple mutations.

As part of the fitness evaluation (see Section~\ref{subsec:fitnessfunc}), we need to select a statistical test to identify mutant detection (see Definition~\ref{def:mutantDetection}) and a distance metric (see Definition~\ref{def:distanceMetric}) to quantify the difference between the original and mutated circuits for each test case. For the statistical test, we have chosen the {\it chi-square} test with a significance level $\alpha$ of 0.05, which is commonly used in the literature~\cite{quitoASE21tool, WangICST2021, Mendiluze2021, MendiluzeUsandizaga2025}. For the distance metric, we opted for the {\it Hellinger distance}, as it is one of the most frequently employed methods in the literature for measuring differences between output distributions~\cite{muqeet2024mitigating,pauliStringsASE2024,pontolillo2025ideal,robustmutationanalysisquantum}. The Hellinger distance ranges from $0$ to $1$, where $0$ indicates matching distributions and $1$ indicates completely different distributions and is defined as follows:
\begin{equation*}
\Hellinger(\quantumcircuit, \mutantcircuit, \testcase)
= \frac{1}{\sqrt{2}} \sqrt{\sum_{\outputstate \in \OutputStates} \left(\sqrt{\Pr(\outputstate \mid \quantumcircuit,\testcase)} - \sqrt{\Pr(\outputstate \mid \mutantcircuit,\testcase)}\right)^2}
\end{equation*}

To account for the stochasticity of \approach and the resource constraints inherent in executing quantum circuits, each experiment was repeated 10 times under identical experimental conditions. Note that the mutant generators in the literature used as baselines (see Section~\ref{subsubsec:baselines}), are deterministic in terms of the number of generated mutants but still contain a stochastic factor related to the gate parameters; thus, they were also executed 10 times for each circuit. Also, as the generation of \testsuite has a stochastic factor in selecting the inputs, we generated and stored 10 different \testsuite to ensure that all the approaches are compared using the same \testsuite in each run.

\subsection{Evaluation Metrics}\label{subsec:metrics}

To answer the research questions presented in Section~\ref{subsec:RQ}, this section describes how the effectiveness of \approach is evaluated.

\subsubsection{Non-equivalent mutants passing \testsuite}

To evaluate the effectiveness of the approaches in generating mutants that represent faults not covered by \testsuite, we measure the number of generated mutants that pass the test suite \testsuite and are classified as non-equivalent according to the validation process described in Section~\ref{subsec:validation}.

We report both the total number of mutants passing \testsuite and the total number of non-equivalent mutants. We also calculate the proportion of non-equivalent mutants among all mutants that pass \testsuite. The total number of non-equivalent mutants obtained provides information about the number of useful mutants generated by each approach, while the proportion of non-equivalent mutants among all mutants that pass \testsuite accounts for the possibility of obtaining equivalent mutants. These metrics are used to answer RQ1.1 and RQ2.1.

\subsubsection{Test case detection ratio}\label{sec:detectionRatio}

To evaluate how challenging the generated mutants are to detect, we measure the proportion of test cases from \completeTS that are able to detect each mutant. For a mutant \mutantcircuit, its detection ratio is defined as:
\begin{equation}
\detectionRatio(\mutantcircuit,\completeTS) = \frac{\testcasedetecting(\mutantcircuit, \completeTS)}{|\completeTS|} \times 100
\end{equation}
where \testcasedetecting is defined as in Equation~\ref{eq:tcd}. Lower detection ratios indicate more challenging mutants, as fewer test cases are able to expose the behavioural differences between the mutant and the original circuit. For each approach, we report the median detection ratio across all generated mutants. This metric is used to answer RQ1.2 and RQ2.1.

\subsubsection{Test suite contribution}

To assess whether the generated mutants contribute to the improvement of \testsuite, we measure the number of additional test cases required to detect the generated mutants per program. Starting from \testsuite, we iteratively select new test cases from \completeTS based on the number of mutants they detect. At each step, the test case detecting the largest number of remaining undetected mutants is selected and added to the test suite until all generated mutants are detected.

The final count of added test cases represents the number required to detect all generated mutants. A larger number of required test cases indicates that the generated mutants represent a more diverse set of faults and provide greater potential for test suite improvement. This metric is used to answer RQ1.3 and RQ2.1.

\subsubsection{Second order mutants characteristics}

To analyse the characteristics associated with successful \highorder mutants, we measure the frequency of each mutation characteristic among the generated mutants. This analysis allows us to identify which mutation characteristics and combinations are more frequently associated with \highorder mutants that satisfy the main objective of \approach, i.e., generating undetected mutants that contribute to the improvement of the test suite. This metric is used to answer RQ2.2.

\subsubsection{Second order mutants test case detection}

To analyse how combining mutations affects mutant detection, we compare the test cases that detect individual mutations with those that detect the corresponding \highorder mutant. Let \tcdetectionsetfom denote the set of unique test cases that detect the constituent \firstorder mutants independently, and let \tcdetectionsethom denote the set of test cases that detect the corresponding \highorder mutant.

By comparing these two sets, we analyse whether the combination of mutations keeps the detection behaviour observed for the individual mutations or whether interactions between mutations lead to different detection sets. This metric is used to answer RQ2.3.

\subsection{Statistical Analysis} \label{subsec:statistics}

To account for the stochastic nature of the search-based approaches, we performed statistical analyses to compare the different mutation generation approaches across the metrics mentioned in Section~\ref{subsec:metrics}. Each approach was executed using 10 independently generated test suites for each subject system. Since these executions represent repeated measurements of the same quantum circuit rather than independent observations, we considered the circuits as the experimental unit. Therefore, for each metric and approach, the results obtained from the 10 runs were aggregated using the median, producing a single representative value per circuit while preserving robustness against variability introduced by the stochastic approaches.

We used the Friedman test~\cite{friedman} to assess whether there were statistically significant differences among the compared mutation generation approaches. The Friedman test is a non-parametric alternative to repeated-measures ANOVA and is appropriate for comparing multiple related samples. Since all approaches were evaluated on the same set of circuits, the results were treated as paired observations. 
When significant differences were found, pairwise comparisons were performed using the Wilcoxon signed-rank test~\cite{wilcoxon} with Holm correction~\cite{holmcorrection}.

For measuring the effect size of the comparison, we utilise the Cliff’s Delta (\(\delta\))~\cite{cliff1993dominance}, which measures how large and in what direction differences occur using probabilistic dominance. This effect size allows us to evaluate how substantial the observed differences are; we define the strength of the effect size as follows~\cite{meissel2024using}:
\begin{enumerate*}[label=(\roman*)]
\item \textbf{Negligible}: \(|\delta|<0.15\)
\item \textbf{Small}: \(0.15 \le |\delta| < 0.33\)
\item \textbf{Medium}: \(0.33 \le |\delta| < 0.47\)
\item \textbf{Large}: \(|\delta| \ge 0.47\)
\end{enumerate*}.
By using this combination of tests, we can assess not only whether differences exist but also how substantial those differences are across the evaluated approaches. For count-based metrics, missing values caused by the absence of generated mutants were replaced by zero because they represent the absence of generated mutants. For test case detection ratio, undefined values caused by zero generated mutants were excluded because they do not represent failed detection but rather the absence of mutants to evaluate.
This ensures that comparisons are performed only on valid paired observations.

\subsection{Execution Environment}\label{subsec:execution}
All original and mutated circuits were executed under the same conditions, utilising the same computational resources. The experiments were conducted on a national high-performance cluster of servers, which included 2 AMD Epyc 7601 processors, 2TB of RAM, an AMD Vega20 GPU, and a high-speed 4TB NVMe drive. All quantum circuits were written in QASM, and we used Qiskit version 1.1.0 to create and modify the circuits as needed. To enhance consistency and reproducibility, a fixed random seed was employed during the simulations.

\section{Results and Analysis}\label{sec:results}

In this section, we present the results from the experiments described in Section~\ref{sec:experiment}. We implement \approach, using four search algorithms, and compare it against two baseline quantum circuit mutation generators. We also examine the effects of introducing either single or multiple mutations in a quantum circuit.

\subsection{Results for RQ1 -- \approach effectiveness}\label{subsec:resRQ1}

In this first research question, we evaluated \approach for finding the desired mutants. In RQ1.1 (Section~\ref{subsubsec:resRQ1.1}), we assessed \approach's ability to identify mutants that pass all tests in \testsuite and verify that the mutants found are actually non-equivalent (i.e., detected by \completeTS). In RQ1.2 (Section~\ref{subsubsec:resRQ1.2}), we analysed whether the obtained mutants are detected by fewer test cases than those generated by the baseline approaches, indicating that they are more challenging mutants. Finally, in RQ1.3 (Section~\ref{subsubsec:resRQ1.3}), we examined whether these mutants improve the given test suite, analysing how many new test cases must be added to \testsuite to detect the generated mutants.

\subsubsection{Results for RQ1.1 -- Non-equivalent mutants that pass \testsuite}\label{subsubsec:resRQ1.1}
We first report the success rates of different approaches at finding at least one mutant that is not detected by \testsuite and is not equivalent for each quantum circuit. The results showed that \muskit and \geneticalgorithm achieved the highest success rates, at \textbf{100\%} (300/300) and \textbf{99.67\%} (299/300), respectively; \geneticalgorithm failed in only one run for the 7-qubit QFT. The other approaches show a success rate of \textbf{86.33\%} (259/300) for \randomsearch, \textbf{83.67\%} (251/300) for \qmutpy, \textbf{82\%} (246/300) for \oneplusone and \textbf{76.33\%} (229/300) for \hillclimbing with the lowest success rate. This success rate reflects the ability of different approaches to find at least one mutant for a quantum circuit, but it does not indicate the number of mutants obtained or their quality. To assess these aspects, we perform more detailed analyses, as explained in the following.

\begin{figure}[!tb]
\centering
\includegraphics[width=\linewidth]{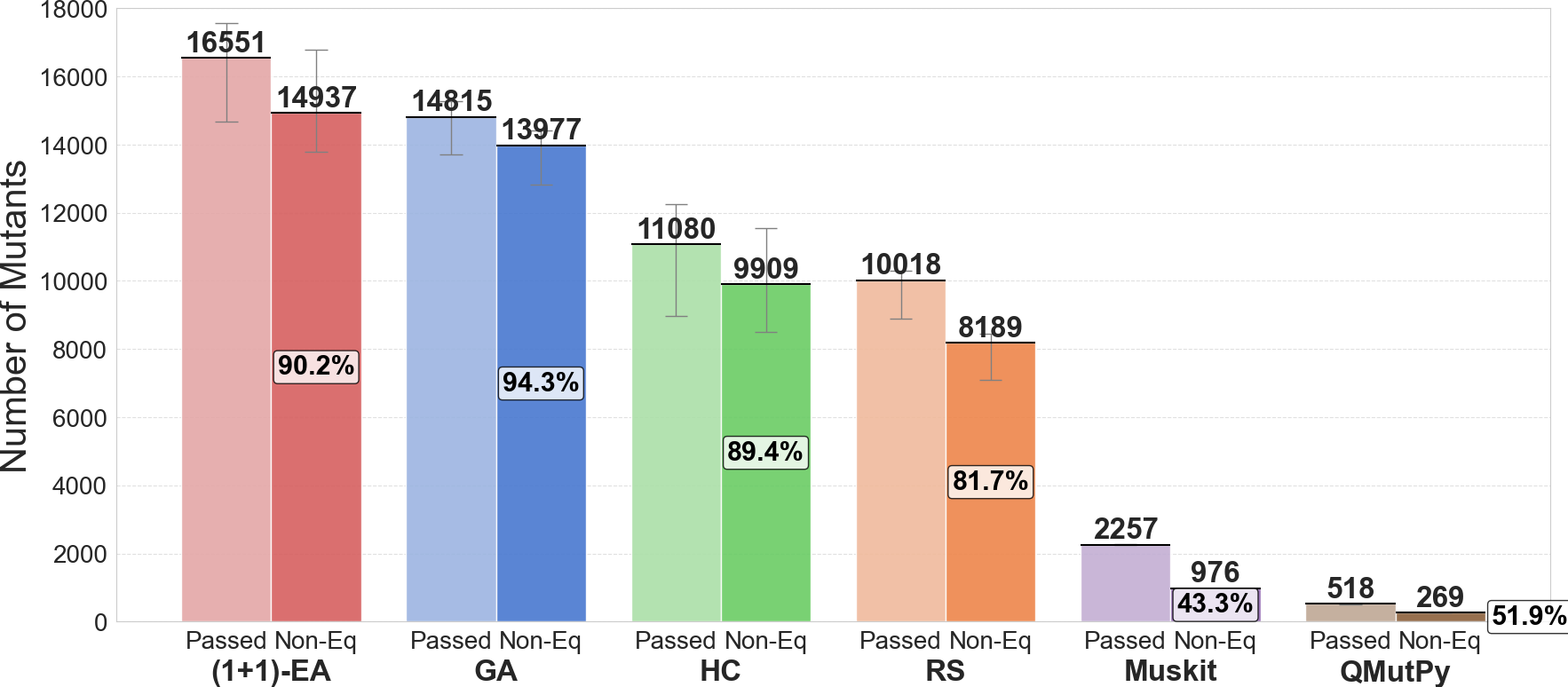}
\caption{RQ1.1 -- Median number of mutants passing and non-equivalent by search across all circuits per run.}
\Description{RQ1.1}
\label{fig:barPlotFoundValidated}
\end{figure}

Figure~\ref{fig:barPlotFoundValidated} shows the median number of mutants obtained from running the approaches for all 30 quantum circuits.
The left bar (lighter colour) represents the number of mutants passing \testsuite, and the right bar (darker colour) shows the number of mutants that are not equivalent based on \completeTS. The whiskers on the top of the bars represent the standard deviation of the values across the runs.

The figure shows that the approaches that obtain the majority of mutants that pass \testsuite are \oneplusone and \geneticalgorithm, while the literature tools \muskit and \qmutpy are those that generate the fewest. The value highlighted in the bar shows the percentage of mutants not detected by \testsuite that were non-equivalent; results indicate that \geneticalgorithm is better at optimising the effect of the mutants to make them pass \testsuite but still be detectable. The literature tools show the worst performance by having only about half of the mutants that pass to be non-equivalent. 

%In Figure~\ref{fig:barPlotFoundValidated}, we decided to aggregate the mutants obtained from all quantum circuits and analyse them as one single experiment, focusing on the total number of mutants. However, because some approaches generate more mutants for a given algorithm or program size, these visualisations might not be sufficiently descriptive.

To further support these results, we employed a statistical test to determine whether the observed differences are significant. First, the Friedman test revealed statistically significant differences among the compared approaches across all evaluated metrics ($p <0.05$), indicating that at least one approach differs from the others. To further identify where these differences occur and quantify their magnitude, we conducted post hoc pairwise comparisons using the Wilcoxon signed-rank test with Holm correction, along with effect size analysis using Cliff's delta. Table~\ref{tab:pairwisestats} reports the pairwise comparison results for all the approaches.
\begin{table}[!t]
\centering 
\caption{RQ1.1 -- Pairwise statistical comparison of the approaches for generated mutants, non-equivalent mutants and non-equivalent ratio. Effect sizes are reported using Cliff's $\delta$ with qualitative interpretation.}
\label{tab:pairwisestats}
\setlength{\tabcolsep}{3pt}
\begin{tabular}{@{}lcccccc@{}}
\toprule
& \multicolumn{2}{c}{\#Passed \testsuite} & \multicolumn{2}{c}{\#Non-equivalent} & \multicolumn{2}{c}{Non-equivalent ratio}\\
\cmidrule(lr){2-3} \cmidrule(lr){4-5}\cmidrule(lr){6-7}
Comparison & $p$-value & Effect size ($\delta$) & $p$-value & Effect size ($\delta$) & $p$-value & Effect size ($\delta$) \\
\midrule
%SORT BY ALGORITHM BASED ON PERFORMANCE
\geneticalgorithm vs \oneplusone & 0.31 & -- & 0.17 & -- & 0.41 & --\\
\geneticalgorithm vs \hillclimbing & $<0.05$ & Medium (0.34) & $<0.05$ & Medium (0.37) & 0.45 & --\\
\geneticalgorithm vs \randomsearch & $<0.05$ & Medium (0.44) & $<0.05$ & Medium (0.46) & $<0.05$ & Medium (0.44) \\
\geneticalgorithm vs \muskit & $<0.05$ & Large (0.52) & $<0.05$ & Large (0.77) & $<0.05$ & Large (1.00) \\
\geneticalgorithm vs \qmutpy & $<0.05$ & Large (0.94)& $<0.05$ & Large (0.97) & $<0.05$ & Large (0.97) \\
\oneplusone vs \hillclimbing & 0.06 & -- & $<0.05$ & Small (0.19) & 0.91 & --\\
\oneplusone vs \randomsearch & $<0.05$ & Small (0.30) & $<0.05$ & Small (0.31) & 0.45 & --\\
\oneplusone vs \muskit & 0.07 & -- & $<0.05$ & Medium (0.44) & $<0.05$ & Large (0.91) \\
\oneplusone vs \qmutpy & $<0.05$ & Large (0.65) & $<0.05$ & Large (0.75) & $<0.05$ & Large (0.81) \\ 
\hillclimbing vs \randomsearch & 0.20 & -- & $<0.05$ & Small (0.13) & 0.45 & --\\
\hillclimbing vs \muskit & 0.31 & -- & 0.07 & -- & $<0.05$ & Large (0.78) \\
\hillclimbing vs \qmutpy & $<0.05$ & Medium (0.36)& $<0.05$ & Large (0.55) & $<0.05$ & Large (0.68) \\ 
\randomsearch vs \muskit & 0.38 & -- & 0.17 & -- & $<0.05$ & Large (0.73) \\
\randomsearch vs \qmutpy & 0.21 & -- & $<0.05$ & Small (0.30)& $<0.05$ & Large (0.68) \\
\muskit vs \qmutpy & $<0.05$ & Large (0.90) & $<0.05$ & Large (0.83) & 0.45 & --\\

% SORT BASED ON BLANK SPACES
% \geneticalgorithm vs \qmutpy & $<0.05$ & Large (0.94)& $<0.05$ & Large (0.97) & $<0.05$ & Large (0.97) \\
% \geneticalgorithm vs \muskit & $<0.05$ & Large (0.52) & $<0.05$ & Large (0.77) & $<0.05$ & Large (1.00) \\
% \oneplusone vs \qmutpy & $<0.05$ & Large (0.65) & $<0.05$ & Large (0.75) & $<0.05$ & Large (0.81) \\ 
% \hillclimbing vs \qmutpy & $<0.05$ & Medium (0.36)& $<0.05$ & Large (0.55) & $<0.05$ & Large (0.68) \\ 
% \geneticalgorithm vs \randomsearch & $<0.05$ & Medium (0.44) & $<0.05$ & Medium (0.46) & $<0.05$ & Medium (0.44) \\
% \muskit vs \qmutpy & $<0.05$ & Large (0.90) & $<0.05$ & Large (0.83) & 0.45 & --\\
% \geneticalgorithm vs \hillclimbing & $<0.05$ & Medium (0.34) & $<0.05$ & Medium (0.37) & 0.45 & --\\
% \oneplusone vs \randomsearch & $<0.05$ & Small (0.30) & $<0.05$ & Small (0.31) & 0.45 & --\\
% \oneplusone vs \muskit & 0.07 & -- & $<0.05$ & Medium (0.44) & $<0.05$ & Large (0.91) \\
% \randomsearch vs \qmutpy & 0.21 & -- & $<0.05$ & Small (0.30)& $<0.05$ & Large (0.68) \\
% \oneplusone vs \hillclimbing & 0.06 & -- & $<0.05$ & Small (0.19) & 0.91 & --\\
% \hillclimbing vs \randomsearch & 0.20 & -- & $<0.05$ & Small (0.13) & 0.45 & --\\
% \hillclimbing vs \muskit & 0.31 & -- & 0.07 & -- & $<0.05$ & Large (0.78) \\
% \randomsearch vs \muskit & 0.38 & -- & 0.17 & -- & $<0.05$ & Large (0.73) \\
% \geneticalgorithm vs \oneplusone & 0.31 & -- & 0.17 & -- & 0.41 & --\\
\bottomrule
\end{tabular}
\end{table}

The statistical tests confirm that \geneticalgorithm significantly outperforms all other approaches, except \oneplusone, in obtaining a higher number of mutants that pass \testsuite and a higher number of non-equivalent mutants. It shows a medium effect against the other search algorithms, \hillclimbing and \randomsearch, and a large effect size against the literature tools \muskit and \qmutpy. Regarding the non-equivalent ratios among the mutants that pass, \geneticalgorithm still achieves a significant difference from \randomsearch and the literature tools, with medium and large effect sizes, respectively. The comparison with \oneplusone and \hillclimbing shows that the differences are not significant, highlighting that the three approaches might have similar non-equivalent ratios, as shown in Figure~\ref{fig:barPlotFoundValidated}. 

Among the rest of the search-based approaches (i.e., apart from \geneticalgorithm), the observed differences are either non-significant or with only small effect sizes, suggesting that these approaches achieve comparable performance.

When compared to the literature tools in terms of the number of mutants passing \testsuite, \muskit exhibits performance comparable to the search-based approaches, as no significant differences were observed. In contrast, \qmutpy differs significantly from \oneplusone and \hillclimbing, with large and medium effect sizes, respectively. A similar trend is observed for the number of non-equivalent mutants: \muskit remains comparable to the search-based approaches, differing significantly only from \oneplusone, with a medium effect size; \qmutpy again shows significant differences against all search-based approaches, with large effect sizes for \oneplusone and \hillclimbing, and a small effect size for \randomsearch. Regarding the ratio of non-equivalent mutants, all search-based approaches significantly outperform both literature-based tools, consistently exhibiting large effect sizes.

Overall, \qmutpy demonstrates the weakest performance, as \muskit also significantly outperforms \qmutpy in terms of both the number of mutants passing \testsuite and the number of non-equivalent mutants, with large effect sizes.

\begin{tcolorbox}[colback=blue!5!white, colframe=white, breakable]
\textbf{Concluding remarks for RQ1.1:} 
% \geneticalgorithm and \muskit achieved the best success rates across all the quantum circuits, generating at least one mutant that passes the test suite and is not equivalent in 99.67\% and 100\% of the cases, respectively. 
\geneticalgorithm and \oneplusone demonstrated the highest effectiveness in generating both the highest number of mutants passing the test suite and the number of them that are not equivalent. All search approaches showed superior performance in terms of the non-equivalent ratio compared to \muskit and \qmutpy.
\end{tcolorbox}

\subsubsection{Results for RQ1.2 -- Test case detection}\label{subsubsec:resRQ1.2}
\begin{figure}[!tb]
\centering
\includegraphics[width=\linewidth]{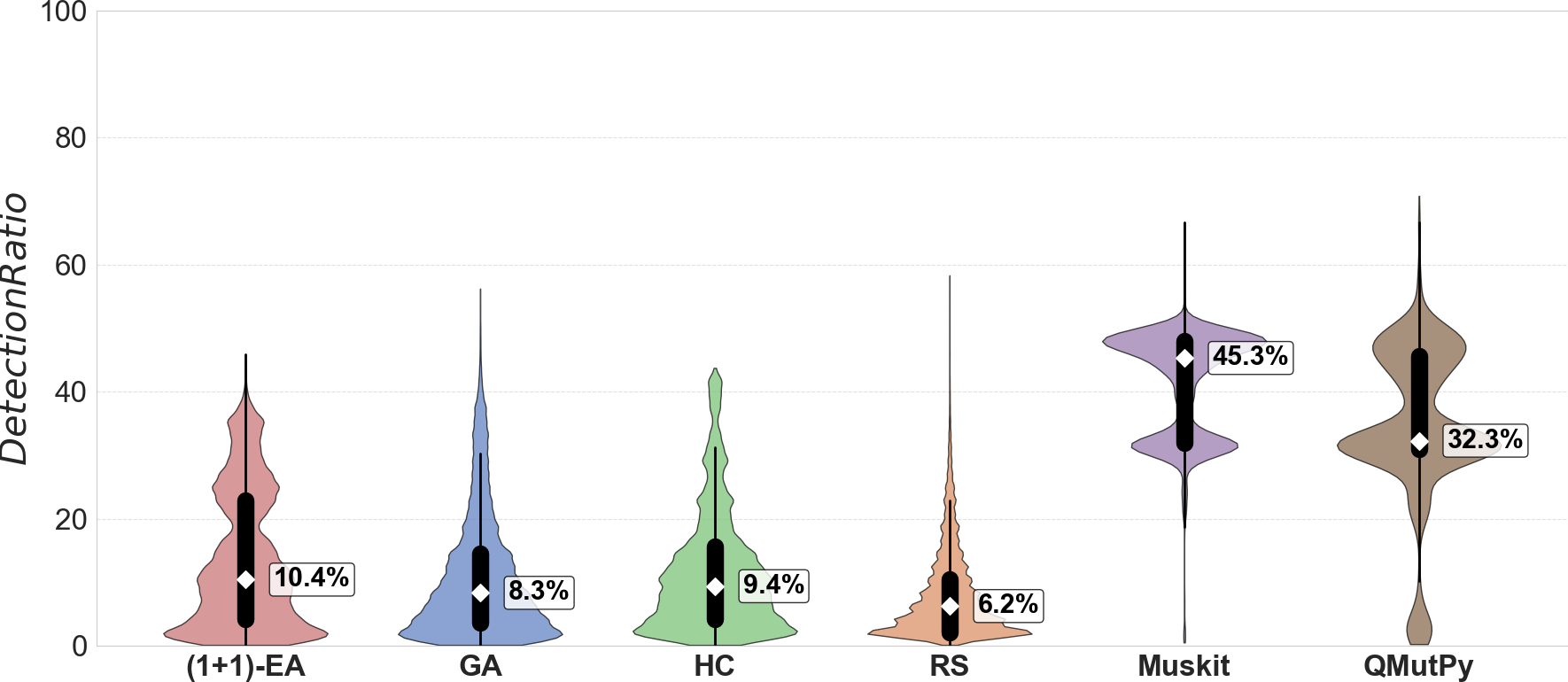}
\caption{RQ1.2 -- \detectionRatio for the mutants generated by each approach.}
\Description{RQ1.2}
\label{fig:TCDetectingBaselines}
\end{figure}

In this RQ, we want to evaluate how challenging the mutants generated by each approach are. For that, we consider only the mutants that pass \testsuite and are not equivalent, as these are the mutants relevant for our study. Figure~\ref{fig:TCDetectingBaselines} shows, for each approach, the distribution of \detectionRatio across the mutants from \completeTS as defined in Section~\ref{sec:detectionRatio}.
In this RQ, we treat each mutant generated across all runs and quantum circuits as a data point. The value highlighted shows the median \detectionRatio obtained.
%The line in the middle shows the median value with the red and blue dots denoting the maximum and minimum values, respectively.

% \begin{figure}[!tb]
% \centering
% \includesvg[width=0.8\linewidth]{BaselinesComparison_boxplot.svg}
% \caption{Comparison of \testcase detecting each of the mutants.}
% \label{fig:TCDetectingBaselines}
% \end{figure}

The figure clearly shows that the four search-based approaches achieve mutants with lower \detectionRatio compared to \muskit and \qmutpy, meaning the obtained mutants are more challenging to detect. Among the search-based approaches, there appears to be some similarity in detection distributions, with all showing triangular shapes that mostly focus on lower \numtestcases detections. To verify the findings, we also conducted the same statistical tests as in the previous experiment; the Friedman test again revealed that at least one approach differs from the others significantly ($p < 0.05$). The results of the pairwise comparison are shown in Table~\ref{tab:pairwisestatsTC}.
\begin{table}[!t]
\centering
\caption{RQ1.2 -- Pairwise statistical comparison of the approaches for the \detectionRatio.}
\label{tab:pairwisestatsTC}
% \setlength{\tabcolsep}{9pt}
% \begin{tabular}{lccc}
% \toprule
% \geneticalgorithm vs & $p$-value & $\delta$ & Effect size\\
% % \geneticalgorithm vs & $p$-value & \multicolumn{2}{c}{Effect size}\\
% % \cmidrule(lr){3-4}
% % & & $\delta$ & category\\
% \midrule
% \qmutpy & $<0.05$ & -0.83 & Large \\
% \muskit & $<0.05$ & -0.99 & Large \\
% \randomsearch & $<0.05$ & 0.27 & Small \\
% \hillclimbing & 0.89 & 0.06 & Negligible \\
% \oneplusone & 0.89 & 0.05 & Negligible \\
\begin{tabular}{@{}lcc|lcc@{}}
\toprule
Comparison & $p$-value & Effect size ($\delta$) & Comparison & $p$-value & Effect size ($\delta$) \\
\midrule
\oneplusone vs \muskit & $<0.05$ & Large (-1.00) & \geneticalgorithm vs \muskit & $<0.05$ & Large (-1.00)\\
\hillclimbing vs \muskit & $<0.05$ & Large (-1.00) & \randomsearch vs \muskit & $<0.05$ & Large (-1.00)\\
\oneplusone vs \qmutpy & $<0.05$ & Large (-0.84) & \randomsearch vs \qmutpy & $<0.05$ & Large (-0.84)\\
\geneticalgorithm vs \qmutpy & $<0.05$ & Large (-0.83) & \hillclimbing vs \qmutpy & $<0.05$ & Large (-0.83)\\
\geneticalgorithm vs \randomsearch & $<0.05$ & Small (0.21) & \hillclimbing vs \randomsearch & $<0.05$ & Small (0.18)\\
\oneplusone vs \randomsearch & $<0.05$ & Negligible (0.13) & \muskit vs \qmutpy & 0.79 & --\\
\geneticalgorithm vs \hillclimbing & 0.79 & -- & \geneticalgorithm vs \oneplusone & 0.79 & --\\
\oneplusone vs \hillclimbing & 0.79 & -- & -- & --\\
\bottomrule
\end{tabular}
\end{table}
The statistical tests confirmed our conclusion from the figure, showing that all search algorithms have significantly lower detection rates for the generated mutants than the literature tools, with a large effect size. Among them, as noted earlier, the statistical tests confirm that there is no big difference in \detectionRatio, with only \randomsearch having a small or negligible effect size relative to the other search approaches. This indicates that the mutants that pass \testsuite and are not equivalent across approaches have similar \detectionRatio.

\begin{tcolorbox}[colback=blue!5!white, colframe=white, breakable]
\textbf{Concluding remarks for RQ1.2:} 
Even though no big differences were observed among the search-based approaches in the difficulty of detecting the mutants, all of them significantly outperformed \muskit and \qmutpy by generating mutants that are detected by fewer test cases. 
\end{tcolorbox}

\subsubsection{Results for RQ1.3 -- Test suite contribution}\label{subsubsec:resRQ1.3}

\begin{figure}[!t]
\centering
\includegraphics[width=\linewidth]{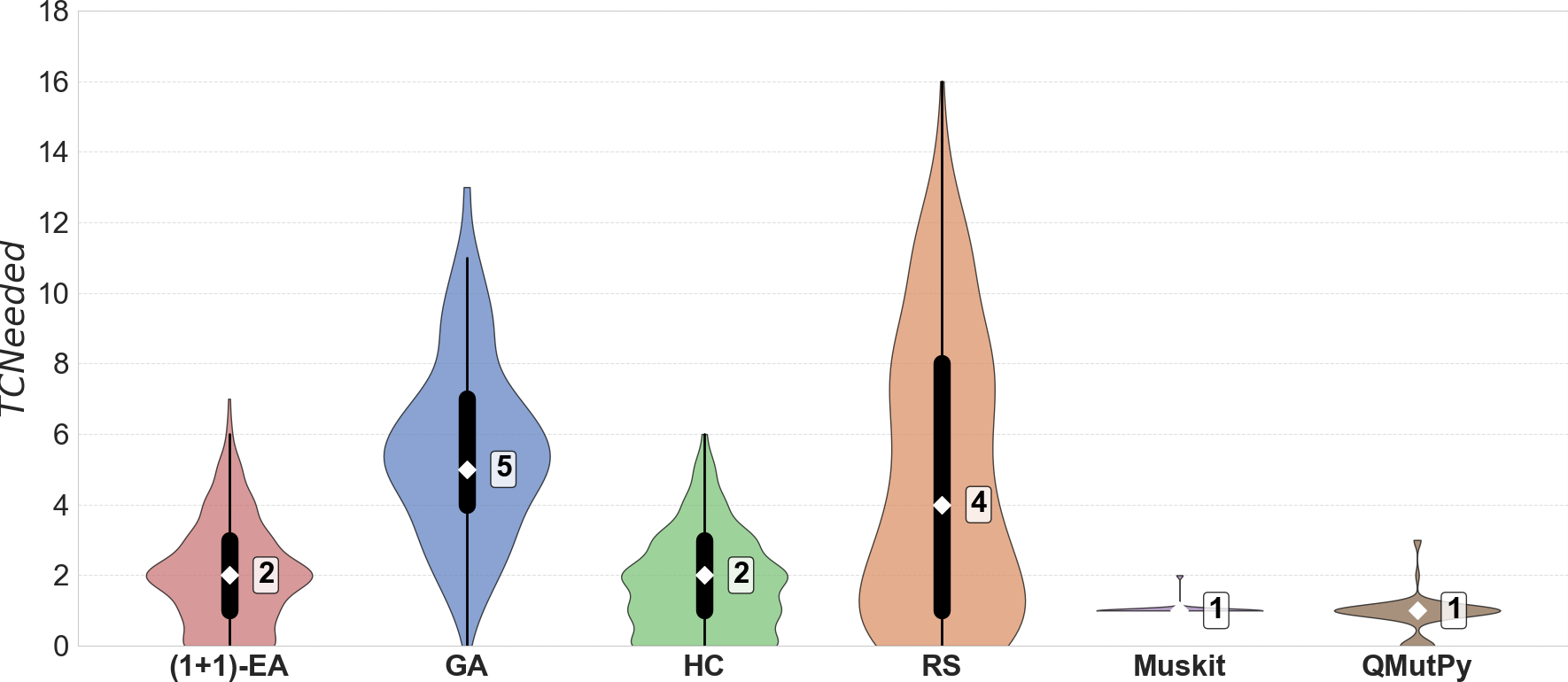}
\caption{RQ1.3 -- New \testcase needed to detect mutants generated.}
\Description{RQ1.3}
\label{fig:TCNeeded}
\end{figure}

In this RQ, we analysed how much the set of mutants generated by each approach contributes to improving \testsuite as defined in Section~\ref{subsec:metrics}.
%We define the improvement of \testsuite as the need to include more test cases in it. The inclusion of more test cases in the test suite means that the generated mutants represent a more diverse set of faults, as more test cases are needed to detect them. Also, this means that there is a need for specific test cases that detect a specific type of mutant representing a concrete fault.
% %meaning that more faults not detected by \testsuite were identified and require more diverse test cases to be detected. 
% We calculate the number of test cases needed to add to detect all the new mutants by selecting from the unique test cases that detect the mutants. We start by adding new test cases to \testsuite from those that detect more mutants, and continue until all mutants are detected.
% the ones that detect most of them, and adding the test cases to \testsuite one by one until all mutants are covered.
% Table~\ref{tab:statsTCNeeded_raw} shows the median, mean, and standard deviation of 
Figure~\ref{fig:TCNeeded} shows the distribution of the median number of test cases needed to be added to \testsuite in order to detect all the generated mutants for each quantum circuit.

The results show that \geneticalgorithm achieves the highest score, requiring the median number of five additional tests to be included in the test suite, followed by \randomsearch, which requires four test cases but shows a wider vertical spread, indicating greater variability. The literature tools seem to perform the worst, as the median indicates that including a single test case in the test suite detects all the mutants generated. These findings are corroborated by the statistical test results shown in Table~\ref{tab:statsTCNeeded}. 
\begin{table}[!tb]
\centering
\caption{RQ1.3 -- Pairwise statistical comparison of the approaches in terms of number of new \testcase needed to detect the generated mutants.}
\label{tab:statsTCNeeded}
% \begin{subtable}[t]{0.48\textwidth}
% \centering
% \caption{Raw Statistics}
% \label{tab:statsTCNeeded_raw}
% \setlength{\tabcolsep}{6pt}
% \begin{tabular}{lccc}
% \toprule
% Approach & Median & Mean & Std \\
% \midrule
% \geneticalgorithm & 5 & 5.59 & 2.49 \\
% \randomsearch & 4 & 4.85 & 3.64 \\
% \oneplusone & 2 & 2.01 & 1.21 \\
% \hillclimbing & 2 & 1.71 & 1.16 \\
% \muskit & 1 & 1.03 & 0.18 \\
% \qmutpy & 1 & 0.92 & 0.41 \\
% \bottomrule
% \end{tabular}
% \end{subtable}
% \hfill
% \begin{subtable}[t]{0.48\textwidth}
% \centering
% \caption{Pairwise comparison}
% \label{tab:statsTCNeeded_pairwise}
% \vspace{3pt}
% \setlength{\tabcolsep}{6pt}
% \begin{tabular}{lccc}
% \toprule
% \geneticalgorithm vs & $p$-value & $\delta$ & Effect size \\
% \midrule
% \randomsearch & $<0.05$ & 0.17 & Small \\
% \oneplusone & $<0.05$ & 0.79 & Large \\
% \hillclimbing & $<0.05$ & 0.83 & Large \\
% \muskit & $<0.05$ & 0.96 & Large \\
% \qmutpy & $<0.05$ & 0.96 & Large \\
% \bottomrule
% \end{tabular}
\begin{tabular}{@{}lcc|lcc@{}}
\toprule
Comparison & $p$-value & Effect size ($\delta$) & Comparison & $p$-value & Effect size ($\delta$) \\
\midrule
\geneticalgorithm vs \oneplusone& $<0.05$ & Large (0.98)& \geneticalgorithm vs \hillclimbing & $<0.05$ & Large (0.98)\\
\geneticalgorithm vs \muskit & $<0.05$ & Large (1.00)& \geneticalgorithm vs \qmutpy & $<0.05$ & Large (0.99)\\
\oneplusone vs \qmutpy & $<0.05$ & Large (0.74)& \oneplusone vs \muskit & $<0.05$ & Large (0.73)\\
\randomsearch vs \qmutpy & $<0.05$ & Large (0.65)& \randomsearch vs \muskit & $<0.05$ & Large (0.62)\\
\randomsearch vs \hillclimbing & $<0.05$ & Large (0.47)& \hillclimbing vs \qmutpy & $<0.05$ & Large (0.61)\\
\hillclimbing vs \muskit & $<0.05$ & Large (0.56) & \oneplusone vs \randomsearch & $<0.05$ & Medium (-0.42)\\
\geneticalgorithm vs \randomsearch & 0.12 & -& \oneplusone vs \hillclimbing & 0.16 & --\\
\muskit vs \qmutpy & 0.20 & -& -- & -- & --\\
\bottomrule
\end{tabular}
% \end{subtable}
\end{table}
The table shows that \geneticalgorithm performs significantly better than all other approaches except for \randomsearch, with which the differences seem not to be significant. Again, all search approaches show significantly better results than \muskit and \qmutpy, all of them with large effect sizes. This demonstrates the superiority of the search algorithms in creating mutants that yield more test cases.

\begin{tcolorbox}[colback=blue!5!white, colframe=white, breakable]
\textbf{Concluding remarks for RQ1.3:} 
\geneticalgorithm provides the mutants that contribute more to improving the test suite, as they require a median of five new tests to be added to \testsuite to detect them. All search approaches significantly outperform \muskit and \qmutpy, showing the need to improve the test suite.
% The mutants generated by \geneticalgorithm seem to be the most diverse, as they require larger improvements to the test suite, which aligns with the main objective of our study.
% These results demonstrate that search-based mutation generation is a highly effective strategy for generating diverse and challenging quantum mutants, with the genetic algorithm providing the best overall balance between effectiveness, robustness, and contribution to test-suite improvement.
\end{tcolorbox}

\subsection{Results for RQ2 -- High order mutants}\label{subsec:resRQ2}
In this RQ, we analyse how different the obtained mutants are when we introduce a single or multiple mutations to the circuit. For this analysis, although all search algorithms generated both \firstorder and \secondorder mutants, we chose to focus only on the mutants produced by \geneticalgorithm, as it achieved the best overall performance in RQ1. First, in RQ2.1 (Section~\ref{sec:resRQ2.1}), we compare \firstorder and \secondorder mutants by utilising the same metrics used in RQ1. Then, in RQ2.2 (Section~\ref{sec:resRQ2.2}), we focus on analysing the characteristics of \secondorder mutants and how the characteristics of mutations affect the generation. Finally, in RQ2.3 (Section~\ref{sec:resRQ2.3}), we examine the interactions among mutations introduced into the same circuit and how these interactions affect mutant detection.

\subsubsection{Results for RQ2.1 -- First order vs second order mutants}\label{sec:resRQ2.1}

\begin{figure}[!tb]
\centering
\begin{subfigure}{0.31\textwidth}
\centering
\includegraphics[width=\linewidth]{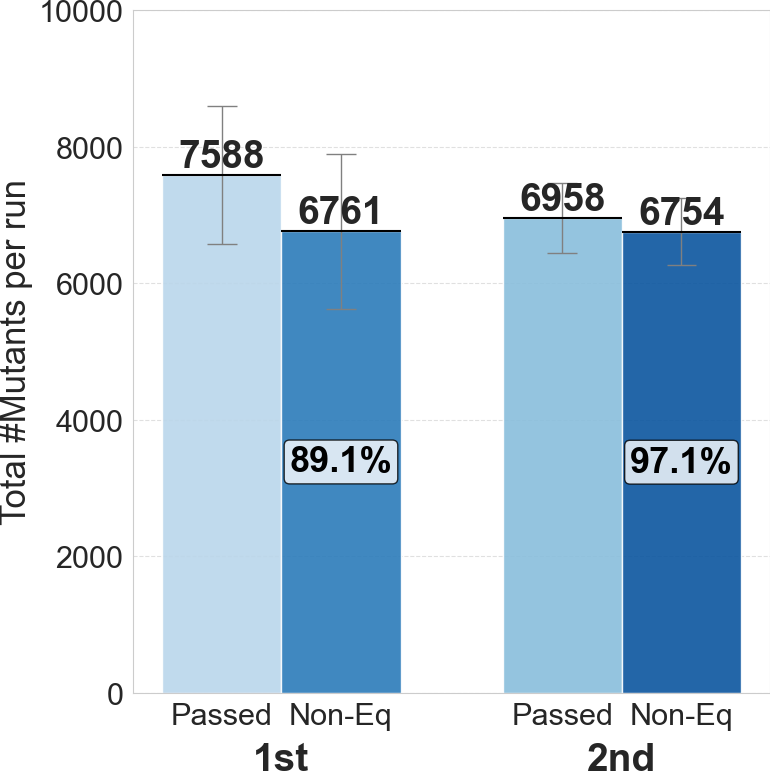}
\caption{Number of mutants generated per run.}
\label{subfig:SurvivedNoneqOrders}
\end{subfigure}
\hfill
\begin{subfigure}{0.31\textwidth}
\centering
\includegraphics[width=\linewidth]{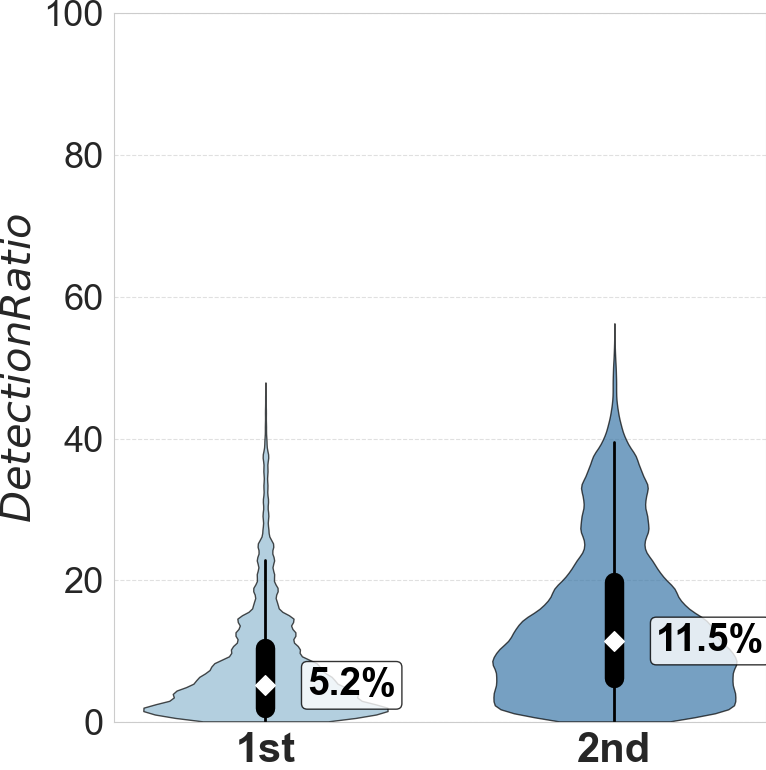}
\caption{\detectionRatio for the mutants generated.}
\label{subfig:TCDetectionOrders}
\end{subfigure}
\hfill
\begin{subfigure}{0.3\textwidth}
\centering
\includegraphics[width=\linewidth]{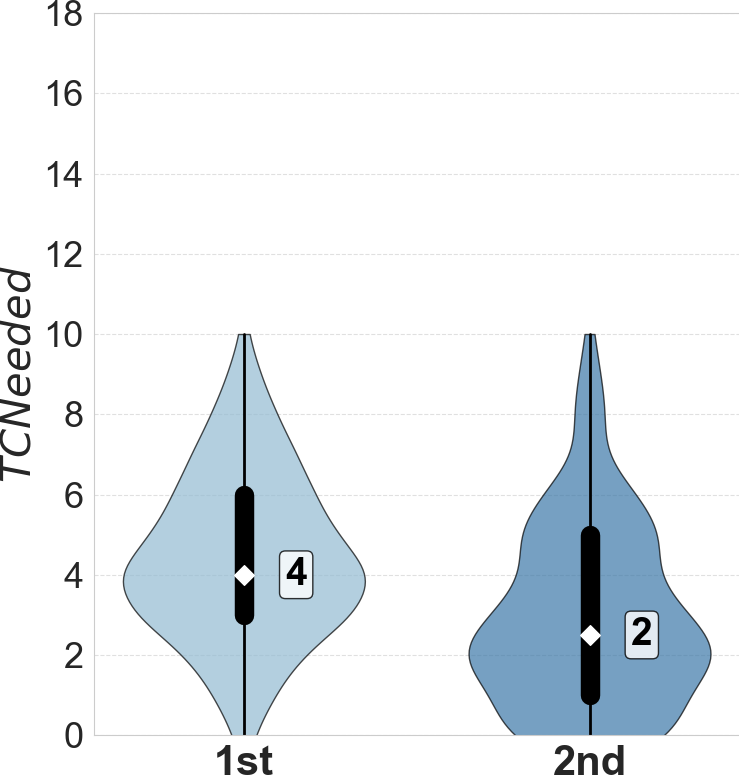}
\caption{Number of new \testcase needed to detect mutants.}
\label{subfig:TCNeededOrders}
\end{subfigure}
\caption{RQ2.1 -- Comparison of \firstorder and \secondorder mutants generated by \geneticalgorithm.}
\Description{RQ2.1}
\label{fig:1stvs2nd}
\end{figure}

Figure~\ref{fig:1stvs2nd} shows the results obtained from comparing \firstorder and \secondorder mutants generated by \geneticalgorithm. 
Figure~\ref{subfig:SurvivedNoneqOrders} shows the total number of mutants obtained across all circuits per run per order. Figure~\ref {subfig:TCDetectionOrders} shows \detectionRatio across the mutants per order. Figure~\ref{subfig:TCNeededOrders} shows the number of new \testcase needed to add to \testsuite to detect the generated mutants per order.

Figure~\ref{subfig:SurvivedNoneqOrders} shows that \geneticalgorithm is able to achieve a slightly higher number of \firstorder mutants than of \secondorder mutants. Second order mutants have a better ratio of non-equivalence, which results in both orders having a similar number of non-equivalent mutants. In Figure~\ref{subfig:TCDetectionOrders}, \detectionRatio results show that, overall, \firstorder mutants achieve a lower detection rate than \secondorder mutants, which achieve a median detection rate that is more than double of that of \firstorder: 5.2\% vs 11.5\%. Thus, we conclude that \firstorder mutants are easier to find during the search, but they are more likely to be equivalent as they have a lower non-equivalent ratio than \secondorder mutants. The higher non-equivalent ratio of \secondorder mutants is also related to the bigger effect they produce in the circuit as shown in the \detectionRatio in Figure~\ref{subfig:TCDetectionOrders}. In Figure~\ref{subfig:TCNeededOrders}, the results show that \secondorder mutants require fewer \testcase to be detected, with a median of 3 compared to 4 for \firstorder. This indicates that the generated \secondorder mutants contribute less to the improvement of the test suite, as they can be detected using fewer \testcase, which is also related to having a larger effect on the circuits.

Table~\ref{tab:stats1stvs2nd} presents the statistical test results comparing the metrics for \firstorder and \secondorder mutants.
\begin{table}[!t]
\centering
\caption{RQ2.1 -- Statistical comparison between \firstorder and \secondorder mutations using the Wilcoxon signed-rank test and Cliff's delta effect size. The sign of the effect size indicates the direction of the effect, with positive values leaning towards \firstorder mutants and negative values towards \secondorder mutants.}
\label{tab:stats1stvs2nd}
\begin{tabular}{@{}lcc@{}}
\toprule
Metric & $p$-value & Effect size ($\delta$) \\
\midrule
\#Passed \testsuite & $<0.05$ & Small (0.18) \\
\#Non-equivalent & 0.07 & --\\%Small (0.16)\\
Non-equivalent ratio & 0.4 & --\\%Small (-0.3)\\
\testcase detection \% & $<0.05$ & Small (-0.22) \\
Number of new \testcase needed & $<0.05$ & Large (0.51)\\
\bottomrule
\end{tabular}
\end{table}
The statistical tests confirm the previous findings and show that both orders generate similar numbers of mutants passing \testsuite and being non-equivalent, with a small difference in terms of the number of mutants passing the test suite and no significant difference in terms of the number of non-equivalent mutants or the non-equivalent ratio. Also, in terms of \testcase detection, there is again a small effect size leaning towards \secondorder mutants, shown by a negative Cliff's $\delta$, indicating \secondorder mutants have slightly more effect on the circuit and thus they are detected by more \testcase. Finally, the biggest difference between the two orders comes from the number of new \testcase needed to detect them, which shows a significant difference with a large effect size in favour of \firstorder mutants. This means that while both orders show similar behaviour, \firstorder mutants are more valuable for the study's main objective, as they are the ones that contribute the most to the improvement of the test suite.

\begin{tcolorbox}[colback=blue!5!white, colframe=white, breakable]
\textbf{Concluding remarks for RQ2.1:} 
Overall, \firstorder and \secondorder mutants show comparable behaviour. While \firstorder mutants have a slightly worse non-equivalent ratio, the non-equivalent mutants obtained are detected by fewer test cases and contribute more to the improvement of the test suite by needing to add more test cases to detect them. 
Thus, we conclude that \firstorder mutants align better with our objective of finding non-equivalent mutants that pass \testsuite and contribute to its improvement. 
%Moreover, the results show that \secondorder mutants have a greater effect on the circuit, as their non-equivalence ratio is slightly higher and they are detected by more test cases.
\end{tcolorbox}

\subsubsection{Results for RQ2.2 -- Second order mutants characteristics}\label{sec:resRQ2.2}
\begin{figure}[tb]
\centering
\includegraphics[width=\linewidth]{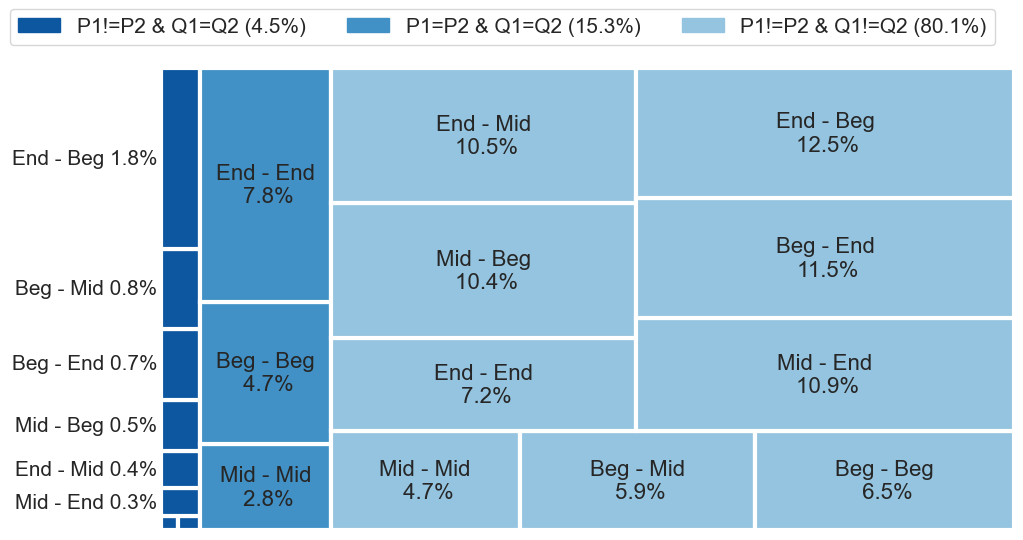}
\caption{RQ2.2 -- Qubits and positions utilised for generating \secondorder mutants.}
\Description{RQ2.2}
\label{fig:treemapPos}
\end{figure}

Following the analysis of \secondorder mutants, we wanted to explore more the characteristics of each mutation and the interactions among them that generate \secondorder mutants. Figure~\ref{fig:treemapPos} shows the proportions of qubit-position combinations utilised in the generated \secondorder mutants. 
The labels on the top divide the possible combinations into three categories:
\begin{enumerate}
\item Different position in the circuit but same qubit: $P1!=P2~\&~Q1=Q2$ (Dark Blue).
\item Same position in the circuit and thus the same qubit: $P1=P2~\&~Q1=Q2$ (Blue).
\item Different position in the circuit and different qubit: $P1!=P2~\&~Q1!=Q2$ (Light Blue).
\end{enumerate}

Each of them is further divided into subsections showing the relative position of the circuit where the mutation is introduced:
\begin{enumerate}
\item Beginning of the circuit, referring to the first third of the circuit (Beg).
\item Middle of the circuit referring to the second third of the circuit (Mid).
\item End of the circuit referring to the last third of the circuit (End).
\end{enumerate}

Figure~\ref{fig:treemapPos} shows that the majority of the generated \secondorder mutants are composed of mutations in different positions and qubits (80.1\%), suggesting that, even when not close to each other, mutations can interact. Figure~\ref{fig:treemapPos} also shows that the least frequent combination is to have mutations in different positions but the same qubit, with only 4.5\% of \secondorder mutants falling into such category; instead, the mutations in the same position (i.e. one right after the other) are slightly more frequent with 15.3\% of the cases. When referring to the relative position in the circuit, the figure shows that the most frequent combinations are mutations at distinct positions, with those combining mutations in the beginning and end of the circuit being the most frequent. Figure~\ref{fig:treemapPos} suggests that mutations introduced far apart (both in terms of relative position and qubits) lead to more subtle changes, as \approach was able to generate more mutants passing \testsuite and being non-equivalent with such characteristics.

\begin{figure}[!tb]
\centering
\includegraphics[width=0.9\linewidth]{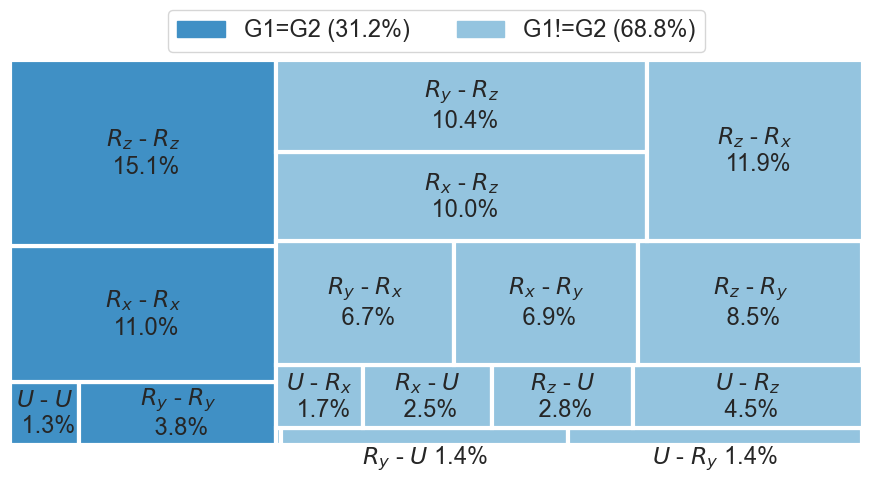}
\caption{RQ2.2 -- Gate combinations utilised for generating \secondorder mutants.}
\Description{RQ2.2}
\label{fig:treemapGate}
\end{figure}

Figure~\ref{fig:treemapGate} shows the proportions of combinations of gates used in each mutation in the generated \secondorder mutants.
The figure is first divided into two categories, depending on whether the gates $G$ used in each mutation are the same type or different, $G1=G2$ and $G1!=G2$ respectively. Then, each of the main categories is further divided into the possible specific combinations that can be obtained utilising our operators, for instance ``\rz - \rz'' or ``\U - \rx''. The figure shows that the majority of the generated \secondorder mutants use different gates for each mutation (68.8\%), compared to those using the same gate (31.2\%). This suggests that mutations utilising gates that modify different bases in the circuit lead to more subtle changes, making them more relevant to our objective of finding mutants that pass \testsuite and are non-equivalent. Regarding specific gates, the \rz gate seems to be the most effective one, as it contributes to the most successful combinations, and when talking about an individual combination, the double \rz seems to be the most efficient, as it individually achieves a higher percentage than the rest of the combinations, 15.1\%.

\begin{tcolorbox}[colback=blue!5!white, colframe=white, breakable]
\textbf{Concluding remarks for RQ2.2:} Mutations applied to different parts of the circuits and affecting different bases represent a larger proportion of the generated \secondorder mutants. Specifically, utilising the \rz gate when generating \secondorder mutations and applying it at the beginning or end of the circuits yields mutants that align more closely with our objective. 
\end{tcolorbox}
%In terms of the mutations applied to form \secondorder mutants, we conclude that the mutations applied to different parts of the circuits and that affect different bases are the easiest to generate, as they represent a bigger amount of the \secondorder mutants generated. We also conclude that the \rz gate is the most used gate when generating \secondorder mutations, and applying it to the beginning or the end of the circuits yields mutants that align more with our objective. 

% \caption{2nd order mutations characteristics interactions}
% \label{fig:charPie}
% \end{figure}

\subsubsection{Results for RQ2.3 -- Second order mutants test case detection}\label{sec:resRQ2.3}

In this research question, we evaluate how interactions between mutations in \secondorder mutants affect their detectability. 
% We first analyse the detection behaviour of each individual mutation independently, recording the set of unique \testcase that detect each \firstorder mutant denoted as \tcdetectionsetfom. Formally, let \tcdetectionsethom denote the set of \testcase that detect the corresponding combined \secondorder mutant. 
We compare the set of unique test cases that detect each \firstorder mutant independently \tcdetectionsetfom, with the set of test cases that detect the corresponding combined \secondorder mutants \tcdetectionsethom. We group them based on their overlap to analyse their interaction effects: 
\begin{enumerate}
\item \firstorder mutations are not detected but the \secondorder mutant is detected.% $\tcdetectionsethom \neq \varnothing,\; \tcdetectionsetfom = \varnothing$.
\item There is no overlap between the two detection sets.% $\tcdetectionsethom \cap \tcdetectionsetfom = \varnothing$
\item There is some overlap between the two detection sets.% $\tcdetectionsethom \cap \tcdetectionsetfom \neq \varnothing$
\item The two detection sets are equal.% $\tcdetectionsethom=\tcdetectionsetfom$
\item \firstorder mutations are detected and \tcdetectionsetfom includes \tcdetectionsethom.% $\tcdetectionsethom\subset\tcdetectionsetfom$
\item \firstorder mutations are detected and \tcdetectionsetfom is included in \tcdetectionsethom.% $\tcdetectionsetfom\subset\tcdetectionsethom$
\end{enumerate}

\begin{figure}[!tb]
\centering
\resizebox{\linewidth}{!}{
% \documentclass[tikz,border=10pt]{standalone}

% \begin{document}

\begin{tikzpicture}[
>=Stealth,
every node/.style={font=\large}
]

%%%%%%%%%%%%%%%%%%%%%%%%%%%%%%%%%%%%%%%%%%%%%
% DONUT CHART
%%%%%%%%%%%%%%%%%%%%%%%%%%%%%%%%%%%%%%%%%%%%%

% \def\R{2.45}
% \def\r{1.2}

% % sector boundaries
% \foreach \a in {90,-13.5,-108,-202.392,-267.624,-268.92,-270}
% {}

% \fill[myblue1]
% (0,0)--(90:\R)
% arc(90:-13.5:\R)
% --(-13.5:\r)
% arc(-13.5:90:\r)
% --cycle;

% \fill[myblue2]
% (0,0)--(-13.5:\R)
% arc(-13.5:-108:\R)
% --(-108:\r)
% arc(-108:-13.5:\r)
% --cycle;

% \fill[myblue3]
% (0,0)--(-108:\R)
% arc(-108:-202.392:\R)
% --(-202.392:\r)
% arc(-202.392:-108:\r)
% --cycle;

% \fill[myblue4]
% (0,0)--(-202.392:\R)
% arc(-202.392:-267.624:\R)
% --(-267.624:\r)
% arc(-267.624:-202.392:\r)
% --cycle;

% \fill[myblue5]
% (0,0)--(-267.624:\R)
% arc(-267.624:-268.92:\R)
% --(-268.92:\r)
% arc(-268.92:-267.624:\r)
% --cycle;

% \fill[myblue6]
% (0,0)--(-268.92:\R)
% arc(-268.92:-270:\R)
% --(-270:\r)
% arc(-270:-268.92:\r)
% --cycle;

% % \fill[myblue7]
% % (0,0)--(-270:\R)
% % arc(-270:-270:\R)
% % --(-270:\r)
% % arc(-270:-270:\r)
% % --cycle;

% \fill[white] (0,0) circle (\r);

% \draw[thick] (0,0) circle (\R);
% \draw[thick] (0,0) circle (\r);

% \node at (0,0.5) {\large High order};
% \node at (0,0) {\large mutants};
% \node at (0,-0.5) {\large (HOM)};

%%%%%%%%%%%%%%%%%%%%%%%%%%%%%%%%%%%%%%%%%%%%%
% HOM ONLY
%%%%%%%%%%%%%%%%%%%%%%%%%%%%%%%%%%%%%%%%%%%%%

\begin{scope}[shift={(-6,1.5)}]

\draw[dashed] (0,0) circle (0.8);
\fill[pattern color=myblue7, pattern=north east lines] (0,0) circle (0.8);

\node[below=25pt]
at (0,0)
{\LARGE 0.36\%};
\node[above=25pt]
at (0,0)
{\LARGE \textbf{$(1)~\tcdetectionsethom \neq \varnothing,\; \tcdetectionsetfom = \varnothing$}};

\end{scope}

%%%%%%%%%%%%%%%%%%%%%%%%%%%%%%%%%%%%%%%%%%%%%
% DISJOINT
%%%%%%%%%%%%%%%%%%%%%%%%%%%%%%%%%%%%%%%%%%%%%

\begin{scope}[shift={(0,1.5)}]

\draw (-1,0) circle (0.8);

\draw[dashed] (1,0) circle (0.8);
\fill[pattern color=myblue7, pattern=north east lines] (1,0) circle (0.8);

\node[below=25pt]
at (0,0)
{\LARGE 0.30\%};
\node[above=25pt]
at (0,0)
{\LARGE \textbf{$(2)~\tcdetectionsethom \cap \tcdetectionsetfom = \varnothing$}};

\end{scope}

%%%%%%%%%%%%%%%%%%%%%%%%%%%%%%%%%%%%%%%%%%%%%
% PARTIAL OVERLAP
%%%%%%%%%%%%%%%%%%%%%%%%%%%%%%%%%%%%%%%%%%%%%

\begin{scope}[shift={(6,1.5)}]

\draw[dashed] (-0.6,0) circle (0.8);
\draw (0.4,0) circle (0.8);
\fill[pattern color=myblue7, pattern=north east lines]
(-0.6,0) circle (0.8);

\node[below=25pt]
at (0,0)
{\LARGE 26.22\%};
\node[above=25pt]
at (0,0)
{\LARGE \textbf{$(3)~\tcdetectionsethom \cap \tcdetectionsetfom \neq \varnothing$}};

\end{scope}
%%%%%%%%%%%%%%%%%%%%%%%%%%%%%%%%%%%%%%%%%%%%%
% SAME TC
%%%%%%%%%%%%%%%%%%%%%%%%%%%%%%%%%%%%%%%%%%%%%

\begin{scope}[shift={(-6,-2)}]

\draw (0,0) circle (0.8);

\fill[pattern color=myblue7, pattern=north east lines] (0,0) circle (0.8);

\node[below=25pt]
at (0,0)
{\LARGE 28.75\%};
\node[above=25pt]
at (0,0)
{\LARGE \textbf{$(4)~\tcdetectionsethom=\tcdetectionsetfom$}};

\end{scope}

%%%%%%%%%%%%%%%%%%%%%%%%%%%%%%%%%%%%%%%%%%%%%
% HOM ⊂ FOM
%%%%%%%%%%%%%%%%%%%%%%%%%%%%%%%%%%%%%%%%%%%%%

\begin{scope}[shift={(0,-2)}]

\draw (0,0) circle (0.8);

\draw[dashed] (0,0) circle (0.5);
\fill[pattern color=myblue7, pattern=north east lines] (0,0) circle (0.5);

\node[below=25pt]
at (0,0)
{\LARGE 26.25\%};
\node[above=25pt]
at (0,0)
{\LARGE \textbf{$(5)~\tcdetectionsethom\subset\tcdetectionsetfom$}};

\end{scope}

%%%%%%%%%%%%%%%%%%%%%%%%%%%%%%%%%%%%%%%%%%%%%
% FOM ⊂ HOM
%%%%%%%%%%%%%%%%%%%%%%%%%%%%%%%%%%%%%%%%%%%%%

\begin{scope}[shift={(6,-2)}]

\draw[dashed] (0,0) circle (0.8);
\fill[pattern color=myblue7, pattern=north east lines]
(0,0) circle (0.8);

\draw (0,0) circle (0.5);

\node[below=25pt]
at (0,0)
{\LARGE 18.12\%};
\node[above=25pt]
at (0,0)
{\LARGE \textbf{$(6)~\tcdetectionsetfom\subset\tcdetectionsethom$}};

\end{scope}

%%%%%%%%%%%%%%%%%%%%%%%%%%%%%%%%%%%%%%%%%%%%%
% NO DETECTION
%%%%%%%%%%%%%%%%%%%%%%%%%%%%%%%%%%%%%%%%%%%%%

% \begin{scope}[shift={(-4.5,4.5)}]

% \draw[dashed] (0,0) circle (0.6);

% \node[below=20pt]
% at (0,0)
% {\large 2.72\%};
% \node[above=20pt]
% at (0,0)
% {\large \textbf{$\tcdetectionsethom = \varnothing,\; \tcdetectionsetfom = \varnothing$}};

% \end{scope}

%%%%%%%%%%%%%%%%%%%%%%%%%%%%%%%%%%%%%%%%%%%%%
% CONNECTORS
%%%%%%%%%%%%%%%%%%%%%%%%%%%%%%%%%%%%%%%%%%%%%
% X, Y -> X, Y

% \draw[->,thick] (-0.3,2.9) -- (-1.5,3.5); %No Detection

% \draw[->,thick] (-0.08,2.4) -- (-1.4,3.1); %HOM only

% \draw[->,thick] (0,2.4) -- (1.3,3.1); %Disjoint

% \draw[->,thick] (1.9,1.5) -- (3.5,1); %Same TC

% \draw[->,thick] (-1.9,1.5) -- (-3.5,1); % FOM subsumed by HOM

% \draw[->,thick] (-1.2,-2.15) -- (-2.3,-2.8); % Partial Overlap

% \draw[->,thick] (1.2,-2.15) -- (2.3,-2.8); % HOM subsumed by FOM

%%%%%%%%%%%%%%%%%%%%%%%%%%%%%%%%%%%%%%%%%%%%%
% LEGEND
%%%%%%%%%%%%%%%%%%%%%%%%%%%%%%%%%%%%%%%%%%%%%

\node[align=left,font=\small]
at (0,-4)
{
Blue Dashed Circle = \highorder detection test set \qquad
White Circle = \firstorder detection test set \qquad
};

\end{tikzpicture}

% \end{document}
}
\caption{RQ2.3 -- Distribution of \secondorder mutants by the relationship between \tcdetectionsetfom and \tcdetectionsethom.}
\Description{RQ2.3}
\label{fig:hom-distribution}
\end{figure}
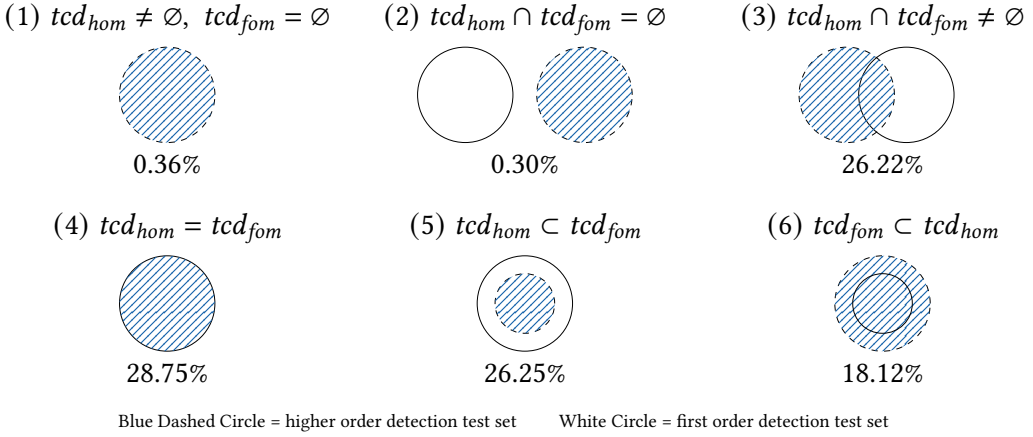

Figure~\ref{fig:hom-distribution} shows the distribution of the different groups of \testcasedetecting that we obtained and their proportions among the generated \secondorder mutants.
The figure shows that most \secondorder generated mutants share some extent of \tcdetectionsethom with \tcdetectionsetfom. In only 0.66\% of cases \tcdetectionsethom is completely independent: in 0.36\%, \firstorder mutations are not detected independently (case (1)), but the \secondorder mutant is; and in 0.3\%, there is no overlap between the two sets (case (2)). Among the other cases, 28.75\% show the exact same set for both \tcdetectionsethom and \tcdetectionsetfom (case (4)), while the rest show some interaction between the two mutations that either increases or decreases their detection. In case of the 18.12\% of mutants where \tcdetectionsetfom is included in \tcdetectionsethom (case (6)), we can conclude that by combining both independent mutations in the same circuit, the effect of the fault was increased and thus the \secondorder mutant was detected by more \testcase. On the contrary, in 26.25\% of the cases \tcdetectionsethom is included in \tcdetectionsetfom (case (5)), meaning that by combining both mutations in the same circuit, the effect of the mutation was reduced, reducing the \testcase detecting them. Lastly, we also have 26.22\% of cases where \testcase detection sets differ but share some tests (case (3)), indicating that combining mutations in the same circuits can cause some \testcase to stop detecting them while others appear.

\begin{tcolorbox}[colback=blue!5!white, colframe=white, breakable]
\textbf{Concluding remarks for RQ2.3:} In 71.25\% of \secondorder mutants, the set of test cases that detect the \secondorder mutant is different from the set of test cases that detect each of the \firstorder mutants. This shows that in most cases, when \firstorder mutants are introduced in the same circuit, they interact with each other and change the behaviour of the mutant. The effect of the mutations was either reduced or increased when another mutation was introduced into the circuit, and in some cases, even mutations that were not detected independently were detected together, highlighting the need for new test cases. 
\end{tcolorbox}

\section{Discussion and Recommendations}\label{sec:discussion}
In this section, we present the findings of our study and reflect on their implications for both researchers and practitioners. We also address the threats to the validity of our study, emphasising its internal, external, and construct validity.

\subsection{Findings}

The findings of this study demonstrate that search-based optimisation is an effective approach for generating challenging quantum circuit mutants that are currently undetected by existing test suites, yet remain not equivalent. Unlike mutation analysis tools in the literature, which often create easily detectable mutants, \approach focuses the search on areas of the mutation space that highlight weaknesses in the current test suite. 

The experiments demonstrated the importance of explicitly considering the effect of the mutation during optimisation. Random search, which lacks guidance, produced a higher proportion of equivalent mutants than the guided search algorithms. This supports the inclusion of the second term in the fitness function (see Equation~\ref{eq:fitnessFunction}), which rewards mutants that remain challenging to detect while producing measurable variations from the original circuit. Without such guidance, the search algorithms naturally tend towards mutations that pass testing because they have little or no impact on the circuit, rather than because they represent genuinely challenging faulty scenarios, resulting in a higher number of equivalent mutants. The experiments also highlight the benefit of the newly introduced parametric mutation operators, as even random search, despite its simplicity, consistently generated more challenging mutants than existing quantum mutation tools. 

%The finding highlight the importance of the newly introduced parametric mutation operators, whose use of parametric quantum gates creates a richer mutation space than the operators used in existing quantum circuit mutation tools. Despite its simplicity, random search consistently generated more challenging mutants than existing quantum mutation tools. 
% A particularly interesting observation is the competitive performance of random search. 

Another important finding is that more challenging mutants alone are insufficient for improving test suites. While more challenging mutants expose weaknesses in existing tests, repeatedly generating similar mutants yields weak results, as they often require the same additional test case to detect them. Our results therefore suggest that diversity should also be considered as an objective for search-based mutant generation. This observation explains the performance of the genetic algorithm, whose exploration maintains greater diversity than the local search approaches. By discovering a broader range of challenging mutants, the genetic algorithm exposes a wider variety of testing deficiencies and, consequently, provides greater opportunities to improve the overall quality of the test suite.

Finally, the analysis of \highorder mutants shows that interactions between \firstorder mutations play an important role in quantum mutation testing. In the experiments, although the generation of \firstorder and \secondorder mutants exhibits comparable performance, the behaviour of \secondorder mutants cannot always be predicted based on their individual \firstorder mutations. We observed instances in which combining mutations increased, decreased, or altered their detectability compared with evaluating the same mutations independently as \firstorder mutants. This indicates that \highorder mutants are particularly valuable for quantum mutation analysis, as they often necessitate new test cases that would not arise from considering mutations independently as \firstorder mutants.

\subsection{Implications for Researchers}
Given the current state of quantum circuit mutation analysis and the limited availability of benchmarks, the introduction of new mutation operators and the representation of undetected faults provide valuable contributions to the research community. Existing quantum mutation approaches are often constrained by predefined mutation operators, potentially leading to an incomplete assessment of test suite effectiveness. By targeting the generation of mutants that remain undetected by existing test suites, \approach enables researchers to explore more challenging mutants and evaluate whether testing techniques can detect subtle faults.

The generated mutants can serve as complementary evaluation benchmarks for quantum software testing approaches. Researchers developing new testing techniques can utilise these mutants to assess the effectiveness of their methods beyond traditional mutation operators. In particular, the ability to generate mutants specifically tailored to the weaknesses of a given test suite provides opportunities to study the limitations of current quantum testing practices and to identify areas where additional testing strategies are required.

Furthermore, the introduction of \highorder mutants into quantum circuits creates new research opportunities for understanding possible fault interactions in quantum programs. Unlike \firstorder mutants, where individual changes may only represent isolated faults, \highorder mutants allow researchers to investigate the combined effects of multiple faults and their potential interactions within quantum circuits. These mutants can be used to analyse complex behaviours, like qubit dependencies or error propagation across entangled qubits. Such analyses may provide deeper insights into the relationship between circuit structure, fault characteristics, and test effectiveness.

Finally, this work highlights the potential of search-based approaches for exploring the quantum mutation space. Rather than relying solely on manually designed mutation operators, future research can investigate adaptive or learning-based approaches for discovering new quantum faults. This opens opportunities for combining mutation testing with optimisation approaches, automated test generation, and fault modelling to improve the robustness and reliability of quantum software testing methodologies.

\subsection{Implications for Practitioners}
Mutation analysis can be costly and time-consuming, requiring significant computational resources. This challenge is amplified in quantum software, where existing mutation tools often produce trivial mutants that add little value. \approach allows practitioners to focus on generating targeted, challenging mutants that highlight weaknesses in their testing techniques. By creating mutants that evade detection by the current test suite, valuable insights can be gained about gaps in the test suite. This targeted methodology shifts mutation analysis from a broad evaluation of test adequacy to a more focused approach, prioritising efforts on faults that expose limitations in the existing test suite.

The incorporation of \highorder mutants further extends the practical applicability of mutation analysis by enabling the evaluation of multiple mutations in the same quantum circuit. In real-world quantum software, faults may not always occur independently, and interactions between multiple faulty components may lead to unexpected behaviours. By considering combinations of mutations, practitioners can evaluate whether their test suites can detect more complex faulty scenarios and improve their coverage of diverse faults.

Furthermore, as \approach iteratively generates mutants based on the current test suite, it can be integrated into existing quantum software testing pipelines. For example, during continuous integration and testing processes, \approach could periodically search for new mutants and use them as indicators of potential weaknesses in the testing technique used. When no additional non-equivalent mutants can be discovered, this provides practitioners with greater confidence that the test suite has been evaluated against a broader and more challenging set of faulty scenarios.

Overall, \approach provides practitioners with a mechanism to complement existing mutation testing practices by prioritising mutant quality over quantity. Rather than generating potentially redundant mutants, practitioners can focus on a set of higher quality mutants that provide deeper insights into test suite effectiveness.

\subsection{Threats to Validity}

In this section, we will discuss the key factors that may affect the validity of our conclusions and outline the measures taken to minimise their impact. We will address concerns related to the validity of our measurements and evaluation criteria, the construct validity, the experimental methodology, internal validity, and the generalisability of our findings beyond the evaluated settings, external validity.

\subsubsection{Construct Validity}

Construct validity assesses whether the metrics and evaluation methods used accurately measure the concepts examined in this study. In our experiments, we assess mutant detection using the chi-square test with a fixed significance threshold, comparing the output distributions of the original and mutated circuits, as in several earlier studies~\cite{quitoASE21tool, WangICST2021, Mendiluze2021, MendiluzeUsandizaga2025}. 
While this offers an objective and reproducible criterion for identifying mutant detection, \approach offers the opportunity to utilise different statistical tests or significance thresholds, which might categorise some mutants differently. 
Since we developed the fitness function in a generic manner, we also needed to choose a specific distance metric to quantify the effect of the mutations in our experiments: the Hellinger distance.
Although the Hellinger distance is well suited for comparing probability distributions~\cite{muqeet2024mitigating,pauliStringsASE2024,pontolillo2025ideal,robustmutationanalysisquantum}, other distance metrics, such as Jensen-Shannon divergence, could alter the optimisation process and potentially lead to different search behaviour.

Another threat concerns the identification of equivalent mutants. Since proving semantic equivalence between two quantum circuits is generally infeasible, \approach assumes that mutants that cannot be distinguished by our comprehensive test suite are equivalent for the purposes of this study. Consequently, some mutants that pass the test suite may still be behaviourally different under inputs or observations not exercised by the test suite. To reduce this threat, we employ an extensive input-output testing strategy using both classical and quantum input states, with measurements performed in the three bases ($X$, $Y$, and $Z$). 
%Although this cannot guarantee the identification of all behavioural differences, it provides broad coverage of circuit behaviour and increases confidence in the equivalence approximation adopted throughout the experiments.

Another aspect related to construct validity concerns the representation of circuit behaviour used for mutant detection. In this study, we utilise output probability distributions as the primary oracle, as they provide an observable representation of quantum circuit execution and are directly applicable when evaluating measurement outcomes. However, output distributions are not the only possible approach for characterising quantum circuit behaviour. Alternative representations, such as quantum state vectors or expectation values, may capture different aspects of circuit execution and could potentially identify behavioural differences that are not reflected in the measured output distributions. Consequently, mutants classified as equivalent under our oracle definition may exhibit differences when evaluated using alternative behavioural representations. Nevertheless, output distributions represent one of the most practical and widely applicable approaches for quantum software testing~\cite{klamroth2025detecting,pauliStringsASE2024,pontolillo2025ideal,muqeet2024mitigating, MendiluzeUsandizaga2025,XinyiLanscape}, particularly when considering execution on current quantum hardware.

Another potential threat concerns the realism of the generated mutants. Although the proposed mutation operators are designed to introduce challenging and behaviourally meaningful faults, they are not derived from existing faults. Therefore, the generated mutants should be interpreted as fault-representative testing artefacts rather than exact replications of real-world quantum programming faults. Nevertheless, the objective of mutation analysis is not necessarily to reproduce every possible real fault, but rather to provide challenging and informative artefacts for evaluating and improving the effectiveness of testing approaches.

\subsubsection{Internal Validity}

Internal validity concerns whether the observed results are influenced by aspects of the experimental methodology rather than the proposed approach itself. Since all evaluated search approaches are stochastic, their performance may vary between executions due to different random initialisations. To reduce the influence of randomness, every experiment was repeated ten times. We chose ten repetitions to balance obtaining representative results while managing computational costs, given the number of circuits, test cases, and evaluations required.
%and the reported results aggregate the outcomes across all executions.

The performance of search-based optimisation approaches is also influenced by their parameter settings, including the population size, mutation probability, and crossover probability. The parameters adopted in this work were selected through preliminary experimentation to provide stable search behaviour across all benchmark circuits. Nevertheless, different parameter configurations may influence convergence speed or the characteristics of the generated mutants.

% Finally, to minimise the risk of biased conclusions favouring a specific optimisation strategy, we evaluated four different search algorithms that represent various search paradigms, including local search and population-based approaches.
To assess whether the effectiveness of the proposed mutation framework depends on a particular optimisation strategy, we evaluated four search algorithms representing different search paradigms, including local search and population-based approaches. The algorithms' consistent ability to generate challenging mutants enhances our confidence that the improvements observed are due to the proposed mutation framework, rather than being a result of a single search algorithm.

\subsubsection{External Validity}

External validity concerns the extent to which the findings generalise beyond the experimental setting. Our evaluation was conducted on six quantum programs comprising a total of thirty quantum circuits spanning different categories and circuit sizes. Although these benchmarks represent a diverse collection of quantum algorithms commonly adopted in the literature, they do not cover the full spectrum of emerging quantum applications.

The experiments were performed using quantum circuit simulation and circuits containing up to seven qubits. This limitation was primarily imposed by the computational cost of repeatedly executing thousands of test cases across multiple generations and independent search runs. Consequently, the reported results may differ for substantially larger quantum circuits or realistic environments. We acknowledge this as a possible limitation of the study; however, given the current state of quantum computers, we believe that the inherent noise in these devices makes it infeasible to evaluate our approaches, as the effect of the mutations can be hidden by noise or even aggravated~\cite{robustmutationanalysisquantum}.

%To partially mitigate this threat, we additionally evaluated \approach on a 24-qubit quantum circuit, representing the largest benchmark supported by the simulator utilised, and successfully executed the generated circuits on a real IBM quantum computer. Although these experiments were not part of the primary empirical evaluation, they demonstrate that \approach can be executed on both larger circuits and real quantum hardware, providing initial evidence of its practical applicability and scalability.
\section{Conclusions and Future Work}\label{sec:Conclusionsandfuture}
In this paper, we presented \approach, a search-based approach for generating quantum circuit mutants that are not detected by a given test suite yet not equivalent. By introducing novel parametric mutation operators and utilising search-based algorithms to explore the resulting mutation space, \approach systematically generates mutants that reveal previously undetected faults and help improve the test suite.

Our evaluation of 30 quantum circuits spanning six quantum algorithms demonstrated that \approach consistently generates more challenging mutants than existing quantum circuit mutation tools. Among the evaluated search algorithms, the genetic algorithm achieved the best overall performance by producing a higher proportion of non-equivalent mutants that required new test cases for detection. Furthermore, we introduced \highorder mutation testing for quantum circuits for the first time, and showed that interactions between multiple mutations produce behaviours that cannot always be inferred from their constituent \firstorder mutants, highlighting the potential of \highorder mutation analysis for understanding fault interactions in quantum software.

Future work will explore several directions to further advance search-based quantum mutation testing. First, we plan to investigate multi-objective search approaches that optimise different mutation objectives simultaneously, rather than combining them into a single fitness function. For example, the objectives of generating undetected mutants and maximising behavioural impact could be optimised independently, allowing the search to identify a broader range of solutions. Mutant diversity could also be incorporated as a separate optimisation objective to encourage the generation of distinct mutants that expose different weaknesses in a test suite. Another promising direction is the design of additional mutation operators capable of representing a wider variety of quantum faults. In particular, we intend to investigate parameterisable mutations involving multi-qubit gates.

Finally, we would also like to investigate alternative mechanisms for mutant detection, including different statistical tests and similarity measures, together with other quantum software testing techniques and oracle definitions, to assess how the effectiveness of search-based mutant generation varies under different testing settings.

%Overall, this work demonstrates that search-based mutation analysis provides an effective way to generate challenging mutants in a targeted manner. Rather than maximising the number of generated mutants, \approach prioritises mutants that expose weaknesses in existing test suites, providing more informative mutants for evaluating and improving quantum software testing.

\begin{acks}
E. Mendiluze Usandizaga is supported by Simula's internal strategic project on quantum software engineering. S. Ali is supported by the Norwegian Quantum Software Center (Project \#361350) funded by the Research Council of Norway and Oslo Metropolitan University's Quantum Hub. P. Arcaini is supported by the ASPIRE grant No. JPMJAP2301, JST. The research presented in this paper has benefited from the Experimental Infrastructure for Exploration of Exascale Computing (eX3), which is financially supported by the Research Council of Norway under contract 270053. This work was supported, in part, by Research Ireland grants 20/FFP-P/8818, and 13/RC/2094\_P2 and co-funded under the European Regional Development Fund through the Southern \& Eastern Regional Operational Programme to Lero - the Research Ireland Research Centre for Software (www.lero.ie). AI-based tools, including ChatGPT, Gemini, and Grammarly, were used to assist with refining and polishing the manuscript text, including grammar and language checks. These tools were also used to support the visual presentation and refinement of plots, figures, and tables. The authors reviewed and validated all AI-assisted content and remain fully responsible for the final content of the publication.

\end{acks}

\bibliographystyle{ACM-Reference-Format}
\bibliography{references}


\begin{thebibliography}{59}

%%% the Web address.

\ifx \showCODEN    \undefined \def \showCODEN     #1{\unskip}     \fi
\ifx \showISBNx    \undefined \def \showISBNx     #1{\unskip}     \fi
\ifx \showISBNxiii \undefined \def \showISBNxiii  #1{\unskip}     \fi
\ifx \showISSN     \undefined \def \showISSN      #1{\unskip}     \fi
\ifx \showLCCN     \undefined \def \showLCCN      #1{\unskip}     \fi
\ifx \shownote     \undefined \def \shownote      #1{#1}          \fi
\ifx \showarticletitle \undefined \def \showarticletitle #1{#1}   \fi
\ifx \showURL      \undefined \def \showURL       {\relax}        \fi
% invisible to TeX
\providecommand\bibfield[2]{#2}
\providecommand\bibinfo[2]{#2}
\providecommand\natexlab[1]{#1}
\providecommand\showeprint[2][]{arXiv:#2}

\bibitem[Abreu et~al\mbox{.}(2026)]%
        {stInQuantumWorldIEEEComputer2026}
\bibfield{author}{\bibinfo{person}{Rui Abreu}, \bibinfo{person}{Shaukat Ali}, \bibinfo{person}{Paolo Arcaini}, \bibinfo{person}{José Campos}, \bibinfo{person}{Michael Felderer}, \bibinfo{person}{Claude Gravel}, \bibinfo{person}{Fuyuki Ishikawa}, \bibinfo{person}{Stefan Klikovits}, \bibinfo{person}{Andriy Miranskyy}, \bibinfo{person}{Anila Mjeda}, \bibinfo{person}{Mohammad~Reza Mousavi}, \bibinfo{person}{Masaomi Yamaguchi}, \bibinfo{person}{Lei Zhang}, {and} \bibinfo{person}{Jianjun Zhao}.} \bibinfo{year}{2026}\natexlab{}.
\newblock \showarticletitle{Software Testing in the Quantum World}.
\newblock \bibinfo{journal}{\emph{Computer}} \bibinfo{volume}{59}, \bibinfo{number}{4} (\bibinfo{year}{2026}), \bibinfo{pages}{135--138}.
\newblock
\href{https://doi.org/10.1109/MC.2026.3655854}{doi:\nolinkurl{10.1109/MC.2026.3655854}}


\bibitem[Abreu et~al\mbox{.}(2022)]%
        {metamorphic}
\bibfield{author}{\bibinfo{person}{Rui Abreu}, \bibinfo{person}{João~Paulo Fernandes}, \bibinfo{person}{Luis Llana}, {and} \bibinfo{person}{Guilherme Tavares}.} \bibinfo{year}{2022}\natexlab{}.
\newblock \showarticletitle{Metamorphic Testing of Oracle Quantum Programs}. In \bibinfo{booktitle}{\emph{2022 IEEE/ACM 3rd International Workshop on Quantum Software Engineering (Q-SE)}}. \bibinfo{pages}{16--23}.
\newblock
\href{https://doi.org/10.1145/3528230.3529189}{doi:\nolinkurl{10.1145/3528230.3529189}}


\bibitem[Ali et~al\mbox{.}(2021)]%
        {testingQuantumICST2021}
\bibfield{author}{\bibinfo{person}{Shaukat Ali}, \bibinfo{person}{Paolo Arcaini}, \bibinfo{person}{Xinyi Wang}, {and} \bibinfo{person}{Tao Yue}.} \bibinfo{year}{2021}\natexlab{}.
\newblock \showarticletitle{Assessing the Effectiveness of Input and Output Coverage Criteria for Testing Quantum Programs}. In \bibinfo{booktitle}{\emph{2021 IEEE 14th International Conference on Software Testing, Validation and Verification (ICST)}}. \bibinfo{pages}{13--23}.
\newblock
\href{https://doi.org/10.1109/ICST49551.2021.00014}{doi:\nolinkurl{10.1109/ICST49551.2021.00014}}


\bibitem[Ammann et~al\mbox{.}(2014)]%
        {AmmannICST2014}
\bibfield{author}{\bibinfo{person}{Paul Ammann}, \bibinfo{person}{Marcio~Eduardo Delamaro}, {and} \bibinfo{person}{Jeff Offutt}.} \bibinfo{year}{2014}\natexlab{}.
\newblock \showarticletitle{Establishing Theoretical Minimal Sets of Mutants}. In \bibinfo{booktitle}{\emph{2014 IEEE Seventh International Conference on Software Testing, Verification and Validation}}. \bibinfo{pages}{21--30}.
\newblock
\href{https://doi.org/10.1109/ICST.2014.13}{doi:\nolinkurl{10.1109/ICST.2014.13}}


\bibitem[Back(1996)]%
        {back1996evolutionary}
\bibfield{author}{\bibinfo{person}{Thomas Back}.} \bibinfo{year}{1996}\natexlab{}.
\newblock \bibinfo{booktitle}{\emph{Evolutionary algorithms in theory and practice: evolution strategies, evolutionary programming, genetic algorithms}}.
\newblock \bibinfo{publisher}{Oxford university press}.
\newblock


\bibitem[Campos and Souto(2021)]%
        {qbugs}
\bibfield{author}{\bibinfo{person}{Jos\'{e} Campos} {and} \bibinfo{person}{Andr{\'e} Souto}.} \bibinfo{year}{2021}\natexlab{}.
\newblock \showarticletitle{{QBugs}: A Collection of Reproducible Bugs in Quantum Algorithms and a Supporting Infrastructure to Enable Controlled Quantum Software Testing and Debugging Experiments}. In \bibinfo{booktitle}{\emph{2021 IEEE/ACM 2nd International Workshop on Quantum Software Engineering (Q-SE)}}. \bibinfo{publisher}{IEEE Computer Society}, \bibinfo{address}{Los Alamitos, CA, USA}, \bibinfo{pages}{28--32}.
\newblock
\href{https://doi.org/10.1109/Q-SE52541.2021.00013}{doi:\nolinkurl{10.1109/Q-SE52541.2021.00013}}


\bibitem[Cliff(1993)]%
        {cliff1993dominance}
\bibfield{author}{\bibinfo{person}{Norman Cliff}.} \bibinfo{year}{1993}\natexlab{}.
\newblock \showarticletitle{Dominance statistics: Ordinal analyses to answer ordinal questions}.
\newblock \bibinfo{journal}{\emph{Psychological bulletin}} \bibinfo{volume}{114}, \bibinfo{number}{3} (\bibinfo{year}{1993}), \bibinfo{pages}{494}.
\newblock
\href{https://doi.org/10.1037/0033-2909.114.3.494}{doi:\nolinkurl{10.1037/0033-2909.114.3.494}}


\bibitem[de~la Barrera et~al\mbox{.}(2022)]%
        {delaBarrera2022}
\bibfield{author}{\bibinfo{person}{Antonio~Garc{\'i}a de~la Barrera}, \bibinfo{person}{Ignacio Garc{\'i}a-Rodr{\'i}guez de Guzm{\'a}n}, \bibinfo{person}{Macario Polo}, {and} \bibinfo{person}{Jos{\'e}~A. Cruz-Lemus}.} \bibinfo{year}{2022}\natexlab{}.
\newblock \bibinfo{booktitle}{\emph{Quantum Software Testing: Current Trends and Emerging Proposals}}.
\newblock \bibinfo{publisher}{Springer International Publishing}, \bibinfo{address}{Cham}, \bibinfo{pages}{167--191}.
\newblock
\showISBNx{978-3-031-05324-5}
\href{https://doi.org/10.1007/978-3-031-05324-5_9}{doi:\nolinkurl{10.1007/978-3-031-05324-5_9}}


\bibitem[de~la Barrera et~al\mbox{.}(2023)]%
        {QST_SOTA}
\bibfield{author}{\bibinfo{person}{Antonio~García de~la Barrera}, \bibinfo{person}{Ignacio García-Rodríguez~de Guzmán}, \bibinfo{person}{Macario Polo}, {and} \bibinfo{person}{Mario Piattini}.} \bibinfo{year}{2023}\natexlab{}.
\newblock \showarticletitle{Quantum software testing: State of the art}.
\newblock \bibinfo{journal}{\emph{Journal of Software: Evolution and Process}} \bibinfo{volume}{35}, \bibinfo{number}{4} (\bibinfo{year}{2023}), \bibinfo{pages}{e2419}.
\newblock
\href{https://doi.org/10.1002/smr.2419}{doi:\nolinkurl{10.1002/smr.2419}}


\bibitem[Dem\v{s}ar(2006)]%
        {friedman}
\bibfield{author}{\bibinfo{person}{Janez Dem\v{s}ar}.} \bibinfo{year}{2006}\natexlab{}.
\newblock \showarticletitle{Statistical Comparisons of Classifiers over Multiple Data Sets}.
\newblock \bibinfo{journal}{\emph{J. Mach. Learn. Res.}}  \bibinfo{volume}{7} (\bibinfo{date}{Dec.} \bibinfo{year}{2006}), \bibinfo{pages}{1--30}.
\newblock
\showISSN{1532-4435}


\bibitem[Fortunato et~al\mbox{.}(2022a)]%
        {QMutPy3}
\bibfield{author}{\bibinfo{person}{Daniel Fortunato}, \bibinfo{person}{Jos{\'e} Campos}, {and} \bibinfo{person}{Rui Abreu}.} \bibinfo{year}{2022}\natexlab{a}.
\newblock \showarticletitle{Mutation Testing of Quantum Programs: A Case Study With {Qiskit}}.
\newblock \bibinfo{journal}{\emph{IEEE Transactions on Quantum Engineering}}  \bibinfo{volume}{3} (\bibinfo{year}{2022}), \bibinfo{pages}{1--17}.
\newblock
\href{https://doi.org/10.1109/TQE.2022.3195061}{doi:\nolinkurl{10.1109/TQE.2022.3195061}}


\bibitem[Fortunato et~al\mbox{.}(2022b)]%
        {QMutPy2}
\bibfield{author}{\bibinfo{person}{Daniel Fortunato}, \bibinfo{person}{Jos\'{e} Campos}, {and} \bibinfo{person}{Rui Abreu}.} \bibinfo{year}{2022}\natexlab{b}.
\newblock \showarticletitle{Mutation Testing of Quantum Programs Written in {QISKit}}. In \bibinfo{booktitle}{\emph{Proceedings of the ACM/IEEE 44th International Conference on Software Engineering: Companion Proceedings}} (Pittsburgh, Pennsylvania) \emph{(\bibinfo{series}{ICSE '22})}. \bibinfo{publisher}{Association for Computing Machinery}, \bibinfo{address}{New York, NY, USA}, \bibinfo{pages}{358--359}.
\newblock
\showISBNx{9781450392235}
\href{https://doi.org/10.1145/3510454.3528649}{doi:\nolinkurl{10.1145/3510454.3528649}}


\bibitem[Fortunato et~al\mbox{.}(2022c)]%
        {QmutPy}
\bibfield{author}{\bibinfo{person}{Daniel Fortunato}, \bibinfo{person}{Jos\'{e} Campos}, {and} \bibinfo{person}{Rui Abreu}.} \bibinfo{year}{2022}\natexlab{c}.
\newblock \showarticletitle{{QMutPy}: A Mutation Testing Tool for Quantum Algorithms and Applications in Qiskit}. In \bibinfo{booktitle}{\emph{Proceedings of the 31st ACM SIGSOFT International Symposium on Software Testing and Analysis}} (Virtual, South Korea) \emph{(\bibinfo{series}{ISSTA 2022})}. \bibinfo{publisher}{Association for Computing Machinery}, \bibinfo{address}{New York, NY, USA}, \bibinfo{pages}{797--800}.
\newblock
\showISBNx{9781450393799}
\href{https://doi.org/10.1145/3533767.3543296}{doi:\nolinkurl{10.1145/3533767.3543296}}


\bibitem[Fortz et~al\mbox{.}(2026)]%
        {robustmutationanalysisquantum}
\bibfield{author}{\bibinfo{person}{Sophie Fortz}, \bibinfo{person}{Eñaut~Mendiluze Usandizaga}, \bibinfo{person}{Shaukat Ali}, \bibinfo{person}{Paolo Arcaini}, {and} \bibinfo{person}{Mohammad~Reza Mousavi}.} \bibinfo{year}{2026}\natexlab{}.
\newblock \bibinfo{title}{Robust Mutation Analysis of Quantum Programs Under Noise}.
\newblock
\showeprint[arxiv]{2605.13279}~[cs.SE]
\urldef\tempurl%
\url{https://arxiv.org/abs/2605.13279}
\showURL{%
\tempurl}


\bibitem[Glantz(2012)]%
        {holmcorrection}
\bibfield{author}{\bibinfo{person}{Stanton~A. Glantz}.} \bibinfo{year}{2012}\natexlab{}.
\newblock \bibinfo{booktitle}{\emph{Primer of Biostatistics} (\bibinfo{edition}{7th edition} ed.)}.
\newblock \bibinfo{publisher}{McGraw Hill}, \bibinfo{address}{New York}.
\newblock
\showISBNx{9780071781503}
\urldef\tempurl%
\url{https://www.accessscience.com/content/book/9780071781503}
\showURL{%
\tempurl}


\bibitem[Greenberger et~al\mbox{.}(1989)]%
        {ghz}
\bibfield{author}{\bibinfo{person}{Daniel~M. Greenberger}, \bibinfo{person}{Michael~A. Horne}, {and} \bibinfo{person}{Anton Zeilinger}.} \bibinfo{year}{1989}\natexlab{}.
\newblock \showarticletitle{Going {Beyond} {Bell}’s {Theorem}}.
\newblock In \bibinfo{booktitle}{\emph{Bell’s {Theorem}, {Quantum} {Theory} and {Conceptions} of the {Universe}}}, \bibfield{editor}{\bibinfo{person}{Menas Kafatos}} (Ed.). \bibinfo{publisher}{Springer Netherlands}, \bibinfo{address}{Dordrecht}, \bibinfo{pages}{69--72}.
\newblock
\showISBNx{978-94-017-0849-4}
\href{https://doi.org/10.1007/978-94-017-0849-4_10}{doi:\nolinkurl{10.1007/978-94-017-0849-4_10}}


\bibitem[Honarvar et~al\mbox{.}(2020)]%
        {honarvar2020property}
\bibfield{author}{\bibinfo{person}{Shahin Honarvar}, \bibinfo{person}{Mohammad~Reza Mousavi}, {and} \bibinfo{person}{Rajagopal Nagarajan}.} \bibinfo{year}{2020}\natexlab{}.
\newblock \showarticletitle{Property-Based Testing of Quantum Programs in {Q\#}}. In \bibinfo{booktitle}{\emph{Proceedings of the IEEE/ACM 42nd International Conference on Software Engineering Workshops}} (Seoul, Republic of Korea) \emph{(\bibinfo{series}{ICSEW'20})}. \bibinfo{publisher}{Association for Computing Machinery}, \bibinfo{address}{New York, NY, USA}, \bibinfo{pages}{430--435}.
\newblock
\showISBNx{9781450379632}
\href{https://doi.org/10.1145/3387940.3391459}{doi:\nolinkurl{10.1145/3387940.3391459}}


\bibitem[Ishimoto et~al\mbox{.}(2025)]%
        {MutationBasedFL}
\bibfield{author}{\bibinfo{person}{Yuta Ishimoto}, \bibinfo{person}{Masanari Kondo}, \bibinfo{person}{Naoyasu Ubayashi}, \bibinfo{person}{Yasutaka Kamei}, \bibinfo{person}{Ryota Katsube}, \bibinfo{person}{Naoto Sato}, {and} \bibinfo{person}{Hideto Ogawa}.} \bibinfo{year}{2025}\natexlab{}.
\newblock \showarticletitle{Evaluating Mutation-based Fault Localization for Quantum Programs}. In \bibinfo{booktitle}{\emph{Proceedings of the 29th International Conference on Evaluation and Assessment in Software Engineering}} \emph{(\bibinfo{series}{EASE '25})}. \bibinfo{publisher}{Association for Computing Machinery}, \bibinfo{address}{New York, NY, USA}, \bibinfo{pages}{666--671}.
\newblock
\showISBNx{9798400713859}
\href{https://doi.org/10.1145/3756681.3757022}{doi:\nolinkurl{10.1145/3756681.3757022}}


\bibitem[Jia and Harman(2011)]%
        {MutationSurvey}
\bibfield{author}{\bibinfo{person}{Yue Jia} {and} \bibinfo{person}{Mark Harman}.} \bibinfo{year}{2011}\natexlab{}.
\newblock \showarticletitle{An Analysis and Survey of the Development of Mutation Testing}.
\newblock \bibinfo{journal}{\emph{IEEE Transactions on Software Engineering}} \bibinfo{volume}{37}, \bibinfo{number}{5} (\bibinfo{year}{2011}), \bibinfo{pages}{649--678}.
\newblock
\href{https://doi.org/10.1109/TSE.2010.62}{doi:\nolinkurl{10.1109/TSE.2010.62}}


\bibitem[Kawai(2022)]%
        {RGatesConversion}
\bibfield{author}{\bibinfo{person}{Ryoichi Kawai}.} \bibinfo{year}{2022}\natexlab{}.
\newblock \bibinfo{title}{A First Step to Quantum Computation with Qiskit}.
\newblock
\urldef\tempurl%
\url{https://kawaihome.link/jbooks/qcomp-short/cover.html}
\showURL{%
\tempurl}


\bibitem[Kintis et~al\mbox{.}(2016)]%
        {KintisSCAM2016}
\bibfield{author}{\bibinfo{person}{Marinos Kintis}, \bibinfo{person}{Mike Papadakis}, \bibinfo{person}{Andreas Papadopoulos}, \bibinfo{person}{Evangelos Valvis}, {and} \bibinfo{person}{Nicos Malevris}.} \bibinfo{year}{2016}\natexlab{}.
\newblock \showarticletitle{Analysing and Comparing the Effectiveness of Mutation Testing Tools: {A} Manual Study}. In \bibinfo{booktitle}{\emph{2016 IEEE 16th International Working Conference on Source Code Analysis and Manipulation (SCAM)}}. \bibinfo{pages}{147--156}.
\newblock
\href{https://doi.org/10.1109/SCAM.2016.28}{doi:\nolinkurl{10.1109/SCAM.2016.28}}


\bibitem[Klamroth et~al\mbox{.}(2025)]%
        {klamroth2025detecting}
\bibfield{author}{\bibinfo{person}{Jonas Klamroth}, \bibinfo{person}{Max Scheerer}, {and} \bibinfo{person}{Oliver Denninger}.} \bibinfo{year}{2025}\natexlab{}.
\newblock \showarticletitle{Detecting and Tolerating Faults in Hybrid Quantum Software Systems Using Architectural Redundancy}. In \bibinfo{booktitle}{\emph{{IEEE} International Conference on Quantum Software, {QSW} 2025, Helsinki, Finland, July 7-12, 2025}}, \bibfield{editor}{\bibinfo{person}{{Rong N.} Chang}, \bibinfo{person}{{Carl K.} Chang}, \bibinfo{person}{Jingwei Yang}, \bibinfo{person}{Nimanthi Atukorala}, \bibinfo{person}{Dan Chen}, \bibinfo{person}{Sumi Helal}, \bibinfo{person}{Sasu Tarkoma}, \bibinfo{person}{Qiang He}, \bibinfo{person}{Tevfik Kosar}, \bibinfo{person}{{Claudio A.} Ardagna}, \bibinfo{person}{Sebastian Feld}, \bibinfo{person}{Elisabetta {Di Nitto}}, {and} \bibinfo{person}{Manuel Wimmer}} (Eds.). \bibinfo{publisher}{{IEEE}}, \bibinfo{address}{Helsinki, Finland}, \bibinfo{pages}{162--172}.
\newblock
\href{https://doi.org/10.1109/QSW67625.2025.00028}{doi:\nolinkurl{10.1109/QSW67625.2025.00028}}


\bibitem[K\"{o}lle et~al\mbox{.}(2025)]%
        {MutationGASynthesis}
\bibfield{author}{\bibinfo{person}{Michael K\"{o}lle}, \bibinfo{person}{Tom Bintener}, \bibinfo{person}{Maximilian Zorn}, \bibinfo{person}{Gerhard Stenzel}, \bibinfo{person}{Leo S\"{u}nkel}, \bibinfo{person}{Thomas Gabor}, {and} \bibinfo{person}{Claudia Linnhoff-Popien}.} \bibinfo{year}{2025}\natexlab{}.
\newblock \showarticletitle{Evaluating Mutation Techniques in Genetic-Algorithm-Based Quantum Circuit Synthesis}. In \bibinfo{booktitle}{\emph{Proceedings of the Genetic and Evolutionary Computation Conference}} (NH Malaga Hotel, Malaga, Spain) \emph{(\bibinfo{series}{GECCO '25})}. \bibinfo{publisher}{Association for Computing Machinery}, \bibinfo{address}{New York, NY, USA}, \bibinfo{pages}{907--915}.
\newblock
\showISBNx{9798400714658}
\href{https://doi.org/10.1145/3712256.3726402}{doi:\nolinkurl{10.1145/3712256.3726402}}


\bibitem[Laurent et~al\mbox{.}(2017)]%
        {LaurentICST2017}
\bibfield{author}{\bibinfo{person}{Thomas Laurent}, \bibinfo{person}{Mike Papadakis}, \bibinfo{person}{Marinos Kintis}, \bibinfo{person}{Christopher Henard}, \bibinfo{person}{Yves~Le Traon}, {and} \bibinfo{person}{Anthony Ventresque}.} \bibinfo{year}{2017}\natexlab{}.
\newblock \showarticletitle{Assessing and Improving the Mutation Testing Practice of {PIT}}. In \bibinfo{booktitle}{\emph{2017 IEEE International Conference on Software Testing, Verification and Validation (ICST)}}. \bibinfo{pages}{430--435}.
\newblock
\href{https://doi.org/10.1109/ICST.2017.47}{doi:\nolinkurl{10.1109/ICST.2017.47}}


\bibitem[Leite~Ramalho et~al\mbox{.}(2025)]%
        {quantumTestingRoadmapTOSEM2025}
\bibfield{author}{\bibinfo{person}{Neilson~Carlos Leite~Ramalho}, \bibinfo{person}{Higor Amario~de Souza}, {and} \bibinfo{person}{Marcos Lordello~Chaim}.} \bibinfo{year}{2025}\natexlab{}.
\newblock \showarticletitle{Testing and Debugging Quantum Programs: The Road to 2030}.
\newblock \bibinfo{journal}{\emph{ACM Trans. Softw. Eng. Methodol.}} \bibinfo{volume}{34}, \bibinfo{number}{5}, Article \bibinfo{articleno}{155} (\bibinfo{date}{May} \bibinfo{year}{2025}), \bibinfo{numpages}{46}~pages.
\newblock
\showISSN{1049-331X}
\href{https://doi.org/10.1145/3715106}{doi:\nolinkurl{10.1145/3715106}}


\bibitem[Li et~al\mbox{.}(2020)]%
        {projectionBased}
\bibfield{author}{\bibinfo{person}{Gushu Li}, \bibinfo{person}{Li Zhou}, \bibinfo{person}{Nengkun Yu}, \bibinfo{person}{Yufei Ding}, \bibinfo{person}{Mingsheng Ying}, {and} \bibinfo{person}{Yuan Xie}.} \bibinfo{year}{2020}\natexlab{}.
\newblock \showarticletitle{Projection-based runtime assertions for testing and debugging Quantum programs}.
\newblock \bibinfo{journal}{\emph{Proc. ACM Program. Lang.}} \bibinfo{volume}{4}, \bibinfo{number}{OOPSLA}, Article \bibinfo{articleno}{150} (\bibinfo{date}{nov} \bibinfo{year}{2020}), \bibinfo{numpages}{29}~pages.
\newblock
\href{https://doi.org/10.1145/3428218}{doi:\nolinkurl{10.1145/3428218}}


\bibitem[Long and Zhao(2024)]%
        {LongTOSEM2024}
\bibfield{author}{\bibinfo{person}{Peixun Long} {and} \bibinfo{person}{Jianjun Zhao}.} \bibinfo{year}{2024}\natexlab{}.
\newblock \showarticletitle{Testing Multi-Subroutine Quantum Programs: From Unit Testing to Integration Testing}.
\newblock \bibinfo{journal}{\emph{ACM Trans. Softw. Eng. Methodol.}} \bibinfo{volume}{33}, \bibinfo{number}{6}, Article \bibinfo{articleno}{147} (\bibinfo{date}{June} \bibinfo{year}{2024}), \bibinfo{numpages}{61}~pages.
\newblock
\showISSN{1049-331X}
\href{https://doi.org/10.1145/3656339}{doi:\nolinkurl{10.1145/3656339}}


\bibitem[Meissel and Yao(2024)]%
        {meissel2024using}
\bibfield{author}{\bibinfo{person}{Kane Meissel} {and} \bibinfo{person}{Esther~S Yao}.} \bibinfo{year}{2024}\natexlab{}.
\newblock \showarticletitle{Using {Cliff’s} delta as a non-parametric effect size measure: an accessible web app and {R} tutorial}.
\newblock \bibinfo{journal}{\emph{Practical Assessment, Research, and Evaluation}} \bibinfo{volume}{29}, \bibinfo{number}{1} (\bibinfo{year}{2024}).
\newblock
\href{https://doi.org/10.7275/pare.1977}{doi:\nolinkurl{10.7275/pare.1977}}


\bibitem[Mendiluze~Usandizaga et~al\mbox{.}(2022)]%
        {Mendiluze2021}
\bibfield{author}{\bibinfo{person}{{Eñaut} Mendiluze~Usandizaga}, \bibinfo{person}{Shaukat Ali}, \bibinfo{person}{Paolo Arcaini}, {and} \bibinfo{person}{Tao Yue}.} \bibinfo{year}{2022}\natexlab{}.
\newblock \showarticletitle{Muskit: A Mutation Analysis Tool for Quantum Software Testing}. In \bibinfo{booktitle}{\emph{Proceedings of the 36th IEEE/ACM International Conference on Automated Software Engineering}} (Melbourne, Australia) \emph{(\bibinfo{series}{ASE '21})}. \bibinfo{publisher}{IEEE Press}, \bibinfo{pages}{1266--1270}.
\newblock
\showISBNx{9781665403375}
\href{https://doi.org/10.1109/ASE51524.2021.9678563}{doi:\nolinkurl{10.1109/ASE51524.2021.9678563}}


\bibitem[Mendiluze~Usandizaga et~al\mbox{.}(2025)]%
        {MendiluzeUsandizaga2025}
\bibfield{author}{\bibinfo{person}{E{\~{n}}aut Mendiluze~Usandizaga}, \bibinfo{person}{Shaukat Ali}, \bibinfo{person}{Tao Yue}, {and} \bibinfo{person}{Paolo Arcaini}.} \bibinfo{year}{2025}\natexlab{}.
\newblock \showarticletitle{Quantum circuit mutants: Empirical analysis and recommendations}.
\newblock \bibinfo{journal}{\emph{Empirical Software Engineering}} \bibinfo{volume}{30}, \bibinfo{number}{4} (\bibinfo{date}{16 Apr} \bibinfo{year}{2025}), \bibinfo{pages}{100}.
\newblock
\showISSN{1573-7616}
\href{https://doi.org/10.1007/s10664-025-10643-z}{doi:\nolinkurl{10.1007/s10664-025-10643-z}}


\bibitem[Mendiluze~Usandizaga et~al\mbox{.}(2026)]%
        {ZenodoRepository}
\bibfield{author}{\bibinfo{person}{Eñaut Mendiluze~Usandizaga}, \bibinfo{person}{Thomas Laurent}, \bibinfo{person}{Paolo Arcaini}, {and} \bibinfo{person}{Shaukat Ali}.} \bibinfo{year}{2026}\natexlab{}.
\newblock \bibinfo{title}{Supplementary material for the paper ``{Search-Based Generation of Undetected Quantum Circuit Mutants}''}.
\newblock
\urldef\tempurl%
\url{https://doi.org/10.5281/zenodo.21718539}
\showURL{%
\tempurl}


\bibitem[Muqeet et~al\mbox{.}(2024a)]%
        {pauliStringsASE2024}
\bibfield{author}{\bibinfo{person}{Asmar Muqeet}, \bibinfo{person}{Shaukat Ali}, {and} \bibinfo{person}{Paolo Arcaini}.} \bibinfo{year}{2024}\natexlab{a}.
\newblock \showarticletitle{Quantum Program Testing Through Commuting {Pauli} Strings on {IBM}'s Quantum Computers}. In \bibinfo{booktitle}{\emph{Proceedings of the 39th IEEE/ACM International Conference on Automated Software Engineering}} (Sacramento, CA, USA) \emph{(\bibinfo{series}{ASE '24})}. \bibinfo{publisher}{Association for Computing Machinery}, \bibinfo{address}{New York, NY, USA}, \bibinfo{pages}{2130--2141}.
\newblock
\showISBNx{9798400712487}
\href{https://doi.org/10.1145/3691620.3695275}{doi:\nolinkurl{10.1145/3691620.3695275}}


\bibitem[Muqeet et~al\mbox{.}(2024b)]%
        {muqeet2024mitigating}
\bibfield{author}{\bibinfo{person}{Asmar Muqeet}, \bibinfo{person}{Tao Yue}, \bibinfo{person}{Shaukat Ali}, {and} \bibinfo{person}{Paolo Arcaini}.} \bibinfo{year}{2024}\natexlab{b}.
\newblock \showarticletitle{Mitigating Noise in Quantum Software Testing Using Machine Learning}.
\newblock \bibinfo{journal}{\emph{IEEE Transactions on Software Engineering}} \bibinfo{volume}{50}, \bibinfo{number}{11} (\bibinfo{year}{2024}), \bibinfo{pages}{2947--2961}.
\newblock
\href{https://doi.org/10.1109/TSE.2024.3462974}{doi:\nolinkurl{10.1109/TSE.2024.3462974}}


\bibitem[Murillo et~al\mbox{.}(2025)]%
        {qseRoadmapTOSEM2025}
\bibfield{author}{\bibinfo{person}{Juan~Manuel Murillo}, \bibinfo{person}{Jose Garcia-Alonso}, \bibinfo{person}{Enrique Moguel}, \bibinfo{person}{Johanna Barzen}, \bibinfo{person}{Frank Leymann}, \bibinfo{person}{Shaukat Ali}, \bibinfo{person}{Tao Yue}, \bibinfo{person}{Paolo Arcaini}, \bibinfo{person}{Ricardo P\'{e}rez-Castillo}, \bibinfo{person}{Ignacio Garc\'{\i}a-Rodr\'{\i}guez~de Guzm\'{a}n}, \bibinfo{person}{Mario Piattini}, \bibinfo{person}{Antonio Ruiz-Cort\'{e}s}, \bibinfo{person}{Antonio Brogi}, \bibinfo{person}{Jianjun Zhao}, \bibinfo{person}{Andriy Miranskyy}, {and} \bibinfo{person}{Manuel Wimmer}.} \bibinfo{year}{2025}\natexlab{}.
\newblock \showarticletitle{Quantum Software Engineering: Roadmap and Challenges Ahead}.
\newblock \bibinfo{journal}{\emph{ACM Trans. Softw. Eng. Methodol.}} \bibinfo{volume}{34}, \bibinfo{number}{5}, Article \bibinfo{articleno}{154} (\bibinfo{date}{May} \bibinfo{year}{2025}), \bibinfo{numpages}{48}~pages.
\newblock
\showISSN{1049-331X}
\href{https://doi.org/10.1145/3712002}{doi:\nolinkurl{10.1145/3712002}}


\bibitem[Nielsen and Chuang(2016)]%
        {nielsen2010quantum}
\bibfield{author}{\bibinfo{person}{{Michael A.} Nielsen} {and} \bibinfo{person}{{Isaac L.} Chuang}.} \bibinfo{year}{2016}\natexlab{}.
\newblock \bibinfo{booktitle}{\emph{Quantum Computation and Quantum Information (10th Anniversary edition)}}.
\newblock \bibinfo{publisher}{Cambridge University Press}, \bibinfo{address}{Cambridge, UK}.
\newblock


\bibitem[Papadakis et~al\mbox{.}(2019)]%
        {MutationSurvey2}
\bibfield{author}{\bibinfo{person}{Mike Papadakis}, \bibinfo{person}{Marinos Kintis}, \bibinfo{person}{Jie Zhang}, \bibinfo{person}{Yue Jia}, \bibinfo{person}{Yves Le~Traon}, {and} \bibinfo{person}{Mark Harman}.} \bibinfo{year}{2019}\natexlab{}.
\newblock \showarticletitle{Chapter Six - Mutation Testing Advances: An Analysis and Survey}.
\newblock \bibinfo{series}{Advances in Computers}, Vol.~\bibinfo{volume}{112}. \bibinfo{publisher}{Elsevier}, \bibinfo{pages}{275--378}.
\newblock
\showISSN{0065-2458}
\href{https://doi.org/10.1016/bs.adcom.2018.03.015}{doi:\nolinkurl{10.1016/bs.adcom.2018.03.015}}


\bibitem[Papadakis and Malevris(2010)]%
        {empiricalfirstandsecondorder}
\bibfield{author}{\bibinfo{person}{Mike Papadakis} {and} \bibinfo{person}{Nicos Malevris}.} \bibinfo{year}{2010}\natexlab{}.
\newblock \showarticletitle{An Empirical Evaluation of the First and Second Order Mutation Testing Strategies}. In \bibinfo{booktitle}{\emph{2010 Third International Conference on Software Testing, Verification, and Validation Workshops}}. \bibinfo{pages}{90--99}.
\newblock
\href{https://doi.org/10.1109/ICSTW.2010.50}{doi:\nolinkurl{10.1109/ICSTW.2010.50}}


\bibitem[Pontolillo et~al\mbox{.}(2025)]%
        {pontolillo2025ideal}
\bibfield{author}{\bibinfo{person}{Gabriel Pontolillo}, \bibinfo{person}{Asmar Muqeet}, \bibinfo{person}{Shaukat Ali}, {and} \bibinfo{person}{Mohammad~Reza Mousavi}.} \bibinfo{year}{2025}\natexlab{}.
\newblock \showarticletitle{From Ideal to Noisy: Adapting Property-Based Testing for Real-World Noisy Quantum Computers}. In \bibinfo{booktitle}{\emph{2025 IEEE International Conference on Quantum Computing and Engineering (QCE)}}, Vol.~\bibinfo{volume}{01}. \bibinfo{pages}{405--416}.
\newblock
\href{https://doi.org/10.1109/QCE65121.2025.00053}{doi:\nolinkurl{10.1109/QCE65121.2025.00053}}


\bibitem[Quetschlich et~al\mbox{.}(2023)]%
        {quetschlich2023mqtbench}
\bibfield{author}{\bibinfo{person}{Nils Quetschlich}, \bibinfo{person}{Lukas Burgholzer}, {and} \bibinfo{person}{Robert Wille}.} \bibinfo{year}{2023}\natexlab{}.
\newblock \showarticletitle{{{MQT Bench}}: {Benchmarking Software and Design Automation Tools for Quantum Computing}}.
\newblock \bibinfo{journal}{\emph{{Quantum}}}  \bibinfo{volume}{7} (\bibinfo{year}{2023}), \bibinfo{pages}{1062}.
\newblock
\showeprint[arxiv]{2204.13719}
\href{https://doi.org/10.22331/q-2023-07-20-1062}{doi:\nolinkurl{10.22331/q-2023-07-20-1062}}
\newblock
\shownote{{{MQT Bench}} is available at \url{https://mqt-bench.app/}}.


\bibitem[R{\o}nnow et~al\mbox{.}(2014)]%
        {quantumSpeedup}
\bibfield{author}{\bibinfo{person}{Troels~F R{\o}nnow}, \bibinfo{person}{Zhihui Wang}, \bibinfo{person}{Joshua Job}, \bibinfo{person}{Sergio Boixo}, \bibinfo{person}{Sergei~V Isakov}, \bibinfo{person}{David Wecker}, \bibinfo{person}{John~M Martinis}, \bibinfo{person}{Daniel~A Lidar}, {and} \bibinfo{person}{Matthias Troyer}.} \bibinfo{year}{2014}\natexlab{}.
\newblock \showarticletitle{Defining and detecting quantum speedup}.
\newblock \bibinfo{journal}{\emph{Science}} \bibinfo{volume}{345}, \bibinfo{number}{6195} (\bibinfo{year}{2014}), \bibinfo{pages}{420--424}.
\newblock
\href{https://doi.org/10.1126/science.1252319}{doi:\nolinkurl{10.1126/science.1252319}}


\bibitem[Schuler and Zeller(2013)]%
        {schuler2013covering}
\bibfield{author}{\bibinfo{person}{David Schuler} {and} \bibinfo{person}{Andreas Zeller}.} \bibinfo{year}{2013}\natexlab{}.
\newblock \showarticletitle{Covering and uncovering equivalent mutants}.
\newblock \bibinfo{journal}{\emph{Software Testing, Verification and Reliability}} \bibinfo{volume}{23}, \bibinfo{number}{5} (\bibinfo{year}{2013}), \bibinfo{pages}{353--374}.
\newblock


\bibitem[Schwarz et~al\mbox{.}(2011)]%
        {schwarz2011breeding}
\bibfield{author}{\bibinfo{person}{Birgit Schwarz}, \bibinfo{person}{David Schuler}, {and} \bibinfo{person}{Andreas Zeller}.} \bibinfo{year}{2011}\natexlab{}.
\newblock \showarticletitle{Breeding High-Impact Mutations}. In \bibinfo{booktitle}{\emph{2011 IEEE Fourth International Conference on Software Testing, Verification and Validation Workshops}}. \bibinfo{pages}{382--387}.
\newblock
\href{https://doi.org/10.1109/ICSTW.2011.56}{doi:\nolinkurl{10.1109/ICSTW.2011.56}}


\bibitem[Silva et~al\mbox{.}(2017)]%
        {SILVA201719}
\bibfield{author}{\bibinfo{person}{Rodolfo~Adamshuk Silva}, \bibinfo{person}{Simone do Rocio~{Senger de Souza}}, {and} \bibinfo{person}{Paulo~Sérgio {Lopes de Souza}}.} \bibinfo{year}{2017}\natexlab{}.
\newblock \showarticletitle{A systematic review on search based mutation testing}.
\newblock \bibinfo{journal}{\emph{Information and Software Technology}}  \bibinfo{volume}{81} (\bibinfo{year}{2017}), \bibinfo{pages}{19--35}.
\newblock
\showISSN{0950-5849}
\href{https://doi.org/10.1016/j.infsof.2016.01.017}{doi:\nolinkurl{10.1016/j.infsof.2016.01.017}}


\bibitem[Sánchez et~al\mbox{.}(2024)]%
        {mutationtestingpractice}
\bibfield{author}{\bibinfo{person}{Ana~B. Sánchez}, \bibinfo{person}{José~A. Parejo}, \bibinfo{person}{Sergio Segura}, \bibinfo{person}{Amador Durán}, {and} \bibinfo{person}{Mike Papadakis}.} \bibinfo{year}{2024}\natexlab{}.
\newblock \showarticletitle{Mutation Testing in Practice: Insights From Open-Source Software Developers}.
\newblock \bibinfo{journal}{\emph{IEEE Transactions on Software Engineering}} \bibinfo{volume}{50}, \bibinfo{number}{5} (\bibinfo{year}{2024}), \bibinfo{pages}{1130--1143}.
\newblock
\href{https://doi.org/10.1109/TSE.2024.3377378}{doi:\nolinkurl{10.1109/TSE.2024.3377378}}


\bibitem[Usandizaga et~al\mbox{.}(2026)]%
        {quantumRepICST2026}
\bibfield{author}{\bibinfo{person}{Eñaut~Mendiluze Usandizaga}, \bibinfo{person}{Thomas Laurent}, \bibinfo{person}{Paolo Arcaini}, {and} \bibinfo{person}{Shaukat Ali}.} \bibinfo{year}{2026}\natexlab{}.
\newblock \showarticletitle{Quantum Circuit Repair by Gate Prioritisation}. In \bibinfo{booktitle}{\emph{2026 IEEE International Conference on Software Testing, Verification and Validation (ICST)}}. \bibinfo{pages}{744--748}.
\newblock
\href{https://doi.org/10.1109/ICST69053.2026.00095}{doi:\nolinkurl{10.1109/ICST69053.2026.00095}}


\bibitem[Visser(2016)]%
        {whathard}
\bibfield{author}{\bibinfo{person}{Willem Visser}.} \bibinfo{year}{2016}\natexlab{}.
\newblock \showarticletitle{What makes killing a mutant hard}. In \bibinfo{booktitle}{\emph{Proceedings of the 31st IEEE/ACM International Conference on Automated Software Engineering}} (Singapore, Singapore) \emph{(\bibinfo{series}{ASE '16})}. \bibinfo{publisher}{Association for Computing Machinery}, \bibinfo{address}{New York, NY, USA}, \bibinfo{pages}{39--44}.
\newblock
\showISBNx{9781450338455}
\href{https://doi.org/10.1145/2970276.2970345}{doi:\nolinkurl{10.1145/2970276.2970345}}


\bibitem[Wang et~al\mbox{.}(2021c)]%
        {WangICST2021}
\bibfield{author}{\bibinfo{person}{Jiyuan Wang}, \bibinfo{person}{Fucheng Ma}, {and} \bibinfo{person}{Yu Jiang}.} \bibinfo{year}{2021}\natexlab{c}.
\newblock \showarticletitle{Poster: {Fuzz} Testing of Quantum Program}. In \bibinfo{booktitle}{\emph{2021 14th IEEE Conference on Software Testing, Verification and Validation (ICST)}}. \bibinfo{pages}{466--469}.
\newblock
\href{https://doi.org/10.1109/ICST49551.2021.00061}{doi:\nolinkurl{10.1109/ICST49551.2021.00061}}


\bibitem[Wang et~al\mbox{.}(2025)]%
        {XinyiLanscape}
\bibfield{author}{\bibinfo{person}{Xinyi Wang}, \bibinfo{person}{Shaukat Ali}, {and} \bibinfo{person}{Davide Taibi}.} \bibinfo{year}{2025}\natexlab{}.
\newblock \showarticletitle{The Landscape of Quantum Software Testing Tools}.
\newblock \bibinfo{journal}{\emph{IEEE Software}} \bibinfo{volume}{42}, \bibinfo{number}{5} (\bibinfo{year}{2025}), \bibinfo{pages}{136--140}.
\newblock
\href{https://doi.org/10.1109/MS.2025.3578154}{doi:\nolinkurl{10.1109/MS.2025.3578154}}


\bibitem[Wang et~al\mbox{.}(2021a)]%
        {CTQuantumQRS2021}
\bibfield{author}{\bibinfo{person}{Xinyi Wang}, \bibinfo{person}{Paolo Arcaini}, \bibinfo{person}{Tao Yue}, {and} \bibinfo{person}{Shaukat Ali}.} \bibinfo{year}{2021}\natexlab{a}.
\newblock \showarticletitle{Application of Combinatorial Testing to Quantum Programs}. In \bibinfo{booktitle}{\emph{2021 IEEE 21st International Conference on Software Quality, Reliability and Security (QRS)}}. \bibinfo{pages}{179--188}.
\newblock
\href{https://doi.org/10.1109/QRS54544.2021.00029}{doi:\nolinkurl{10.1109/QRS54544.2021.00029}}


\bibitem[Wang et~al\mbox{.}(2021b)]%
        {genTestsQPSSBSE2021}
\bibfield{author}{\bibinfo{person}{Xinyi Wang}, \bibinfo{person}{Paolo Arcaini}, \bibinfo{person}{Tao Yue}, {and} \bibinfo{person}{Shaukat Ali}.} \bibinfo{year}{2021}\natexlab{b}.
\newblock \showarticletitle{Generating Failing Test Suites for Quantum Programs With Search}. In \bibinfo{booktitle}{\emph{Search-Based Software Engineering}}, \bibfield{editor}{\bibinfo{person}{Una-May O'Reilly} {and} \bibinfo{person}{Xavier Devroey}} (Eds.). \bibinfo{publisher}{Springer International Publishing}, \bibinfo{address}{Cham}, \bibinfo{pages}{9--25}.
\newblock
\showISBNx{978-3-030-88106-1}
\href{https://doi.org/10.1007/978-3-030-88106-1\_2}{doi:\nolinkurl{10.1007/978-3-030-88106-1\_2}}


\bibitem[Wang et~al\mbox{.}(2022a)]%
        {quitoASE21tool}
\bibfield{author}{\bibinfo{person}{Xinyi Wang}, \bibinfo{person}{Paolo Arcaini}, \bibinfo{person}{Tao Yue}, {and} \bibinfo{person}{Shaukat Ali}.} \bibinfo{year}{2022}\natexlab{a}.
\newblock \showarticletitle{Quito: A Coverage-Guided Test Generator for Quantum Programs}. In \bibinfo{booktitle}{\emph{Proceedings of the 36th IEEE/ACM International Conference on Automated Software Engineering}} (Melbourne, Australia) \emph{(\bibinfo{series}{ASE '21})}. \bibinfo{publisher}{IEEE Press}, \bibinfo{pages}{1237--1241}.
\newblock
\showISBNx{9781665403375}
\href{https://doi.org/10.1109/ASE51524.2021.9678798}{doi:\nolinkurl{10.1109/ASE51524.2021.9678798}}


\bibitem[Wang et~al\mbox{.}(2024)]%
        {qucatASE23tool}
\bibfield{author}{\bibinfo{person}{Xinyi Wang}, \bibinfo{person}{Paolo Arcaini}, \bibinfo{person}{Tao Yue}, {and} \bibinfo{person}{Shaukat Ali}.} \bibinfo{year}{2024}\natexlab{}.
\newblock \showarticletitle{{QuCAT}: A Combinatorial Testing Tool for Quantum Software}. In \bibinfo{booktitle}{\emph{Proceedings of the 38th IEEE/ACM International Conference on Automated Software Engineering}} (Echternach, Luxembourg) \emph{(\bibinfo{series}{ASE '23})}. \bibinfo{publisher}{IEEE Press}, \bibinfo{pages}{2066--2069}.
\newblock
\showISBNx{9798350329964}
\href{https://doi.org/10.1109/ASE56229.2023.00062}{doi:\nolinkurl{10.1109/ASE56229.2023.00062}}


\bibitem[Wang et~al\mbox{.}(2022b)]%
        {mutation-based}
\bibfield{author}{\bibinfo{person}{Xinyi Wang}, \bibinfo{person}{Tongxuan Yu}, \bibinfo{person}{Paolo Arcaini}, \bibinfo{person}{Tao Yue}, {and} \bibinfo{person}{Shaukat Ali}.} \bibinfo{year}{2022}\natexlab{b}.
\newblock \showarticletitle{Mutation-Based Test Generation for Quantum Programs with Multi-Objective Search}. In \bibinfo{booktitle}{\emph{Proceedings of the Genetic and Evolutionary Computation Conference}} (Boston, Massachusetts) \emph{(\bibinfo{series}{GECCO '22})}. \bibinfo{publisher}{Association for Computing Machinery}, \bibinfo{address}{New York, NY, USA}, \bibinfo{pages}{1345--1353}.
\newblock
\showISBNx{9781450392372}
\href{https://doi.org/10.1145/3512290.3528869}{doi:\nolinkurl{10.1145/3512290.3528869}}


\bibitem[Wilcoxon(1945)]%
        {wilcoxon}
\bibfield{author}{\bibinfo{person}{Frank Wilcoxon}.} \bibinfo{year}{1945}\natexlab{}.
\newblock \showarticletitle{Individual Comparisons by Ranking Methods}.
\newblock \bibinfo{journal}{\emph{Biometrics Bulletin}} \bibinfo{volume}{1}, \bibinfo{number}{6} (\bibinfo{year}{1945}), \bibinfo{pages}{80--83}.
\newblock
\showISSN{00994987}
\urldef\tempurl%
\url{http://www.jstor.org/stable/3001968}
\showURL{%
\tempurl}


\bibitem[Woodward(1993)]%
        {MutationOrigin}
\bibfield{author}{\bibinfo{person}{M.R. Woodward}.} \bibinfo{year}{1993}\natexlab{}.
\newblock \showarticletitle{Mutation testing—its origin and evolution}.
\newblock \bibinfo{journal}{\emph{Information and Software Technology}} \bibinfo{volume}{35}, \bibinfo{number}{3} (\bibinfo{year}{1993}), \bibinfo{pages}{163--169}.
\newblock
\showISSN{0950-5849}
\href{https://doi.org/10.1016/0950-5849(93)90053-6}{doi:\nolinkurl{10.1016/0950-5849(93)90053-6}}


\bibitem[Yanofsky and Mannucci(2008)]%
        {yanofsky2008quantum}
\bibfield{author}{\bibinfo{person}{Noson~S Yanofsky} {and} \bibinfo{person}{Mirco~A Mannucci}.} \bibinfo{year}{2008}\natexlab{}.
\newblock \bibinfo{booktitle}{\emph{Quantum computing for computer scientists}}.
\newblock \bibinfo{publisher}{Cambridge University Press}.
\newblock


\bibitem[Ye et~al\mbox{.}(2024)]%
        {QuraTestASE2023}
\bibfield{author}{\bibinfo{person}{Jiaming Ye}, \bibinfo{person}{Shangzhou Xia}, \bibinfo{person}{Fuyuan Zhang}, \bibinfo{person}{Paolo Arcaini}, \bibinfo{person}{Lei Ma}, \bibinfo{person}{Jianjun Zhao}, {and} \bibinfo{person}{Fuyuki Ishikawa}.} \bibinfo{year}{2024}\natexlab{}.
\newblock \showarticletitle{{QuraTest}: Integrating Quantum Specific Features in Quantum Program Testing}. In \bibinfo{booktitle}{\emph{Proceedings of the 38th IEEE/ACM International Conference on Automated Software Engineering}} (Echternach, Luxembourg) \emph{(\bibinfo{series}{ASE '23})}. \bibinfo{publisher}{IEEE Press}, \bibinfo{pages}{1149--1161}.
\newblock
\showISBNx{9798350329964}
\href{https://doi.org/10.1109/ASE56229.2023.00196}{doi:\nolinkurl{10.1109/ASE56229.2023.00196}}


\bibitem[Zhao(2020)]%
        {zhao2021LandscapesAndHorizons}
\bibfield{author}{\bibinfo{person}{Jianjun Zhao}.} \bibinfo{year}{2020}\natexlab{}.
\newblock \showarticletitle{Quantum Software Engineering: Landscapes and Horizons}.
\newblock \bibinfo{journal}{\emph{CoRR}}  \bibinfo{volume}{abs/2007.07047} (\bibinfo{year}{2020}).
\newblock
\showeprint[arXiv]{2007.07047}
\urldef\tempurl%
\url{https://arxiv.org/abs/2007.07047}
\showURL{%
\tempurl}


\bibitem[Zhao et~al\mbox{.}(2021)]%
        {zhao2021bugs4q}
\bibfield{author}{\bibinfo{person}{Pengzhan Zhao}, \bibinfo{person}{Jianjun Zhao}, \bibinfo{person}{Zhongtao Miao}, {and} \bibinfo{person}{Shuhan Lan}.} \bibinfo{year}{2021}\natexlab{}.
\newblock \showarticletitle{{Bugs4Q}: A Benchmark of Real Bugs for Quantum Programs}. In \bibinfo{booktitle}{\emph{2021 36th IEEE/ACM International Conference on Automated Software Engineering (ASE)}}. \bibinfo{pages}{1373--1376}.
\newblock
\href{https://doi.org/10.1109/ASE51524.2021.9678908}{doi:\nolinkurl{10.1109/ASE51524.2021.9678908}}


\end{thebibliography}

\end{document}